\documentclass[12pt,a4paper]{article}

\usepackage{jheppub}
\graphicspath{{Img/}}

\usepackage{color}
\usepackage[dvipsnames, svgnames, x11names]{xcolor}
\usepackage[utf8]{inputenc}
\usepackage{float}
\usepackage{graphicx}
\usepackage{setspace}
\usepackage{subfigure}
\usepackage{bm}
\usepackage{bbm}
\usepackage{booktabs}
\usepackage{tikz}
\usepackage{amsmath}
\usepackage{amssymb}
\usepackage{amsbsy}
\usepackage{amsthm}
\usepackage{amsfonts}
\usepackage{dsfont}
\usepackage{braket}
\usepackage{multirow}
\usepackage{ulem}
\usepackage{slashed}
\usepackage{tikzducks}
\usepackage{dsfont}
\usepackage{csquotes}
\usepackage{array}
\usepackage[compat=1.1.0]{tikz-feynman}

\usepackage{enumitem}
\newlist{senum}{enumerate}{1} 
\setlist[senum]{leftmargin=4.5cm ,align=left, label=\textbf{\Roman*}}

\def\nn{\nonumber}
\def\Tr{\text{Tr}}

\def\bra#1{\left\langle#1\right|}
\def\ket#1{\left|#1\right\rangle}

\def\abs#1{\left|#1\right|}

\def\be{\begin{equation}}       \def\ee{\end{equation}}
\def\bea{\begin{eqnarray}}      \def\eea{\end{eqnarray}}
\def\ba{\begin{array}}
	\def\ea{\end{array}}
\def\bnum{\begin{enumerate} }
	\def\enum{\end{enumerate}}

\def\nn{\nonumber}

\def\=>{\Rightarrow}
\def\>{\rightarrow}

\def\eye2{Fathbb{I}}

\def\Tr{\mathrm{Tr}}

\begin{document}
\begin{flushright}
  YITP-26-61
\end{flushright}

\title{Complexity Inequalities for Quantum Subsystems}

\author{Pawel Caputa,$^{1,2,3}$}

\author{Giuseppe Di Giulio,$^{1}$}

\author{Tran Quang Loc$^{1}$}

\affiliation[1]{
The Oskar Klein Centre and Department of Physics, Stockholm University, AlbaNova, 106 91 Stockholm, Sweden}
\affiliation[2]{Yukawa Institute for Theoretical Physics, Kyoto University, Kitashirakawa Oiwakecho, Sakyo-ku, Kyoto 606-8502, Japan}
\affiliation[3]{Faculty of Physics, University of Warsaw, Pasteura 5, 02-093 Warsaw, Poland}

\abstract{Motivated by the role of the holographic entropy cone in constraining the entanglement structure of states with classical gravitational duals, we investigate combinations of subsystem complexities associated with reduced density matrices in multipartite quantum systems. Focusing on subsystems composed of three disjoint regions, we introduce two quantities: a tripartite complexity, inspired by the tripartite information, and a complexity gap, designed to characterize emergent complexity in the full quantum state beyond that of its constituents.
We study the sign structure of these quantities in three selected approaches to subsystem complexity. In holography, we employ the complexity=volume proposal in AdS spacetimes; for Gaussian many-body states, we use Fisher–Rao subsystem complexity; and we further develop a Krylov-space inspired, effective framework for reduced density matrices, which we test in few-qubit systems and coherent-state dynamics.
Across all three approaches, we find that the tripartite complexity is not sign-definite in general. By contrast, the complexity gap exhibits a definite sign in every example we analyze, although the sign itself depends on the underlying notion of subsystem complexity. Our results suggest that the complexity gap could be a natural candidate building block for a prospective hierarchy of subsystem complexity inequalities.}

\sloppy
\maketitle

\newpage

\section{Introduction}

Complexity is deeply connected to emergence: the appearance of collective phenomena whose behaviour cannot be reduced to a simple sum of microscopic constituents. This principle lies at the heart of modern physics, from quantum many-body systems and critical phenomena to spacetime itself in holography \cite{Maldacena:1997re}. In strongly interacting quantum systems, entanglement and dynamics generate new effective degrees of freedom, phases of matter, and collective geometrical descriptions. Yet despite its central role across these areas, quantum complexity still lacks a precise and universally accepted definition. Existing approaches, including circuit complexity, geometric formulations à la Nielsen, Krylov and operator complexity, and algorithmic notions such as Kolmogorov complexity, capture different aspects of quantum structure and dynamics (see review \cite{Baiguera:2025dkc} and references therein), but no unified framework has emerged, especially for subsystems and mixed states.

This ambiguity becomes particularly important in the context of AdS/CFT, one of the most prominent examples of emergence in theoretical physics. In holography, a higher-dimensional gravitational spacetime with black holes emerges from a lower-dimensional quantum field theory without gravity, suggesting that geometry itself is encoded in quantum information. A major milestone in this direction was the work of Ryu and Takayanagi \cite{Ryu:2006bv} (RT), which identified entanglement entropy as a geometric observable and inaugurated a fruitful program exploring the relationship between quantum information and spacetime structure. Over the past two decades, entanglement-based methods have revealed deep constraints on holographic states, starting from strong sub-additivity \cite{Headrick:2007km} through monogamy of holographic information \cite{Hayden:2011ag} and constraints on multipartite entanglement \cite{Balasubramanian:2025hxg,Balasubramanian:2026chr}, to the Markov gap \cite{Hayden:2021gno},  contributing to the holographic entropy cone program \cite{Bao:2015bfa,Hubeny:2018trv,HernandezCuenca:2019wgh,Grado-White:2024gtx,Grimaldi:2025jad,Czech:2023xed,Bao:2024azn,Czech:2025jnw,Naskar:2026zka,Ju:2026zbu}, where multipartite entanglement inequalities follow from geometric properties of RT surfaces. Nevertheless, entanglement entropy alone appears insufficient to fully characterise holographic emergence, particularly the encoding of black hole interiors and late-time gravitational dynamics.

One of the original motivations for holographic complexity was precisely the expectation that complexity probes aspects of bulk geometry inaccessible to entanglement \cite{Susskind:2014moa}. In particular, suitable notions of complexity are expected to track the growth of Einstein–Rosen bridges and the structure of black hole interiors over timescales far beyond entanglement saturation. So far, strong evidence for this picture was provided in lower-dimensional holographic models, such as double-scaled Sachdev–Ye–Kitaev (SYK) and Jackiw–Teitelboim (JT) gravity \cite{Lin:2022rbf,Rabinovici:2023yex}. At the same time, the broader search for a quantum notion of complexity has driven significant developments in quantum many-body physics and quantum information theory \cite{Baiguera:2025dkc}. Concepts such as operator growth, Spread/Krylov complexity, geometric circuit complexity, and information-theoretic measures now provide new probes of scrambling, chaos, and non-equilibrium dynamics. Together, these developments suggest that complexity captures structural and dynamical features of quantum systems fundamentally distinct from those probed by entanglement entropy. 

Clearly, a central challenge in this program is the formulation of subsystem complexity. Since entanglement measures are intrinsically defined for subregions, any meaningful comparison between entanglement and complexity requires a corresponding notion of complexity for reduced density matrices. Such a framework could clarify which information about a quantum system is visible to entanglement measures and which requires more refined complexity-based tools. In holography, this question is directly tied to the relation between entanglement wedges, bulk reconstruction, and black hole interiors, and may ultimately require extending the standard thermodynamic paradigm underlying the Ryu–Takayanagi formula toward genuinely dynamical notions tight to complexity. Motivated by these questions, in this work we develop and systematically compare several approaches to subsystem complexity from the complementary perspectives of holography, quantum many-body physics, and geometric complexity theory. Our broader goal is to put constraints on quantum subsystem complexity valid only for holographic states. Progress in this direction was already initiated in \cite{Haah:2025hyf,Fan:2025moc}, and here we aim to build upon these developments from a complementary perspective.

This paper is organised as follows. In Sec.~\ref{sec:Preliminaries}, we discuss combinations of subsystem complexities associated with subregions composed of disjoint components, which can be used to characterize complexity in multipartite settings. Focusing on subsystems made of three disjoint components, we introduce the notions of {\it tripartite complexity} $M_3$ and {\it complexity gap} $\Delta^{(3)}_C$, which quantify the extent to which the complexity of a composite quantum system differs from the combined complexities of its constituent subsystems. Finally, we apply these measures to static holographic setups at the end of this section. In Sec.~\ref{sec:subsystemKrylov}, we introduce a generalization of the moment recursion method tailored to subsystem dynamics. This construction allows us to define a notion of Krylov subsystem complexity, which we test in examples including few-qubit systems and coherent-state dynamics.
In Sec.~\ref{sec:SRSpreadComplexity}, we employ the Fisher–Rao subsystem complexity introduced in \cite{DiGiulio:2020hlz} to investigate the dynamics of mutual complexity, tripartite complexity, and the complexity gap in systems of coupled harmonic oscillators. In Sec.~\ref{sec:Conclusions}, we discuss our findings and outline several directions for future research. Additional computational details supporting the results presented in the main text are provided in Appendices~\ref{app:globalAdS3}, \ref{app:RDMevolution}, \ref{app:subsystemKrylovdetails} and \ref{app:QuasiParticle}.

While in preparation we became aware of related work \cite{Fujiki:2026ucr} where a similar quantity to our complexity gap is discussed from the perspective of multipartite entanglement. We thank Tadashi Takayanagi and Nicolò Zenoni for sharing their draft, discussions and comments on relations between our works.

\section{A quest for subsystem complexity}\label{sec:Preliminaries}

In this section, we review some of the measures of multipartite correlations in quantum systems, such as mutual and tripartite information. Inspired by these quantities and by the role they play in holography, we introduce analogous quantities involving subsystem complexity. We then discuss their properties as computed using the {\it complexity=volume} (CV) holographic subsystem complexity proposal \cite{Alishahiha:2015rta}. This analysis serves as guidance for our quest for a definition of subsystem complexity in many-body quantum systems.

\subsection{From entanglement entropy to complexity inequalities}\label{subsec:Inequalities}

The entanglement structure of holographic conformal field theories (CFTs) is constrained by a family of quantities, constructed from von Neumann entropies involving several subsystems, that possess definite signs \cite{Bao:2015bfa}. The resulting inequalities define the so-called holographic entropy cone, which characterizes the entanglement patterns compatible with the existence of a smooth classical dual spacetime. Inspired by this analysis, one may wonder whether subsystem complexity, being associated with bulk volumes rather than areas, satisfies a similar set of inequalities, and whether these can help constrain the properties that a definition of subsystem complexity should satisfy in the boundary theory.
This discussion generalizes the study of super- and subadditivity properties of existing subsystem complexity proposals \cite{Alishahiha:2015rta,Carmi:2016wjl,Agon:2018zso,Camargo:2018eof} to complexity-based quantities defined for subregions composed of multiple disjoint components.

Consider a system bipartite into $A\cup B$,
with a factorization of the corresponding Hilbert spaces, i.e. $\mathcal{H}=\mathcal{H}_A\otimes\mathcal{H}_B$.
Given a quantum state $\ket{\psi}\in \mathcal{H}$, a measure of quantum correlations between $A$ and $B$ is given by the entanglement entropy $S_A=-\Tr\left(\rho_A\ln\rho_A\right)$, where the reduced density matrix $\rho_A$ is given by the partial trace of the full density matrix over the subsystem $B$, i.e. $\rho_A= \Tr_B (\ket{\psi}\bra{\psi}) $.

We assume that the subsystem $A$ is further divided into $A=A_1\cup A_2$ and that $\mathcal{H}_A=\mathcal{H}_{A_1}\otimes\mathcal{H}_{A_2}$. The mutual information, which quantifies the correlations shared between $A_1$ and $A_2$, is defined as
\be
\label{eq:mutualinfo}
I(A_1:A_2)\equiv S_{A_1}+S_{A_2}-S_{A_1\cup A_2}\,,
\qquad 
I(A_1:A_2)\geq 0
\,.
\ee
Now, we consider $A$ to be made up of three subsystems, i.e. $A=A_1\cup A_2\cup A_3$, such that $\mathcal{H}_A=\mathcal{H}_{A_1}\otimes\mathcal{H}_{A_2}\otimes\mathcal{H}_{A_3}$, and we define the tripartite information as
\bea
\label{eq:tripartiteinfo}
I_3(A_1:A_2:A_3)&\equiv& S_{A_1}+S_{A_2}+S_{A_3}-S_{A_1\cup A_2}-S_{A_1\cup A_3}-S_{A_2\cup A_3}+S_{A_1\cup A_2\cup A_3}\nn\\
&=&I(A_1:A_2)+I(A_1:A_3)-I(A_1:A_2\cup A_3)\,.
\eea
This quantity measures how much information $A_1$ shares with $A_2$ and $A_3$ jointly versus separately.
In general, the tripartite mutual information does not have a definite sign. However, it was shown that $I_3(A_1:A_2:A_3)\leq 0$ in holographic CFTs \cite{Hayden:2011ag}.
In this case, having access to $A_2$ and $A_3$ separately would not give us as much information as having access to them jointly. If this is the case, the mutual information is said to be monogamous.

In most equilibrium and out-of-equilibrium settings, the tripartite information has been found to be negative. These include two-dimensional free Dirac QFTs in their ground state \cite{Casini:2008wt}, quench of free fermionic models \cite{Maric:2020dpw,Maric:2022rsc,Caceffo:2023hns}, and time-dependent holographic scenarios \cite{Allais:2011ys,Balasubramanian:2011at,Asadi:2018ijf}. On the other hand, in a few cases, including the two-dimensional free boson QFT with large compactification radius \cite{Hayden:2011ag}, and for higher dimensional CFTs \cite{Agon:2021lus}, the tripartite information is positive.
Interestingly, there are cases where the sign of $I_3(A_1:A_2:A_3)$ changes across the time evolution: this happens in holographic scenarios, but only if the null energy condition is violated \cite{Allais:2011ys}.

In the spirit of extending similar considerations to quantum complexity, one must first confront the difficulty of defining a notion of subsystem complexity, namely a notion of complexity associated with the reduced density matrix of a given subregion of a system. This problem shares many of the ambiguities already present in the definition of pure state complexity, but is intrinsically more challenging since it requires dealing with the structure of quantum mixed states. Indeed, when thinking of constructing a mixed quantum state through fundamental operations, the set of unitary gates alone is no longer sufficient, and more general quantum operations must also be considered. For this reason, only a few definitions of subsystem complexity have been proposed so far, and none of them has clearly emerged as the preferred framework \cite{Caceres:2019pgf,Ruan:2020vze,DiGiulio:2020hlz}.

In the following, we assume the existence of a generic notion of subsystem complexity: given a subsystem $A$, we denote by $C_A$ the subsystem complexity associated with that region. We now introduce a set of quantities that may prove useful for investigating and comparing different notions of subsystem complexity and, later in the manuscript, we test these ideas to several concrete proposals.

Paralleling the setups discussed above for the entanglement measures, we consider a subsystem $A$ divided into two parties $A_1$ and $A_2$ and we define the mutual complexity \cite{Alishahiha:2018lfv, Caceres:2019pgf}
\be
\label{eq:mutualcomplexity}
M(A_1:A_2)\equiv C_{A_1}+C_{A_2}-C_{A_1\cup A_2}
\,.
\ee
This quantity has been analyzed for several holographic proposals of subsystem complexity. It was found that $M(A_1:A_2)$ is always negative for the CV, whereas it does not possess a definite sign for the complexity=action (CA) proposal. \cite{Carmi:2016wjl,Agon:2018zso,Alishahiha:2018lfv,Auzzi:2019vyh}.

For a configuration where $A$ is divided into $A_1$, $A_2$, and $A_3$, it is insightful to introduce the {\it tripartite complexity}
\bea
\label{eq:tripartitecomplexity}
M_3(A_1:A_2:A_3)&\equiv& C_{A_1}+C_{A_2}+C_{A_3}-C_{A_1\cup A_2}-C_{A_1\cup A_3}-C_{A_2\cup A_3}+C_{A_1\cup A_2\cup A_3}\nn\\
&=&M(A_1:A_2)+M(A_1:A_3)-M(A_1:A_2\cup A_3)\,.
\eea
This quantity, defined in analogy with the tripartite information \eqref{eq:tripartiteinfo}, indicates whether the complexity of a state restricted to a tripartition is predominantly determined by the mutual complexity of pairs of subsystems or by the contribution associated with all three subsystems simultaneously.

Another complexity probe for a region composed of three subsystems can be defined as
\begin{equation}
\label{eq:Mtilde}
\Delta^{(3)}_C(A_1:A_2:A_3)\equiv C_{A_1\cup A_2\cup A_3}- C_{A_1}-C_{A_2}-C_{A_3}\,,
\end{equation}
and it compares the complexity of the individual subsystems with that of the full region $A$ in a manner different from \eqref{eq:tripartitecomplexity}. We call this quantity {\it complexity gap} in the rest of this manuscript. More generally, given a subsystem with $n$ components $A=A_1\cup\dots\cup A_n$, we can define the $n$-partite complexity gap as
 \be
\Delta^{(n)}_C(A_1:\dots:A_n)\equiv C_{A_1\cup\dots\cup A_n}-\sum^n_{i=1}C_{A_i}\,,
 \ee
 but we will only need the $n=3$ version. By definition, complexity gap is closely tied to the intuitions behind emergence, where a non-trivial phenomena in total state emerge from simpler subsystems, and we hope to quantify this below.
 Investigations in a similar spirit have been carried out in the context of unitary dynamics in Krylov space \cite{Caputa:2025mii,Caputa:2025ozd}. In those works, the central question was how the complexity of states and operators emerges from their simpler constituents, given their fixed-charge components.
 
In the spirit of \cite{Hubeny:2018ijt,He:2020xuo}, one could say that a putative inequality involving only $\Delta^{(3)}_C(A_1:A_2:A_3)$ would be {\it balanced}, in the sense that the UV divergences in the complexities of disjoint subsystems appearing in \eqref{eq:Mtilde} cancel exactly. On the other hand, such an inequality would not be {\it superbalanced}, since this cancellation no longer occurs when considering the purification of the inequality\footnote{We thank Dimitrios Patramanis for discussions on this point.}.
Although superbalanced inequalities play a crucial role in holographic entanglement, it is not obvious whether the same is true for complexity inequalities. It would be interesting to revisit this issue in future investigations.

 To the best of our knowledge, the quantities \eqref{eq:tripartitecomplexity} and \eqref{eq:Mtilde} have not been considered previously in the literature. Our goal is to understand whether these quantities, together with \eqref{eq:mutualcomplexity}, exhibit characteristic properties when evaluated in holographic and many-body frameworks for subsystem complexity. In particular, we ask whether they satisfy universal constraints, such as definite-sign inequalities. Establishing such relations would provide a first set of candidate inequalities for a {\it subsystem complexity cone}, taking an initial step toward a more systematic characterization of quantum subsystem complexity.

\subsection{Insights from holography}\label{subsec:holography}

In this subsection, we discuss some properties of the holographic CV proposal for subsystem complexity in both static and time-dependent spacetimes. We also evaluate the mutual complexity \eqref{eq:mutualcomplexity}, the tripartite complexity \eqref{eq:tripartitecomplexity}, and the complexity gap \eqref{eq:Mtilde} in examples of a static spacetimes.

\subsubsection{Static spacetimes}\label{subsec:static}

In time-independent holographic scenarios, the entanglement entropy of a spatial region $A$ of the boundary CFT is described in the gravitational AdS bulk by the Ryu–Takayanagi (RT) formula \cite{Ryu:2006bv}
\begin{equation}
\label{eq:RTformula}
S_A = \frac{\text{Area}(\gamma_A)}{4G_N}\,,
\end{equation}
where $\gamma_A$ is the minimal-area (RT) surface in the bulk anchored on the boundary of $A$ and homologous to $A$, and $G_N$ is Newton’s constant.

One of the most promising bulk proposals for the holographic complexity of a CFT pure state is proportional to the regularized volume of AdS slices anchored to the boundary Cauchy slice on which the state is defined \cite{Stanford:2014jda,Susskind:2014jwa,Belin:2021bga}. Instead, the information about boundary reduced density matrices is encoded, from the bulk perspective, in the entanglement wedge, which is bounded by the RT surface \cite{Czech:2012bh}.
Thus, it was proposed that the holographic subsystem complexity associated with a spatial boundary subsystem $A$ is proportional to the volume $V(\gamma_A)$ enclosed by the RT surface, namely \cite{Alishahiha:2015rta}
\be
\label{eq:holo_subsystemC}
C_A=\frac{V(\gamma_A)}{8\pi G_N R}\,,
\ee
where $R$ denotes the AdS radius.

To illustrate some insightful examples of \eqref{eq:holo_subsystemC}, we consider a $(d+2)$-dimensional AdS space, whose metric in Poincaré coordinates reads
\be
ds^2=\frac{R^2}{r^2}\left(-dt^2+dr^2+d\eta^2+\eta^2 d\Omega^2_{d-1}\right)\,,
\ee
where $r\in[0,\infty)$, with the boundary located at $r\to 0$, $t\in\mathbb{R}$, $\eta$ is a non-negative radial coordinate, and $d\Omega^2_{d-1}$ denotes the line element of a $(d-1)$-dimensional sphere.
In AdS$_3$, i.e. $d=1$, the RT formula \eqref{eq:RTformula} reproduces the well-known CFT expression \cite{Holzhey:1994we,Calabrese:2004eu}
\begin{equation}
S_A = \frac{c}{3}\ln\left(\frac{\ell}{\epsilon}\right)\,,
\end{equation}
where $A$ is an interval of length $\ell$ in the boundary CFT and $\epsilon$ is a UV cutoff.

The volume contained within the RT surface can be computed for a boundary subsystem $A$ given by a spherical region of radius $\ell/2$. In this example, the RT surface is described by $\eta=f(r)=\sqrt{\ell^2/4-r^2}$. Thus, the corresponding volume reads
\cite{Alishahiha:2015rta}
\be
V(\gamma_A)=\Omega_{d-1}R^{d+1}\int_{\eta\leq f(r)}d\eta dr \frac{\eta^{d-1}}{r^{d+1}}=\frac{\Omega_{d-1}R^{d+1}}{d}\int^{\ell/2}_\epsilon dr\frac{(\ell^2/4-r^2)^{d/2}}{r^{d+1}}\,.
\ee
For $d=1$, the integral can be performed analytically and, plugging the result into \eqref{eq:holo_subsystemC}, we obtain the holographic CV subsystem complexity of an interval of size $\ell$ in a 1+1-dimensional CFT \cite{Alishahiha:2015rta}
\be
C_A=\frac{R}{4\pi G_N}\left[\sqrt{\frac{\ell^2}{4\epsilon^2}-1}-\arccos^{-1}\left(\frac{2\epsilon}{\ell}\right)\right]\,.
\ee
The parameter $\epsilon$ is again the UV cutoff, which can be chosen to be small compared to the subsystem size $\ell$, leading to
\be
\label{eq:2dCFT1int_holocomp}
C_A=\frac{R}{8\pi G_N}\left[\frac{\ell}{\epsilon}-\pi+O(\epsilon)\right]
=
\frac{R}{8\pi G_N}\left(\frac{\ell}{\epsilon}+\pi w_A\right)
\,,
\qquad
w_A\equiv -1+O(\epsilon)\,,
\ee
where the terms that vanish when the cutoff is removed have been included in the finite contribution and will be neglected from now on.

\begin{figure}[b!]
\centering
\includegraphics[width=0.295\textwidth]{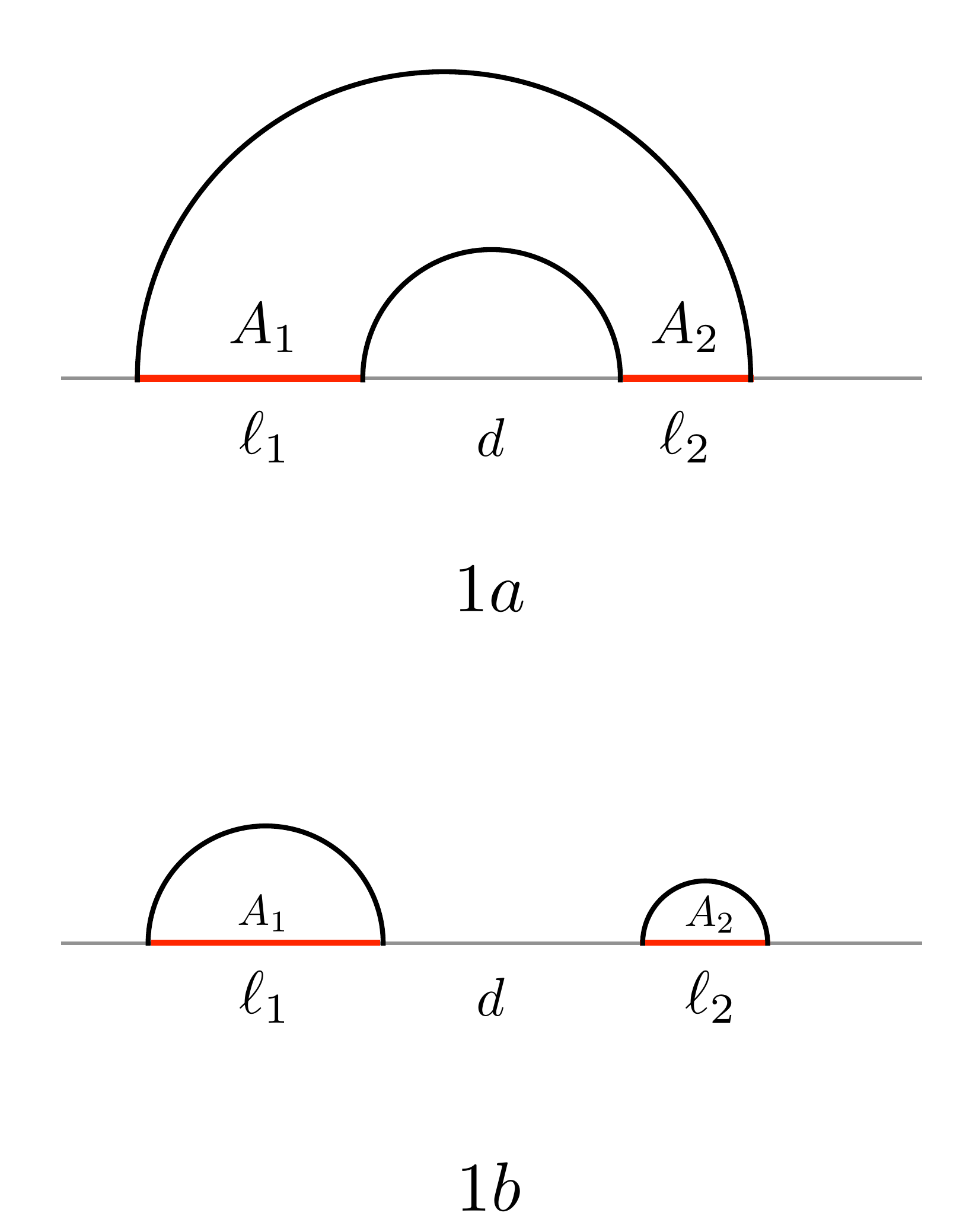}
\hspace{19.5pt}
\includegraphics[width=0.495\textwidth]{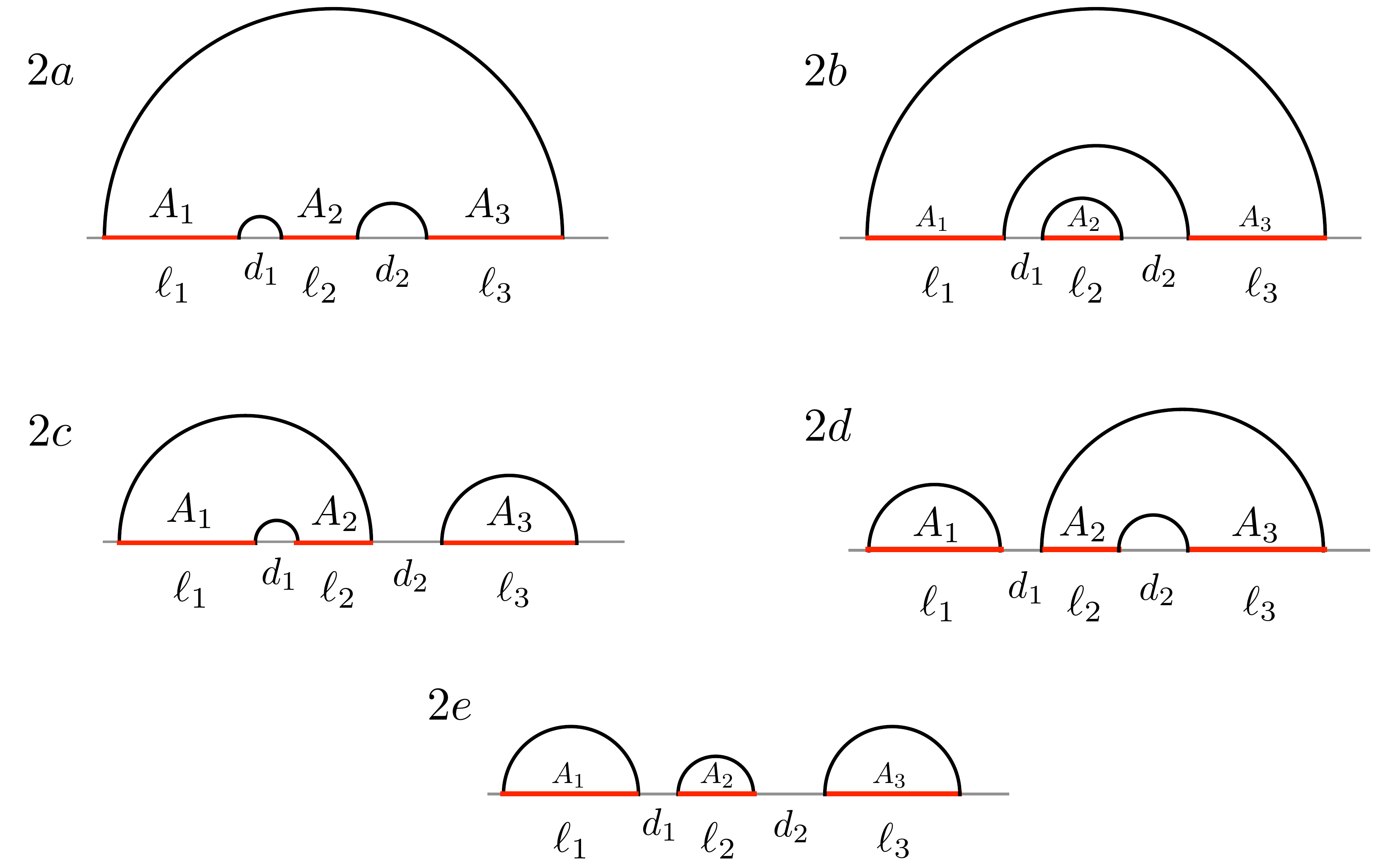}
\caption{
Possible configurations of RT surfaces contributing to the entanglement entropy and subsystem complexity in different regions of parameter space for boundary subregions in a $(1+1)$-dimensional CFT. In the left panels, the subsystem consists of two disjoint intervals, while in the right panels it consists of three disjoint intervals.
}
\label{fig:RT}
\end{figure}
Continuing to focus on the AdS$_3$ case, we study the CV complexity of the subsystem in the corresponding boundary CFT consisting of two disjoint intervals $A_1$ and $A_2$ with lengths $\ell_1$ and $\ell_2$, respectively, and separation $d$. The RT surface (given by geodesics in this case) has two possible configurations, schematically represented in the left panel of Fig.\,\ref{fig:RT}. When $A_1$ and $A_2$ are sufficiently close, namely $d<d_c$, the RT surface connects the two intervals (configuration 1a in the figure). On the other hand, when $d>d_c$, the RT surface consists of two disconnected components (configuration 1b). The critical distance is given in terms of the interval sizes as \cite{Ryu:2006bv}
\begin{equation}
\label{eq:critical distance}
d_c(\ell_1,\ell_2)
=
\frac{
\sqrt{(\ell_1+\ell_2)^2+4\ell_1\ell_2}
-\ell_1-\ell_2
}{2}.
\end{equation}
This implies two regimes for the entanglement entropy and, as a consequence, also for the subsystem complexity, namely
\begin{equation}
\label{eq:subsyste_comple_2int}
 C_{A_1\cup A_2}=
 \frac{R}{8\pi G_N}
\left(
\frac{\ell_1+\ell_2}{\epsilon}+\pi\, w_{12}(d)\right)\,,
\qquad
  w_{12}(d)\equiv  \begin{cases}
     0\,,   & d<d_c\,,
        \\
     -2\,,   & d>d_c\,.
    \end{cases}
\end{equation}
We can use this definition of subsystem complexity to compute the mutual complexity \eqref{eq:mutualcomplexity}, finding
\begin{equation}
  M(A_1: A_2)=  \begin{cases}
      -\frac{R}{4 G_N}
\,,   & d<d_c\,,
        \\
     \,\,\,\,\,\,\,\,\,0\,,   & d>d_c\,.
    \end{cases}
\end{equation}
This piecewise-constant behaviour confirms that the holographic mutual complexity computed through the CV proposal is non-positive, as discussed in \cite{Agon:2018zso}.

\begin{figure}[t!]
    \centering
    \begin{tabular}{cc}
        \includegraphics[width=0.38\textwidth]{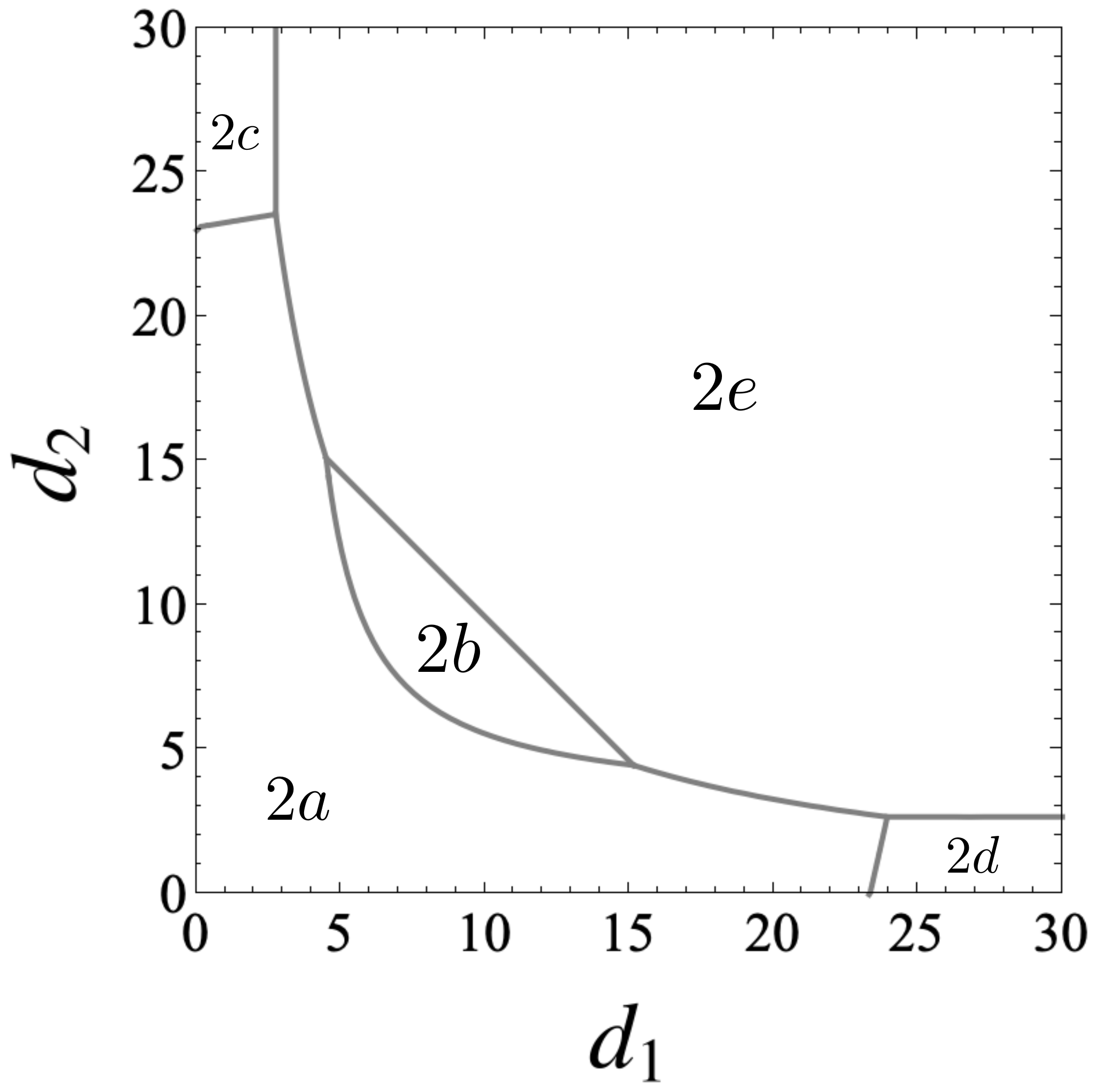} &
        \includegraphics[width=0.43\textwidth]{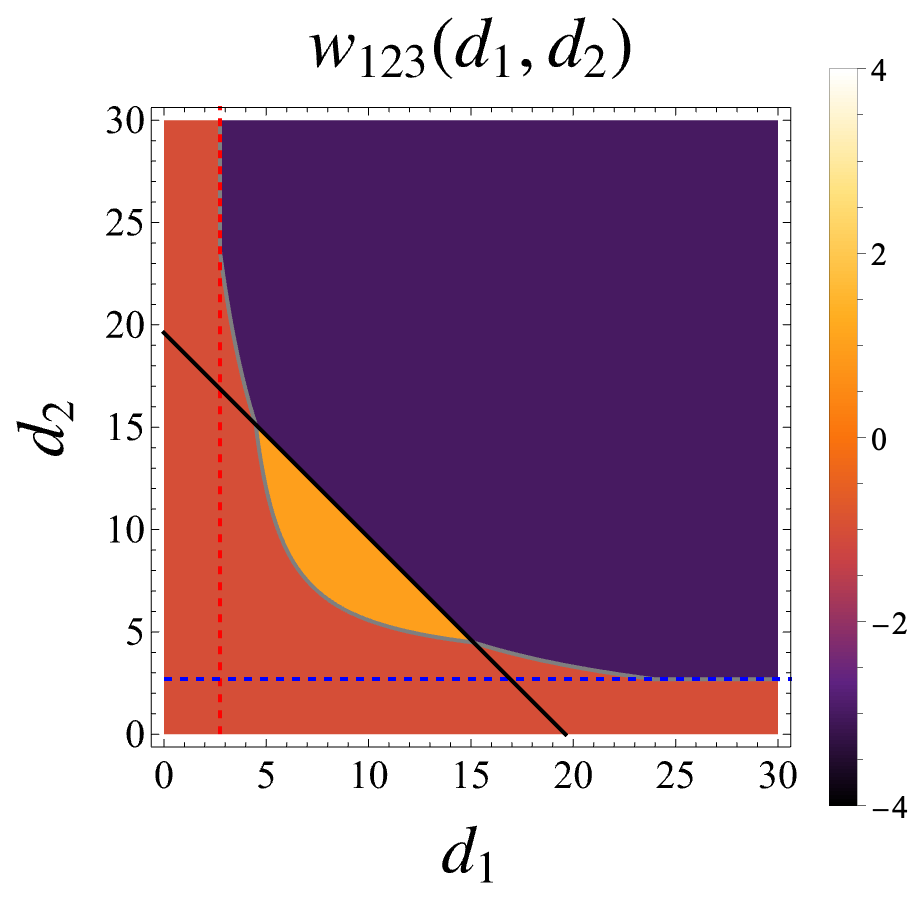} \\[-6pt]
        \includegraphics[width=0.38\textwidth]{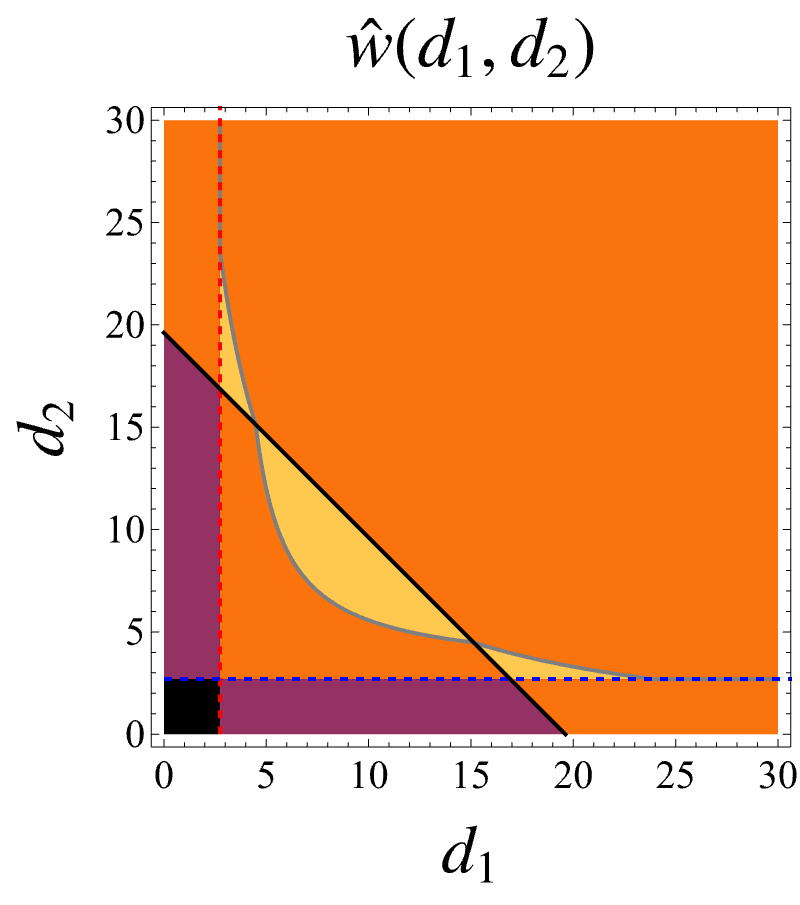} &
        \includegraphics[width=0.38\textwidth]{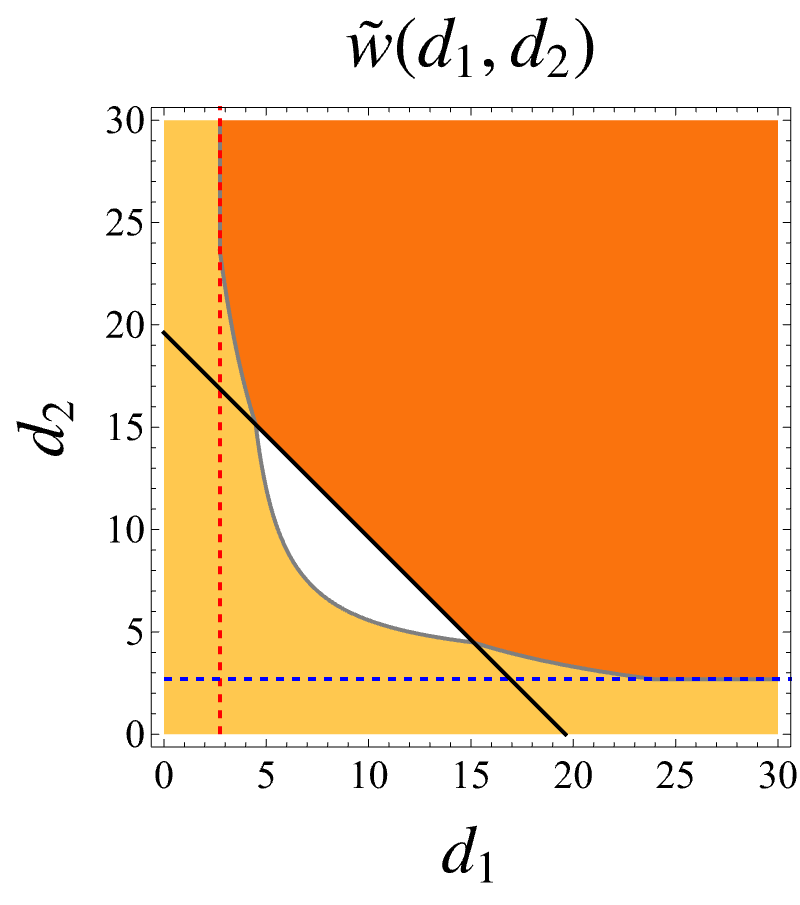}
    \end{tabular}
\caption{
The plot in the top-left panel shows the different regimes in which the finite contribution to the subsystem complexity \eqref{eq:subssystm_Compl_3int} changes value as a function of $d_1$ and $d_2$. The gray curves are obtained by minimizing the lengths of the AdS geodesics shown in the right panel of Fig.~\ref{fig:RT}. Each region is labeled by the RT configurations (cf. Fig.~\ref{fig:RT}) that dominate for the corresponding choice of parameters.
The plots in all panels are obtained for the parameter choice $\ell_1 = 60$, $\ell_2 = 3$, and $\ell_3 = 50$. In the top-right panel, we show $w_{123}$ in \eqref{eq:subssystm_Compl_3int} as a function of $d_1$ and $d_2$. The same is done for $\hat{w}$ (bottom-left) and $\tilde{w}$ (bottom-right), whose expressions are given in \eqref{eq:what} and \eqref{eq:wtilde}, respectively. The red and blue dashed lines correspond to $d_1 = d_c(\ell_1,\ell_2)$ and $d_2 = d_c(\ell_2,\ell_3)$, respectively, where $d_c$ is defined in \eqref{eq:critical distance}, while the solid black line corresponds to \eqref{eq:criticalline_3int}.
}
\label{fig:mutualcomplexity}
\end{figure}

For a subsystem made up of three disjoint intervals $A_1$, $A_2$, and $A_3$, with lengths $\ell_1$, $\ell_2$, and $\ell_3$, respectively, and separations $d_1$ and $d_2$, the analysis becomes somewhat more intricate.
To determine which RT surface dominates, we must compare the lengths of the geodesics corresponding to the five allowed configurations schematically reported in the right panel of Fig.\,\ref{fig:RT} \cite{Hayden:2011ag}. The resulting phases are organized as a function of the separations $d_1$ and $d_2$ in the diagram shown in the top-left panel of Fig.\,\ref{fig:mutualcomplexity} (using the notation of Fig.\,\ref{fig:RT}).
The various regimes are bounded by critical separation curves given by
\begin{equation}
\label{eq:criticalline_3int}
d_1+d_2
=
\frac{
\sqrt{(\ell_1+\ell_3)^2+4\ell_1\ell_3}
-\ell_1-\ell_3
}{2}
-\ell_2\,,
\end{equation}
and other more complicated conditions that come from the minimization of the RT surfaces.

Properly combining \eqref{eq:2dCFT1int_holocomp} to compute the volume inside the RT surfaces in the five configurations, the holographic CV subsystem complexity of $A=A_1\cup A_2\cup A_3$ is found to be
\begin{equation}
\label{eq:subssystm_Compl_3int}
C_{A_1\cup A_2\cup A_3}
=
\frac{R}{8\pi G_N}
\left(
\frac{\ell_1+\ell_2+\ell_3}{\epsilon}
+\pi\, w_{123}(d_1,d_2)
\right)\,,
\end{equation}
where the piece-wise constant function $w_{123}(d_1,d_2)$ is shown in the top-right panel of Fig.\,\ref{fig:mutualcomplexity} as a function of $d_1$ and $d_2$.

Combining \eqref{eq:subssystm_Compl_3int}, \eqref{eq:subsyste_comple_2int}, and \eqref{eq:2dCFT1int_holocomp} with the definitions in \eqref{eq:tripartitecomplexity} and \eqref{eq:Mtilde}, we obtain the holographic tripartite complexity and the complexity gap. Remarkably, for both quantities the extensive contributions to the subsystem complexity cancel exactly, leaving behind only a finite contribution of order $O(\epsilon^0)$. In particular, we obtain
\begin{equation}
M_3(A_1:A_2:A_3)= \frac{R\,\hat{w}(d_1,d_2)}{8 G_N}\,,
\qquad
\Delta^{(3)}_C(A_1:A_2:A_3)=\frac{R\, \tilde{w}(d_1,d_2)}{8 G_N}\,,
\end{equation}
where
\begin{equation}
\label{eq:what}
\hat{w}(d_1,d_2)\equiv w_{A_1}+w_{A_2}+w_{A_3}-w_{12}(d_1)-w_{13}(d_1+d_2+\ell_2)-w_{23}(d_2)+w_{123}(d_1,d_2)\,,
\end{equation}
and
\begin{equation}
\label{eq:wtilde}
\tilde{w}(d_1,d_2)\equiv w_{123}(d_1,d_2)-w_{A_1}-w_{A_2}-w_{A_3}\,,
\end{equation}
with $w_{ij}(d)$ being the function defined in \eqref{eq:subsyste_comple_2int} computed for the two intervals $A_i$ and $A_j$.
The integer-valued piecewise-constant functions $\hat{w}$ and $\tilde{w}$ are plotted in the bottom-left and bottom-right panels of Fig.\,\ref{fig:mutualcomplexity} as functions of $d_1$ and $d_2$, respectively. Interestingly, we observe that $\hat{w}$, and consequently $M_3(A_1:A_2:A_3)$, do not possess a definite sign. In contrast, $\Delta^{(3)}_C(A_1:A_2:A_3)$ remains non-negative throughout the entire allowed parameter range. This suggests that the complexity gap (for three and more subsystems) may be a natural candidate for characterizing a complexity cone.

\subsubsection{Emergence and complexity gap in holography}
\label{subsec:emergence}

\begin{figure}[t!]
\centering
\includegraphics[width=0.395\textwidth]{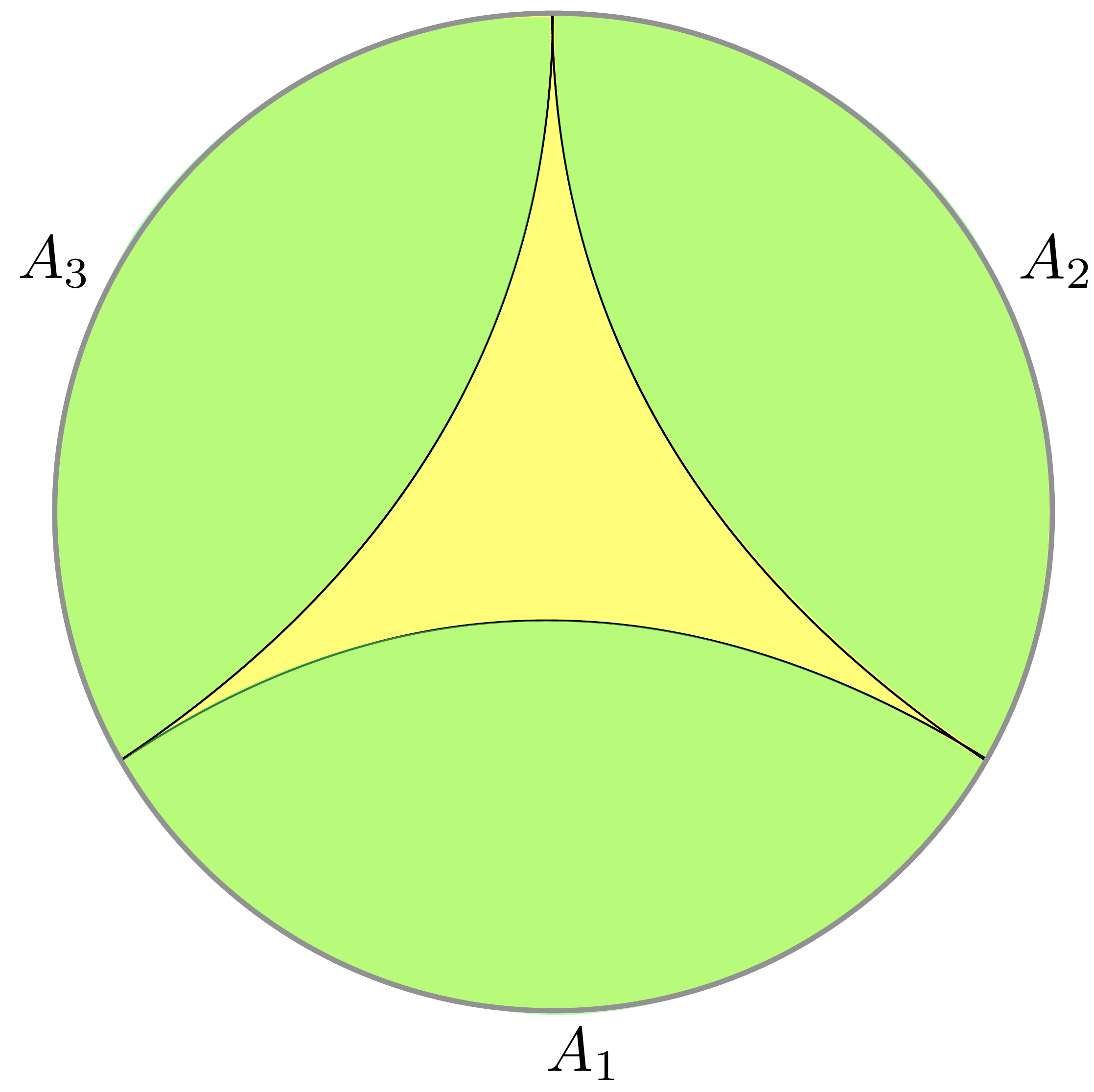}
\caption{
Schematic representation of a fixed-time slice of AdS$_3$ with the boundary (gray circle) partitioned into $A_1 \cup A_2 \cup A_3$. The complexity gap \eqref{eq:Mtilde}, computed using the holographic CV complexity proposal, is obtained by subtracting the volumes of the green regions associated with the subsystem complexities of the individual regions $A_1$, $A_2$, and $A_3$ from the full regularized volume of the time slice. The resulting complexity gap is proportional to the volume of the yellow region.
}
\label{fig:globalAds}
\end{figure}

As we mentioned before, our main purpose of defining the complexity gap is to quantify emergent complexity in the total state. This can be seen very intuitively in the holographic setting without technical computations. For simplicity, consider AdS spacetime in global coordinates dual to the ground state of a CFT defined on a circle.

Similarly to the previous section, in Fig.~\ref{fig:globalAds}, we schematically represent a constant-time slice of AdS$_3$, with the boundary theory defined on the gray circle partitioned into the three regions $A_1 \cup A_2 \cup A_3$. Using the CV holographic subsystem complexity proposal in the definition of the complexity gap, the first term, $C_{A_1\cup A_2 \cup A_3}$, is given by the (regularized) volume of the entire time slice. The subsystem complexities associated with the individual components, which are subtracted from $C_{A_1\cup A_2 \cup A_3}$, instead correspond to the volumes of the bulk regions shown in green in Fig.~\ref{fig:globalAds}. As a consequence, the complexity gap $\Delta^{(3)}_C(A_1:A_2:A_3)$ is positive and admits an elegant geometric interpretation as the volume of the yellow region displayed in the figure.
In Appendix \ref{app:globalAdS3}, we prove this explicitly and generalize the computation to the case in which the bipartition of the boundary CFT is such that $A = A_1 \cup A_2 \cup A_3$ and $B \neq \emptyset$, also studying $M_3(A_1:A_2:A_3)$ defined in \eqref{eq:tripartitecomplexity}.
The expressions, as well as the signs, of the tripartite complexity and the complexity gap turn out to be the same as those discussed in Sec.~\ref{subsec:static}; therefore, we do not comment on them further.

The positive sign of $\Delta^{(3)}_C(A_1:A_2:A_3)$ found also in Sec.~\ref{subsec:static} is consistent with its geometric interpretation as the emergent complexity/volume of a bulk region that is only visible for the total state. This is strikingly similar to previous discussions about bulk reconstructions and quantum error corrections \cite{Dong:2016eik} so it would be interesting to further investigate this connection.

When preparing this manuscript, we became aware of related work by \cite{Fujiki:2026ucr} where similar regions are discussed from the multipartite entanglement  perspective.

\subsubsection{Time-dependent spacetimes}\label{subsec:timedependent}

We now move to analyze a time-dependent scenario.
Consider a holographic 1+1-dimensional CFT on a circle of length $L$. We assume that the initial state is of the form
\begin{equation}
\label{eq:CFT quench initial state}
  \ket{\psi(0)}=e^{-\tau_0 H } \ket{B} \,,
\end{equation}
where $\ket{B}$ is a conformal boundary state regularized via a parameter $\tau_0$, which induces a small amount of initial entanglement. We then consider the subsystem $A$ given by an interval of length $\ell < L/2$.
In the regime $\ell, t \gg \tau_0$, and in the limit $L/\ell \to \infty$, the growth of the entanglement entropy with respect to the initial state is \cite{Calabrese:2005in}
\begin{equation}
   \Delta S_A(t)= s\times \begin{cases}
     2t & t<\ell/2\,,
     \\  
     \ell&t>\ell/2\,.
   \end{cases}
\end{equation}
where $s$ is the entropy density of the asymptotic steady state. This behaviour has also been reproduced via a holographic computation using the time-dependent generalization of the RT formula \cite{Hubeny:2007xt,Hartman:2013qma}.

The subsystem complexity for this setup, using the time-dependent holographic CV proposal \cite{Carmi:2016wjl}, has been studied in \cite{Fan:2025moc, Haah:2025hyf}.
It has been found that the complexity of the subsystem $A$, by subtracting its value in the thermal state, is 
\begin{equation}
\label{eq:holocompl_timedep}
   C_A(t)= \frac{s}{4\tau_0}\times \begin{cases}
    \ell t & t<\ell/2\,,
     \\  
   0&t>\ell/2\,,
   \end{cases}
\end{equation}
while, for the complementary region $B$ of length $L-\ell$,
\begin{equation}
\label{eq:holocompl_timedep_B}
   C_B(t)= \frac{s}{4\tau_0}\times \begin{cases}
     (L-\ell)t & t<\ell/2\,,
     \\  
     L t&t>\ell/2\,.
   \end{cases}
\end{equation}
Interestingly, $C_A(t)+C_B(t)=sLt/(4\tau_0)$ for any value of time\footnote{We thank Evita Verheijden for discussions on this point.} and this corresponds to the complexity of the full pure state of the holographic CFT using the CV proposal  \cite{Susskind:2014moa,Susskind:2014rva,Stanford:2014jda}. Consequently, the mutual complexity vanishes, $M(A_1:A_2)=0$, for all times.

It is insightful to ask whether a similar sum rule holds for some proposal of subsystem complexity in many-body quantum systems and quantum field theory, or within specific classes of such systems. In this spirit, one could use such a relation to place constraints on the space of complexity proposals for holographic boundary theories.

We remark that, from a many-body systems perspective, a behaviour similar to \eqref{eq:holocompl_timedep} and \eqref{eq:holocompl_timedep_B} has been observed via bounds on the subsystem circuit complexity in dynamics modeled by random unitary circuits \cite{Fan:2025moc, Haah:2025hyf}. However, a fully quantitative match with calculations in CFT remains elusive. Finally, to the best of our knowledge, the quantities \eqref{eq:tripartitecomplexity} and \eqref{eq:Mtilde} have not yet been studied in time-dependent holographic settings. We leave this interesting problem for future investigation.

\section{A Krylov space approach}\label{sec:subsystemKrylov}

\subsection{General framework}\label{subsec:KrylovSubsyst_generalframework}

Consider a general bipartite Hilbert space $\mathcal{H}=\mathcal{H}_A\otimes\mathcal{H}_B$ and an initial state $\vert\psi(0)\rangle$ in $\mathcal{H}$ for which we consider the reduced density matrices $\rho_A(0) $ and $\rho_B(0)$ associated to the two parties. Although the factorization of the Hilbert space can be general, we imagine it corresponds to a spatial bipartition of the system and we refer to $A$ and $B$ as the subsystems.  
If the initial state evolves unitarily in time under a time-independent Hamiltonian $H$:
\begin{equation}
  \vert\psi(t)\rangle=  e^{-{\rm i}Ht}\vert\psi(0)\rangle,
\end{equation}
we have a corresponding non-unitary evolution for the reduced density matrices $\rho_A(t) $ and $\rho_B(t)$.
Once the reduced density matrices are obtained, one might hope to treat their evolution and quantify their size/complexity as that of operators undergoing Lindbladian dynamics \cite{Parker:2018yvk}.
In Appendix \ref{app:RDMevolution}, we show that this is generally not possible and that the partial trace induces a highly non-trivial dynamics for the reduced density matrix, governed by the integro-differential {\it Nakajima-Zwanzig (NZ) equation} (see \eqref{eq:NZequation}). Instead, in the following, we employ Krylov-space methods to represent the resulting non-unitary and non-Lindbladian evolution of reduced density matrices. Within this framework, our main idea is to regard the autocorrelation function of reduced density matrices as the central object for encoding their dynamics as well as for extracting quantum complexity. We will make this precise below.

In the standard Krylov-space approach to unitary dynamics, the autocorrelation functions of evolving operators (or, in the vectorized picture, the return amplitudes of evolving states) play a central role \cite{LanczosBook,Parker:2018yvk}. Starting from these quantities, the full dynamics in the Krylov basis can be reconstructed iteratively through the moments recursion method. This procedure is equivalent to applying the Lanczos algorithm to the time-evolving states (or vectorized operators). Representing unitary dynamics in the Krylov basis provides a useful framework for studying the growth of evolving operators and the spreading of time-dependent states across the Hilbert space of the system (see the reviews \cite{Nandy:2024evd,Rabinovici:2025otw} for details and applications).

Motivated by this idea, we start from the autocorrelation functions of reduced density matrices at different times and treat it as the primary object to extract quantum subregion complexity. More precisely, we will apply the moments recursion method borrowed from the unitary setting. We stress, however, that there is an intrinsic difference with respect to the standard Krylov space framework. Since the vectorized reduced density matrix does not satisfy a Schrödinger-like evolution equation, the corresponding moments recursion method cannot be interpreted as a Gram-Schmidt orthonormalisation procedure (in the usual sense).
Nevertheless, in the following, we describe this construction in detail and discuss several examples supporting the utility of our effective approach.

More precisely, we consider the normalized autocorrelation function of the reduced density matrix
\begin{equation}
R_A(t)=\frac{\textrm{Tr}\left(\rho_A(t)\rho_A(0)\right)}{\textrm{Tr}\left(\rho^2_A(0)\right)}\,.
\label{eq:generaldef_RL}
\end{equation}
In principle, one can vectorise $\rho_A(t)$ and introduce a suitable scalar product such that \eqref{eq:generaldef_RL} can be rewritten as the overlap between the initial and the time-evolved vectorized operators. For this reason, throughout the manuscript we refer to $R_A(t)$ as the {\it subsystem return amplitude}.
Using basic properties of the trace and of the reduced density matrix, it is straightforward to see that $R^*_A(t)=R_A(t)$.

In the following, we use $R_A(t)$ to define an effective Krylov-space dynamics encoding the evolution of the reduced density matrix $\rho_A(t)$. This construction holds for a generic subsystem $A$, which can  be comprised of an arbitrary number of subregions. Formally, the quantities in \eqref{eq:generaldef_RL} can be expanded in time in terms of their moments
\begin{equation}
\mu_n^{(A)}=\partial_t^n R_A(t)\big\vert_{t=0}\,. 
\label{eq:subsystem moments}
\end{equation}
Then, we can follow the moment recursion method and iteratively extract Lanczos coefficients directly from $R_A(t)$ \cite{LanczosBook}. Crucially, such Lanczos coefficients do not have the usual interpretation in terms vector's normalizations $(b_n)$ and expectation values of the evolving Hamiltonian $(a_n)$, so we do not know a priori if they are real. Indeed, from the first relation between moments and Lanczos coefficients, we have 
\begin{equation}
\label{eq:a0_general}
    a^{(A)}_0=-{\rm i}\mu_1^{(A)}\,,
\end{equation}
which implies, since $\mu_1^{(A)}\in\mathbb{R}$, that $ a^{(A)}_0 $ is purely imaginary. 
By iterating the algorithm further, the expressions of the Lanczos coefficients in terms of the initial state become increasingly intricate. For instance, the second Lanczos coefficient is given by
\begin{equation}
\label{eq:b1_generalprocedure}
 \left(b_1^{(A)}\right)^2
=
\left(\mu_1^{(A)}\right)^2-  \mu_2^{(A)}
 \,.
\end{equation}
The square of this coefficient can be shown to be positive, namely $b_1^{(A)}$ is real, in a special case (see Appendix \ref{app:Lanczos general} for a detailed derivation). Moreover, in all the other examples analyzed in this manuscript, $\left(b_1^{(A)}\right)^2$ is also found to be positive, leading us to expect that $b_1^{(A)}$ is real in full generality.
As for the remaining coefficients, we can follow the moment recursion method (see the discussion in \cite{Balasubramanian:2022tpr}) and prove by induction that all the $a_n^{(A)}$ are purely imaginary, while $\left(b_n^{(A)}\right)^2$ are real. Indeed, taking \eqref{eq:a0_general} and \eqref{eq:b1_generalprocedure} as the base cases, one can show inductively that ${\rm i}a_n^{(A)}$ is a rational function of ${\rm i}a_{n-1}^{(A)},\ldots,{\rm i}a_0^{(A)}$, ${\rm i}b_n^{(A)},\ldots,{\rm i}b_1^{(A)}$, and $\mu_{2n+1}^{(A)},\ldots,\mu_1^{(A)}$. All these quantities are real, either by construction or by the induction hypothesis, implying that $a_n^{(A)}$ is purely imaginary. An analogous inductive argument shows that $\left(b_n^{(A)}\right)^2$ is real for all $n$.
At present, however, we are unable to establish the sign of $\left(b_n^{(A)}\right)^2$ in full generality. Based on all the examples discussed in this manuscript, we conjecture that $\left(b_n^{(A)}\right)^2>0$ for every $n$, and hence that all the coefficients $b_n^{(A)}$ are real.
An interesting special case is given by when $R_A(t) $ is an even function of time. If this happens, $\mu_{2n-1}^{(A)}=0$, and, as we know from the usual moment recursion method, $a^{(A)}_n =0$.

The iterative procedure stops when we encounter a vanishing $b^{(A)}_n$.
The outcome of the algorithm is the set of coefficients 
\begin{equation}
a^{(A)}_n,\;
n=0,\dots\mathcal{K}_A
\,,
\qquad
b^{(A)}_n\;,
n=1,\dots\mathcal{K}_A
\,,
\end{equation}
which we dub {\it  subsystem Lanczos coefficients.}
A priori, $\mathcal{K}_A$ can be finite or infinite. Notice that, given two complementary subsystems $A$ and $B$, in general $\mathcal{K}_A\neq \mathcal{K}_B$. There are, however, situations in which the two sets of subsystem Lanczos coefficients associated with $A$ and $B$ coincide, and therefore $\mathcal{K}_A=\mathcal{K}_B$. This happens, for example, whenever $R_A(t)=R_B(t)$. Although this equality does not hold in general, in Appendix \ref{app:RARB} we discuss a condition under which it is satisfied in the case Hilbert spaces $\mathcal{H}_A$ and $\mathcal{H}_B$ with the same finite dimension.

By analogy with the standard Krylov approach, we can define an effective evolution equation with the subsystem Lanczos coefficients as
\begin{equation}
\label{eq:evolutionLRequation}
    \mathrm{i} \partial_t \psi^{(A)}_n(t)=a^{(A)}_n \psi^{(A)}_n(t)+b^{(A)}_{n+1} \psi^{(A)}_{n+1}(t)+b^{(A)}_n \psi^{(A)}_{n-1}(t)\,.
\end{equation}
We remark a crucial difference between this prescription and the standard Krylov-space approach. In the latter, evolution equations of the form \eqref{eq:evolutionLRequation} arise by projecting the Schrödinger equation satisfied by the evolving state (or vectorized operator) onto the Krylov basis. On the other hand, in the present construction for reduced density matrix dynamics, the validity of \eqref{eq:evolutionLRequation} must be imposed as an assumption. In this way, the evolution of $\rho_A(t)$ is recast into an effective Schrödinger-like dynamics that encodes the properties of the NZ equation discussed in Appendix \ref{subapp:ZNequation}, which is, a priori, very different from a Schrödinger equation.

More precisely, the equations \eqref{eq:evolutionLRequation} have the following interpretation. Using the Lanczos algorithm, we try to represent the non-Lindbladian dynamics of reduced density matrices as a Schrödinger-like dynamics with wavefunctions $\psi_n^{(A)}(t)$.
This effective dynamics is not necessarily unitary. Indeed, if $a^{(A)}_n$ are non vanishing, since they are purely imaginary, the dynamics will be non-unitary.
This should not come as a surprise, as it is well known, and reviewed in Appendix \ref{subapp:primeexample}, that the time evolution of reduced density matrices is effectively non-unitary.
On the other hand, having $a^{(A)}_n$ equal to zero and $b^{(A)}_n$ real corresponds to a scenario where we can recast the reduced density matrix dynamics into an effective unitary dynamics (resembling a purification).
The non-unitarity of the dynamics implies that, in general, the amplitudes $\psi^{(A)}_n$ do not preserve the probability, i.e.
$\sum_n\vert\psi^{(A)}_n(t)\vert^2\neq 1$.

As a result, the effective dynamics induced by the subsystem Lanczos coefficients can be represented in terms of fictitious Krylov basis vectors as 
\begin{equation}
 \vert \tilde{\rho}_A(t)\rangle  =  e^{-{\rm i}\tilde{H}_At}\vert \tilde{\rho}_A(0)\rangle\,, 
  \label{eq:fictitious dynamics}
\end{equation}
where $\tilde{H}_A$ is a tridiagonal matrix constructed via the subsystem Lanczos coefficients. Notice that the dimension of $\vert \tilde{\rho}_A(t)\rangle $ as a vector might be, in general, different from the dimensionality of the original Hilbert space $\mathcal{H}_A$, where the reduced density matrix $\rho_A$ acts. Interestingly, we will show examples where $\mathcal{K}_A $ is larger than $\dim\mathcal{H}_A$.  The way in which \eqref{eq:fictitious dynamics} encodes the dynamics of the reduced density matrices is by sharing the same return amplitudes
\begin{equation}
 \langle\tilde{\rho}_A(t)\vert \tilde{\rho}_A(0)\rangle  = \left(\psi^{(A)}_0(t)\right)^*=R_A(t) \,, 
  \label{eq:fictitious dynamics2}
\end{equation}
which can be proven by reversing the procedure through which we have constructed the tridiagonal Hamiltonian $\tilde{H}_A$.

If we can solve \eqref{eq:evolutionLRequation} and obtain the amplitudes $\psi_n^{(A)}(t)$, we can quantify the spread of this effective dynamics using generalization of the Krylov/spread complexity
\begin{equation}
  \label{eq:subsystemKrylovcomplexity}
  C_A(t)=\frac{\sum_{n=0}^{\mathcal{K}_A} n \vert\psi_n^{(A)}(t) \vert^2}{\sum_{n=0}^{\mathcal{K}_A} \vert\psi_n^{(A)}(t) \vert^2} 
  \,,
\end{equation}
where the denominators take into account the possible non-conservation of the probability due to the non-unitary dynamics. We refer to \eqref{eq:subsystemKrylovcomplexity} as {\it subsystem Krylov complexity}.

Given a bipartition into $A$ and its complement $B$, following the procedure outlined above, we can compute the subsystem Krylov complexity for both subsystems, obtaining $C_A$ and $C_B$. In general, the two complexities are different, i.e. $C_A \neq C_B$. This follows directly from the fact that there is no a priori relation between the Krylov space dimensionalities $\mathcal{K}_A$ and $\mathcal{K}_B$, and therefore, generically, $\mathcal{K}_A \neq \mathcal{K}_B$.
Interestingly, in cases where $R_A = R_B$, the full effective Krylov dynamics of the two reduced density matrices coincides and, as a consequence, $C_A = C_B$.

In the remainder of this section, we study this quantity in several examples to argue about some of its general properties. Moreover, in all the instances analyzed, we compare the subsystem Krylov complexity with the spread complexity $C_S(t)$ of the full state $\vert\psi(t)\rangle$, and with the Krylov complexity $C_K(t)$ of the full density matrix $\rho(t)=\vert\psi(t)\rangle\langle\psi(t)\vert$ before the partial trace, together with the corresponding Krylov-space dimensionalities, $\mathcal{K}_S$ and $\mathcal{K}_K$, respectively.
We do not provide further details on how to compute these quantities, referring instead to the standard literature for the relevant technical tools \cite{LanczosBook,Parker:2018yvk,Caputa:2021sib,Balasubramanian:2022tpr}. For future convenience, we simply recall that the Krylov dynamics of the pure state and of the pure density matrix can be determined from the return amplitude $R_S$ and the autocorrelation function $R_K$, respectively, defined as
\begin{equation}
\label{eq:returnamplitudes}
R_S(t)= \langle\psi(t)\vert\psi(0)\rangle\,,
\qquad
R_K(t)=\Tr\left[\rho(t)\rho(0)\right] \,.
\end{equation}
We stress that the choice, in this section, of starting from the subsystem return amplitude \eqref{eq:generaldef_RL} to construct the Krylov dynamics of the reduced density matrix is motivated by the central role played by \eqref{eq:returnamplitudes} in the Krylov methods for pure state evolution.

\subsection{Examples with a few qubits}\label{subsec:fewspins}

In the previous section, we described the general method for representing the evolution of reduced density matrices in terms of an effective Krylov-space dynamics. In what follows, in order to investigate the properties of this framework and of the resulting subsystem complexity, we apply the method to several examples involving a small number of qubits.

\subsubsection{Two-qubits example}\label{subsec:2spins}

We begin our analysis with the simplest example of a bipartite Hilbert space, namely that of two two-dimensional Hilbert spaces representing a pair of qubits.
We consider the initial state given by the tensor product of two qubit states
\begin{equation}
\label{eq:unentangled initialstate}
    \vert\psi(0)\rangle=\vert \psi_1\rangle\otimes \vert \psi_2\rangle\,,
\end{equation}
where
\begin{equation}
\label{eq:generalqubit}
    \vert  \psi_i\rangle=
\binom{\cos \frac{\theta_i}{2}}{e^{\mathrm{i} \varphi_i} \sin \frac{\theta_i}{2}}\,,
\end{equation}
with $\varphi_i \in[0,2 \pi)$ and $\theta_i \in[0, \pi)$, and we let the initial state evolve with the Hamiltonian \begin{equation}
\label{eq:evolutionXXX}
    H=-2(X_A X_B+Y_AY_B+Z_AZ_B)\,.
\end{equation}
We further consider the natural bi-partition into $A$ given by the first qubit and $B$ by the second qubit, i.e.
$\vert\psi(0)\rangle\in\mathcal{H}_A\otimes \mathcal{H}_B$. The time-evolved reduced density matrix $\rho_A(t) $ is computed straightforwardly by taking the partial trace on the density matrix of $\vert\psi(t)\rangle $. We are not interested in the explicit expressions of its entries, but rather in the subsystem return amplitude \eqref{eq:generaldef_RL}, which reads
\begin{equation}
\begin{aligned}
R_A(t)
=\frac{1}{4}\left(3 + \cos(8t) + 2\sin^2(4t)\left(\cos\theta_1\cos\theta_2 + \cos(\varphi_1 - \varphi_2)\sin\theta_1\sin\theta_2\right)\right)\,.
\end{aligned}
\label{eq:returnamplitudeLeftunentangled}
\end{equation}
If we repeat the same computation for the second qubit, i.e. the subsystem $B$, we find $R_A(t)=R_B(t)$, as we expected from the condition proven in Appendix \ref{app:RARB}.

Applying to this function the moment-recursion procedure outlined in Sec.\,\ref{subsec:KrylovSubsyst_generalframework}, we extract the non-vanishing Lanczos coefficients. The only non-vanishing $b_{n}^{(A)}$ are $b_{1}^{(A)}=b_-$ and $b_{2}^{(A)}=b_+$, where
\begin{align}
b_{\pm}
&=4\sqrt{1 - \cos\theta_1\cos\theta_2 \pm \cos(\varphi_1 - \varphi_2)\sin\theta_1\sin\theta_2}\,,
\end{align}
while all $a^{(A)}_n$ vanish due to the time-reflection symmetry of \eqref{eq:returnamplitudeLeftunentangled}.

The effective Schrödinger-like Krylov space dynamics is taking place in a three-dimensional Krylov space, i.e. $\mathcal{K}_A=3$. In addition, due to $a^{(A)}_n=0$, the non-unitary evolution of $\rho_A(t)$ can be recast as a unitary evolution in Krylov space. The knowledge of the subsystem Lanczos coefficients allows to recursively compute the amplitudes $\psi_n^{(A)}(t)$ from \eqref{eq:evolutionLRequation} and the subsystem Krylov complexity \eqref{eq:subsystemKrylovcomplexity} reads
\begin{equation}
\begin{aligned}
C_A(t)=C_B(t)&=
\frac{1}{4}\Bigl[
1 - \cos\theta_1 \cos\theta_2 
- \cos(\varphi_1 - \varphi_2)\,\sin\theta_1 \sin\theta_2
\Bigr] \\&
\times \Bigl[
\sin^2(8t)
+ 2 \sin^4(4t)\bigl(
3 + \cos\theta_1 \cos\theta_2 
+ \cos(\varphi_1 - \varphi_2)\,\sin\theta_1 \sin\theta_2
\bigr)
\Bigr] \,,
\end{aligned}
\label{eq:left-rightcomplex_2sp_unent}
\end{equation}
where the equality $C_A=C_B$ comes from the same subsystem return amplitudes.

\begin{figure}[b!]
\centering
\includegraphics[width=0.495\textwidth]{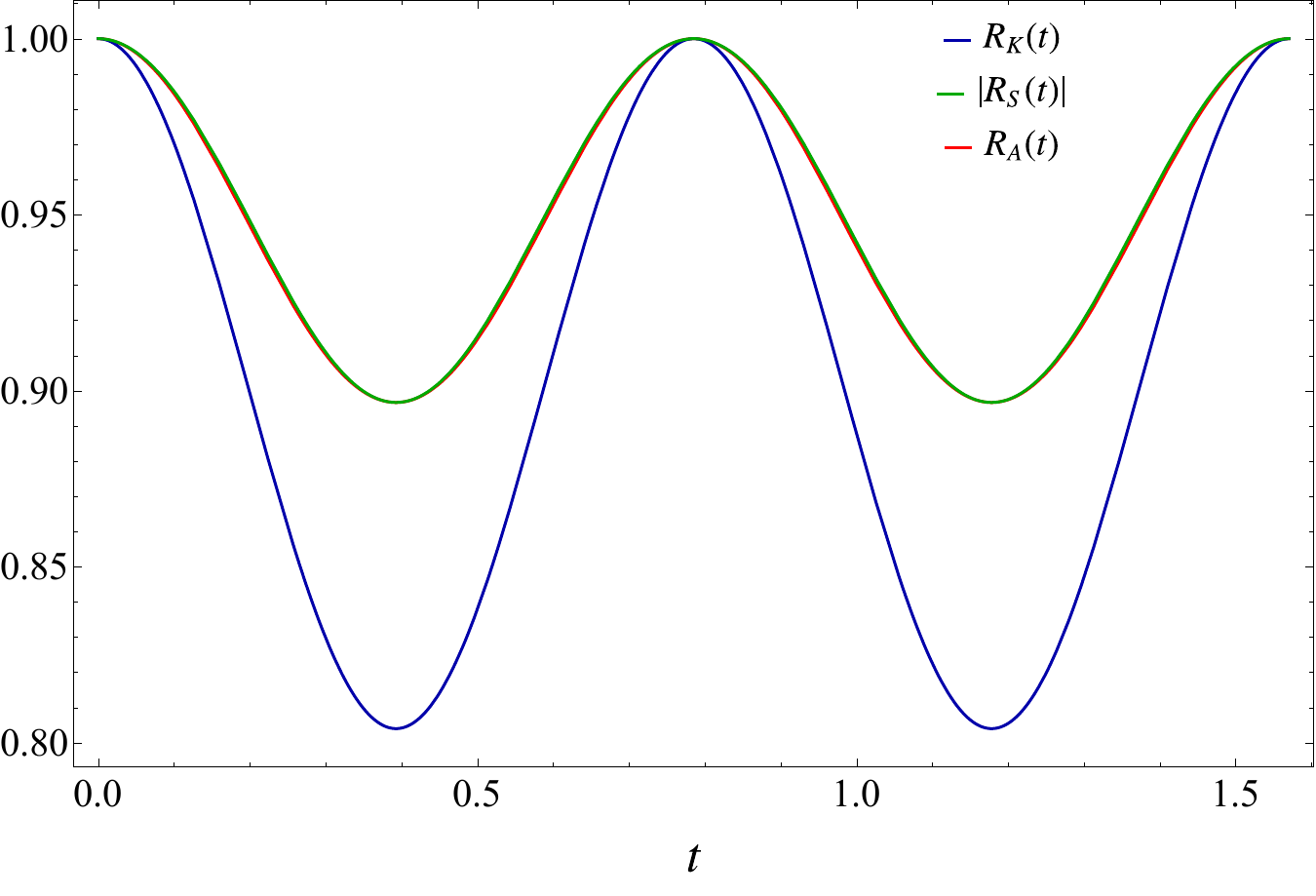}
\includegraphics[width=0.495\textwidth]{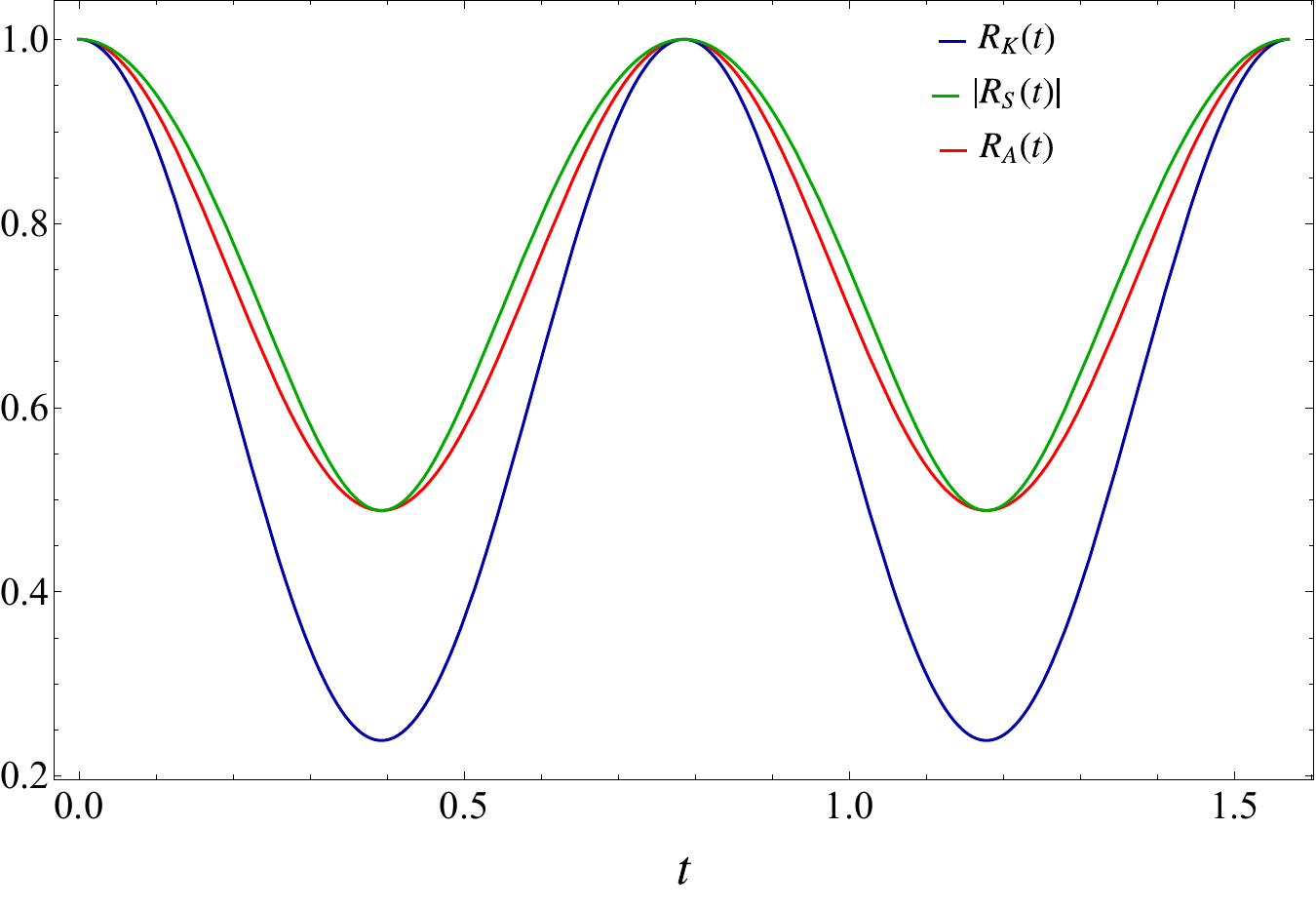}

\caption{
Comparison between the subsystem return amplitude \eqref{eq:returnamplitudeLeftunentangled} for a subsystem consisting of a single qubit embedded in a two-qubit system (red), the absolute value of the return amplitude \eqref{eq:spreadautocorr_2spin_unent} (green), and the autocorrelation function of the corresponding pure state density matrix \eqref{eq:Krylovautocorr_2spin_unent} (blue). The initial state is \eqref{eq:unentangled initialstate} and evolves in time under the Hamiltonian \eqref{eq:evolutionXXX}. In the left panel, we choose $(\theta_1,\theta_2;\varphi_1,\varphi_2) = \left( \frac{2\pi}{3}, \frac{7\pi}{8}; 0, 0 \right)$, while in the right panel we take $(\theta_1,\theta_2;\varphi_1,\varphi_2) = \left( \frac{4\pi}{15}, \frac{3\pi}{5}; \frac{5\pi}{3}, \frac{\pi}{12} \right)$.
}
\label{fig:2}
\end{figure}

\begin{figure}[t!]
\centering
\includegraphics[width=0.495\textwidth]{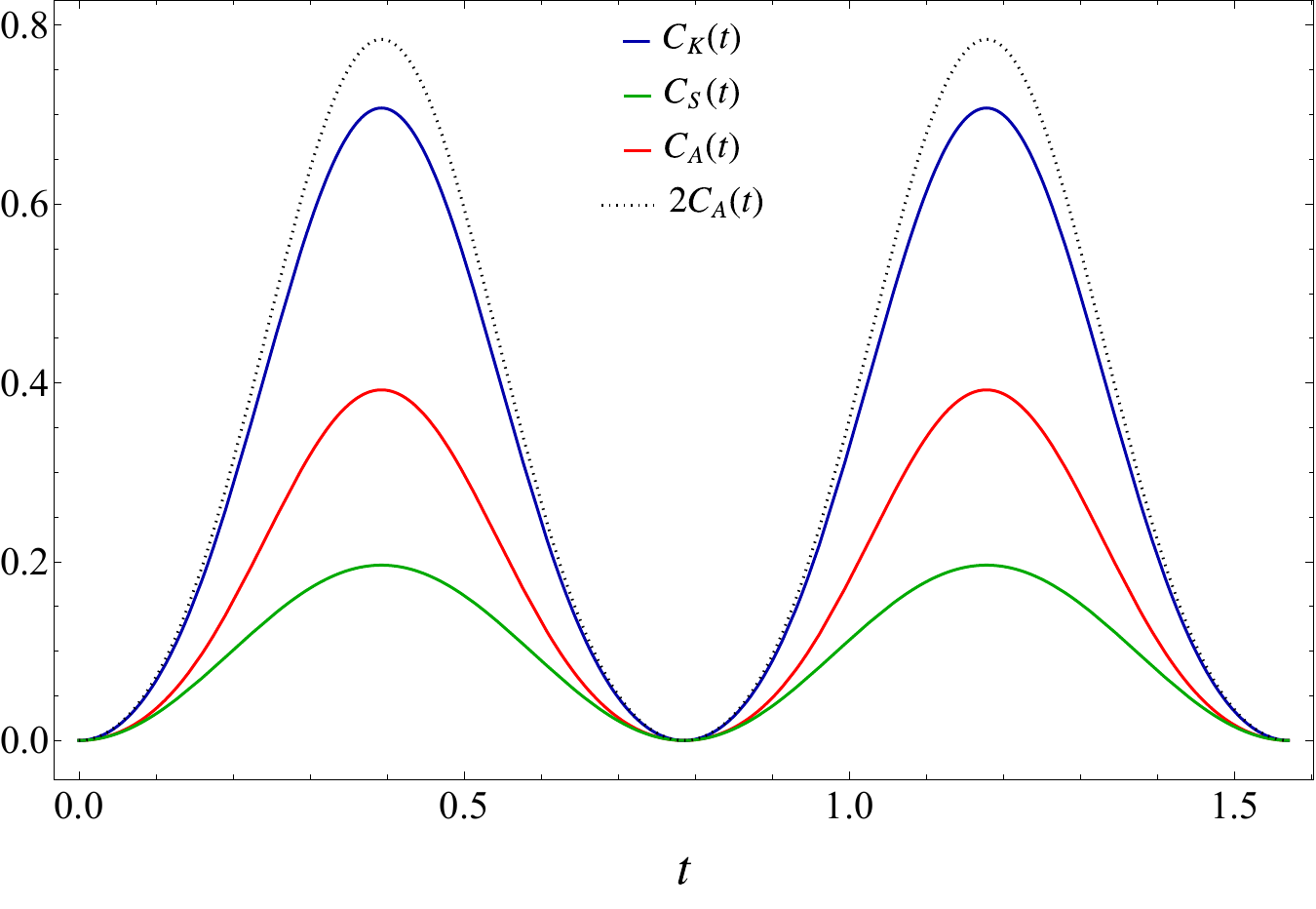}
\includegraphics[width=0.495\textwidth]{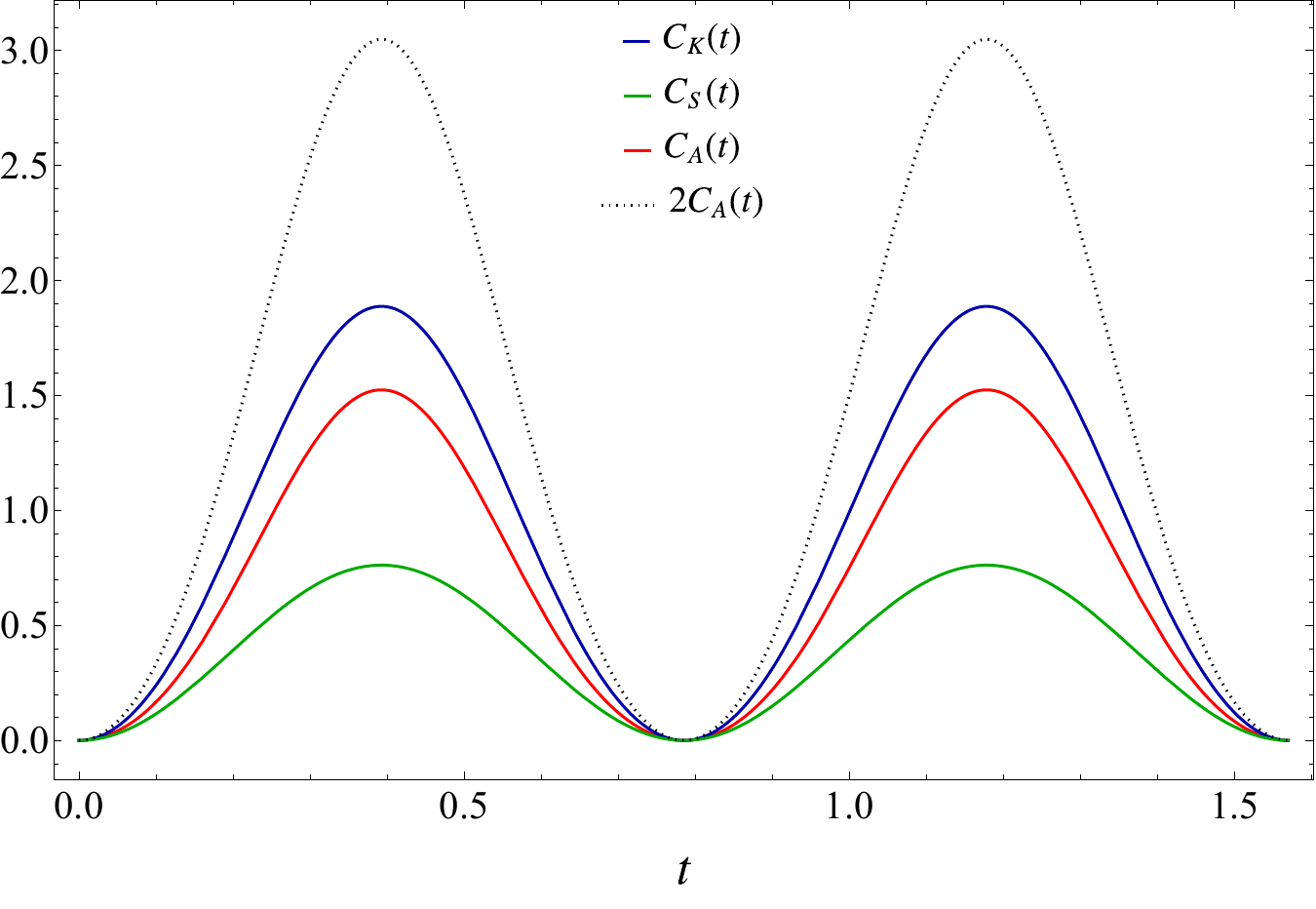}

\caption{Comparison between the subsystem Krylov complexity \eqref{eq:left-rightcomplex_2sp_unent} for a subsystem consisting of a single qubit embedded in a two-qubit system (red), the spread complexity of the pure state initialized in \eqref{eq:unentangled initialstate} and evolved under the Hamiltonian \eqref{eq:evolutionXXX} (green), and the Krylov complexity of the corresponding density matrix (blue). In the left panel, we choose $(\theta_1,\theta_2;\varphi_1,\varphi_2) = \left( \frac{2\pi}{3}, \frac{7\pi}{8}; 0, 0 \right)$, while in the right panel we take $(\theta_1,\theta_2;\varphi_1,\varphi_2) = \left( \frac{4\pi}{15}, \frac{3\pi}{5}; \frac{5\pi}{3}, \frac{\pi}{12} \right)$.}
\label{fig:3}
\end{figure}

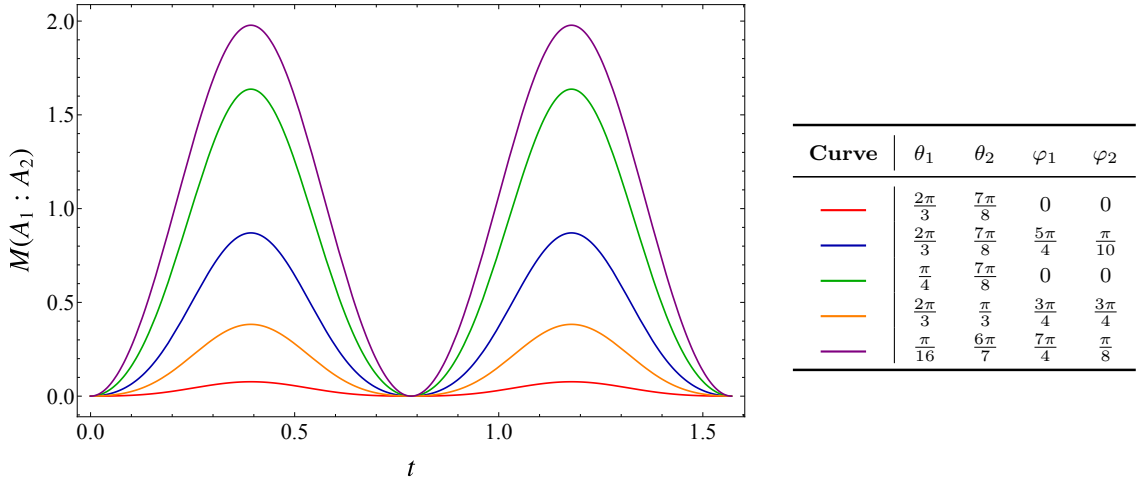
\begin{figure}[b!]
\centering
\begin{minipage}{0.65\textwidth}
    \includegraphics[width=\linewidth]{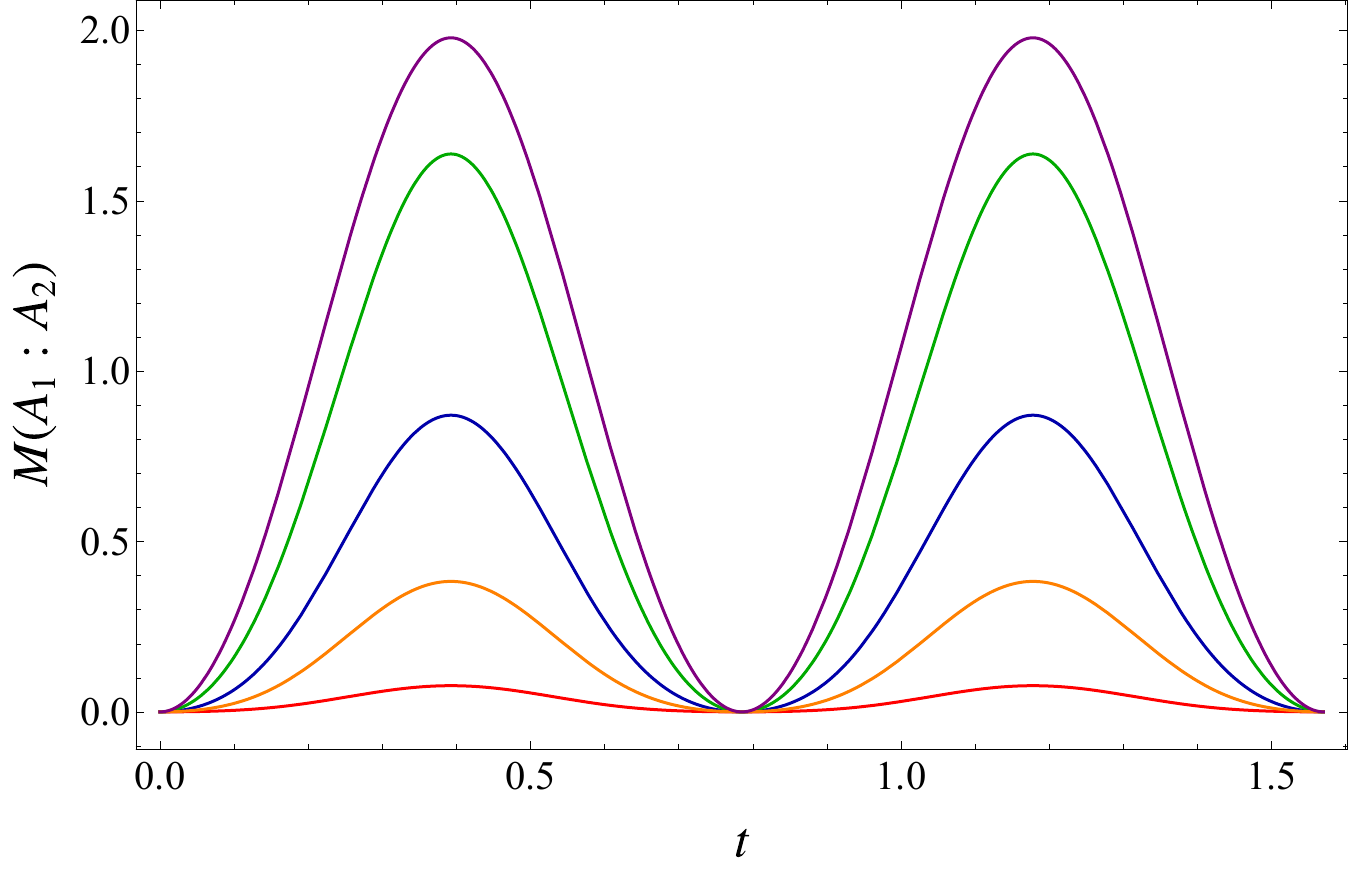}
\end{minipage}
\hfill
\begin{minipage}{0.3\textwidth}
\raggedright
\scriptsize
\renewcommand{\arraystretch}{1.4} 
\begin{tabular}{c|cccc}
\toprule
\textbf{Curve} & $\theta_{1}$ & $\theta_{2}$ & $\varphi_{1}$ & $\varphi_{2}$ \\
\midrule

\tikz{\draw[thick,
color={rgb,255:red,255;green,0;blue,0}]
(0,0) -- (0.6,0);}
& $\frac{2\pi}{3}$ & $\frac{7\pi}{8}$ & $0$ & $0$ \\

\tikz{\draw[thick,
color={rgb,255:red,0;green,0;blue,170}]
(0,0) -- (0.6,0);}
& $\frac{2\pi}{3}$ & $\frac{7\pi}{8}$ & $\frac{5\pi}{4}$ & $\frac{\pi}{10}$ \\

\tikz{\draw[thick,
color={rgb,255:red,0;green,170;blue,0}]
(0,0) -- (0.6,0);}
& $\frac{\pi}{4}$ & $\frac{7\pi}{8}$ & $0$ & $0$ \\

\tikz{\draw[thick,
color={rgb,255:red,255;green,128;blue,0}]
(0,0) -- (0.6,0);}
& $\frac{2\pi}{3}$ & $\frac{\pi}{3}$ & $\frac{3\pi}{4}$ & $\frac{3\pi}{4}$ \\

\tikz{\draw[thick,
color={rgb,255:red,128;green,0;blue,128}]
(0,0) -- (0.6,0);}
& $\frac{\pi}{16}$ & $\frac{6\pi}{7}$ & $\frac{7\pi}{4}$ & $\frac{\pi}{8}$ \\

\bottomrule
\end{tabular}

\end{minipage}
\caption{Mutual complexity \eqref{eq:mutualcomplexity} for a bipartition of a two-qubit system into its individual qubits as a function of time. The quantity is computed using the subsystem complexity \eqref{eq:left-rightcomplex_2sp_unent} and the Krylov complexity of the corresponding pure state density matrix. The initial state is \eqref{eq:unentangled initialstate} with parameters given in the table on the right, while the evolution Hamiltonian is given by \eqref{eq:evolutionXXX}.}
\label{fig:4}

\end{figure}

At this stage, we can compare the subsystem complexity \eqref{eq:left-rightcomplex_2sp_unent} with the spread complexity $C_S$ and the Krylov complexity $C_K$ associated with the corresponding unitarily evolving pure state. In Appendix \ref{app:2q}, we report the expressions for the autocorrelation function $R_K$, the return amplitude $R_S$, and the corresponding Lanczos coefficients.
The number of non-vanishing Lanczos coefficients determines the (maximal) dimensionalities of the Krylov spaces associated with the spreading of the state vector, $\mathcal{K}_S$, and with the growth of the full density matrix, $\mathcal{K}_K$. We find
$\mathcal{K}_S=2$, while $\mathcal{K}_K=\mathcal{K}_A=\mathcal{K}_B=3$.
Moreover, we use the expressions reported in Appendix \ref{app:2q} to compare with them the subsystem return amplitude and the subsystem complexity shown in Fig.\,\ref{fig:2} and Fig.\,\ref{fig:3} respectively.
 All the quantities plotted in these figures exhibit oscillations with the same frequency, which can be extracted from the corresponding analytical expressions. In Fig.\,\ref{fig:2}, and for several other choices of parameters not reported here, we observe $R_K \leq R_A \leq |R_S|$ for all times. Moreover, the complexities shown in Fig.\,\ref{fig:3} appear to satisfy a fixed ordering for all times and for the explored parameters, namely $C_S \leq C_A \leq C_K$. One may wonder how general this hierarchy is; the results reported in the next subsection show that this is not the case in general.

From Fig.\,\ref{fig:3}, it is already evident that the mutual complexity \eqref{eq:mutualcomplexity}, computed using the subsystem Krylov complexity and the pure state Krylov complexity as the last term, is positive for this choice of parameters. To further investigate the sign of this quantity, in Fig.\,\ref{fig:4} we report the mutual complexity for several different choices of parameters.
The oscillatory behaviour also characterizes the evolution of $M(A_1:A_2)$, and is indeed inherited by the complexities in Fig.\,\ref{fig:3}. Interestingly, in this case, the mutual complexity is positive for all values of time and for all parameter choices considered (including more cases than those reported here for conciseness).

\subsubsection{Three-qubits example}\label{subsec:3spins}

In the previous two-qubit example, there was a natural bipartition to consider. In the next example, we consider a system composed of three qubits. This allows for different possible partitions of the system and for studying their relations in terms of the effective Krylov dynamics of the corresponding reduced density matrices.

Consider an initial state given by the tensor product of three qubits
\begin{equation}    \label{eq:initst_3q}
\vert\Psi(0)\rangle=\vert \psi_1\rangle\otimes \vert\psi_2\rangle\otimes \vert\psi_3\rangle\,,
\end{equation}
where $\vert \psi_i\rangle$ is the vector in \eqref{eq:generalqubit}, and the time evolution of $\vert\Psi(0)\rangle$ is induced by the Hamiltonian 
\begin{equation}
\label{eq:Ham_3q}
     H=-(X_1 X_2+X_2 X_3+Y_1Y_2+Y_2Y_3+Z_1Z_2+Z_2Z_3)\,,
\end{equation}
which generalizes \eqref{eq:evolutionXXX} to three qubits.

This system allows various bipartitions and therefore various combinations for studying the subsystem Krylov complexity. We consider the bipartition where the first qubit is in the subsystem $A$ and the second and the third are in $B$. The subsystem return amplitude \eqref{eq:generaldef_RL} can be determined as a general function of time and the parameters $\theta_i$ and  $\phi_i$. The general expression is not particularly illuminating so we do not report it here. However, it is sufficient to focus on a particular, generic choice of parameters: $\theta_1=\frac{\pi}{4} $, $\theta_2=0 $, $\theta_3=\frac{\pi}{2} $, $\varphi_1= \frac{\pi}{2}$, $\varphi_2=0 $, $\varphi_3= 0$, for which  we find
\begin{align}
\nonumber
R_{A}(t) &= \frac{1}{72} \Big( 50 + 4 \sqrt{2} + 12 \cos(2t) + 6 \cos(4t) + 4 \cos(6t) - 4 \sqrt{2} \cos(6t) \\
&\quad + 3 \sqrt{2} \sin(2t) + 3 \sqrt{2} \sin(4t) - 3 \sqrt{2} \sin(6t) \Big)\,,
\\
\label{eq:2spinreducedreturnampli}
\nonumber
R_{B}(t) &= \frac{1}{72}\Big(38 + 3\sqrt{2} + 3(4 + \sqrt{2})\cos(2t) - 3(-2 + \sqrt{2})\cos(4t) + 16\cos(6t)\\ &- 3\sqrt{2}\cos(6t) - 3\sqrt{2}\sin(2t) - 3\sqrt{2}\sin(4t) + 3\sqrt{2}\sin(6t)\Big)\,.
\end{align}
Let us discuss the properties of these functions. First, we notice that, differently from the two-qubits case in Sec.\,\ref{subsec:2spins}, $R_A(t)\neq R_B(t)$ and corresponding subsystem Krylov complexities will also be different, $C_A(t)\neq C_B(t)$. As expected, the two subsystem return amplitudes are real, but not time-reversal symmetric. As discussed in Sec.\,\ref{subsec:KrylovSubsyst_generalframework}, this implies that some of the Lanczos coefficients $a^{(A)}_n$ and $a^{(B)}_n$ will not vanish.

For general values of the parameters, we can apply the moment recursion method detailed in Sec.\,\ref{subsec:KrylovSubsyst_generalframework} and determine the Lanczos coefficients associated with this subsystem dynamics. The number of non-vanishing coefficients depend on the parameters of the initial state. The largest Krylov space dimension that we find, occurring for a wide range of parameters, is $\mathcal{K}_{A}=\mathcal{K}_{B}=7 $. 
Also in this example, we find that all $b^{(A)}_n$ coefficients are real, while all $a^{(A)}_n$ are purely imaginary, as anticipated in Sec.\,\ref{subsec:KrylovSubsyst_generalframework}.

From the knowledge of the subsystem Lanczos coefficients, we can compute the subsystem Krylov complexity for any choice of parameters. The general expression is complicated, so we just analyze directly the plots displayed in this section. Similar analysis can be carried out for other choices of bipartitions of the three-qubits system. The same findings are also  obtained for the Krylov space dimension $\mathcal{K}_{A}$.

Also in this example, we can compare $\mathcal{K}_{A}$ and the corresponding subsystem complexity with $\mathcal{K}_{S}$ and $\mathcal{K}_{K}$, together with the associated spread and Krylov complexities.
The return amplitude of the evolving state and the autocorrelation function of the pure state density matrix can be computed for arbitrary choices of parameters. Their explicit expressions are rather cumbersome and we therefore do not report them here. Nevertheless, we can analyze the non-vanishing Lanczos coefficients extracted from these functions, from which we find that the largest Krylov space dimensions compatible with this model are $\mathcal{K}_{S}=3$ and $\mathcal{K}_{K}=\mathcal{K}_{A}=7$.

\begin{figure}[b!]
\centering
\includegraphics[width=0.495\textwidth]{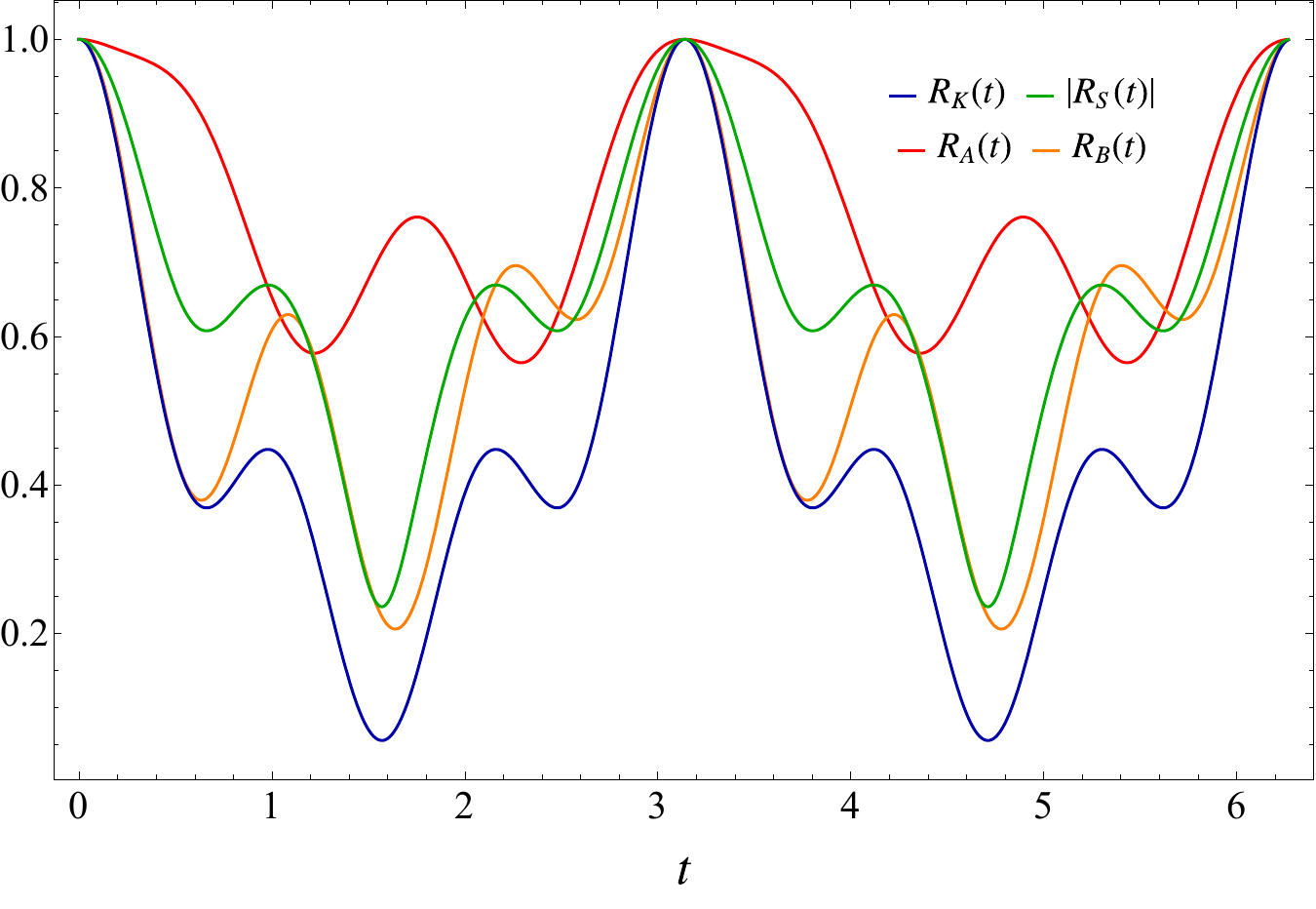}
\includegraphics[width=0.495\textwidth]{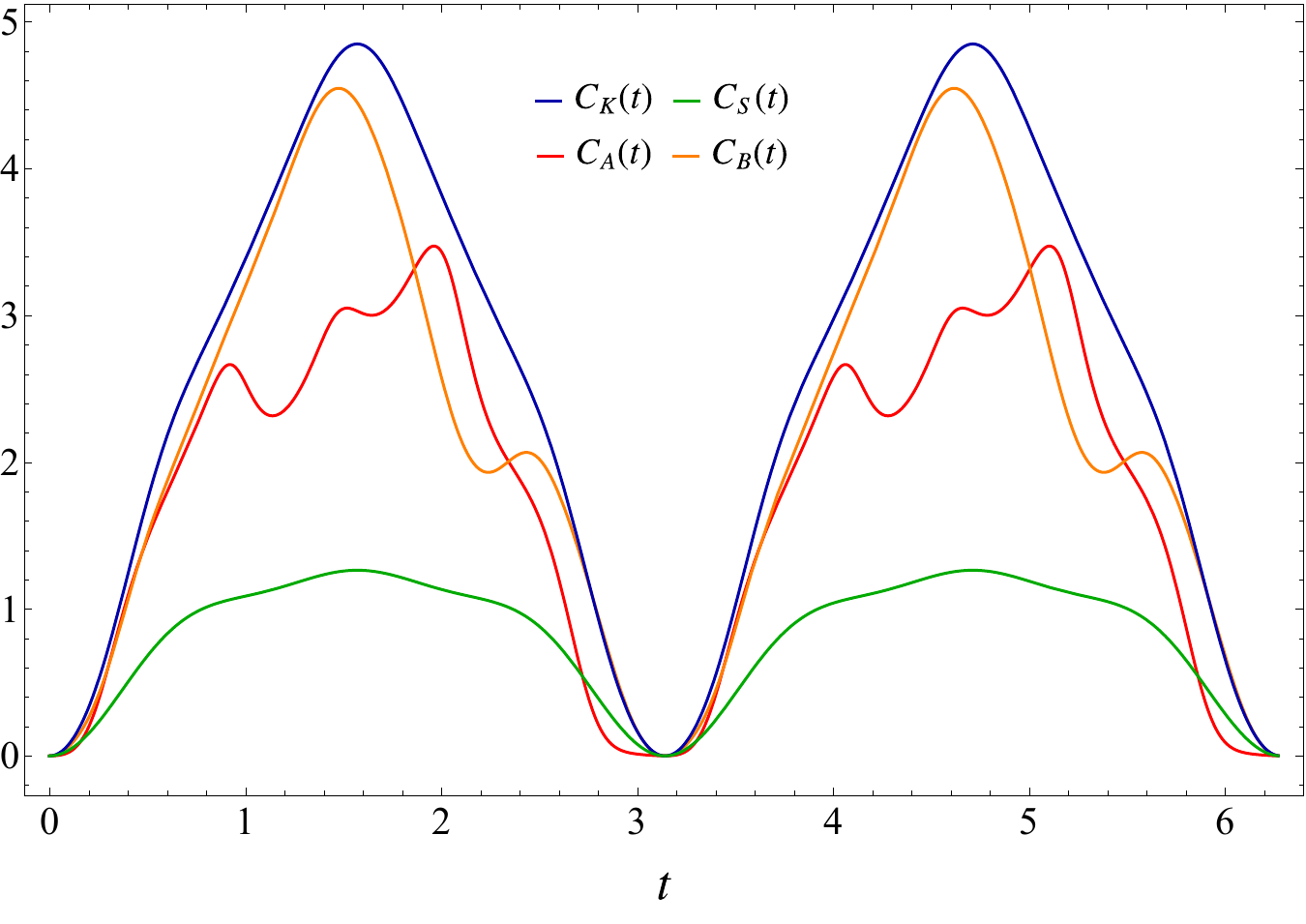}
\vspace{-0.7cm}
\caption{The initial state \eqref{eq:initst_3q} with parameters
$(\theta_1, \theta_2, \theta_3; \varphi_1, \varphi_2, \varphi_3)
= \left(\frac{\pi}{4}, 0, \frac{\pi}{2}; \frac{\pi}{2}, 0, 0\right)$ is evolved via the Hamiltonian \eqref{eq:Ham_3q}. The system is bipartitioned into a subsystem $A$ consisting of the first qubit and a complementary subsystem $B$ containing the remaining two qubits. In the left panel, we plot the return amplitude and the autocorrelation function of the pure state density matrix defined in \eqref{eq:returnamplitudes}, together with the subsystem return amplitudes \eqref{eq:generaldef_RL} associated with the two complementary subsystems, as functions of time. In the right panel, we show the corresponding complexities.} 
\label{fig:5}
\end{figure}

Examples of the return amplitudes $R_{A}$, $R_{K}$, and $R_{S}$, together with the corresponding complexities, are shown in Fig.\,\ref{fig:5} for a representative choice of parameters. In both panels, we observe oscillatory curves. The Krylov complexity of the full density matrix is the largest quantity for all times, while there is no universal ordering among the remaining complexities, in contrast with what was found in Fig.\,\ref{fig:3}.
For most values of time, $C_S$ is again the smallest quantity.
\begin{figure}[t!]
\centering
\includegraphics[width=0.495\textwidth]{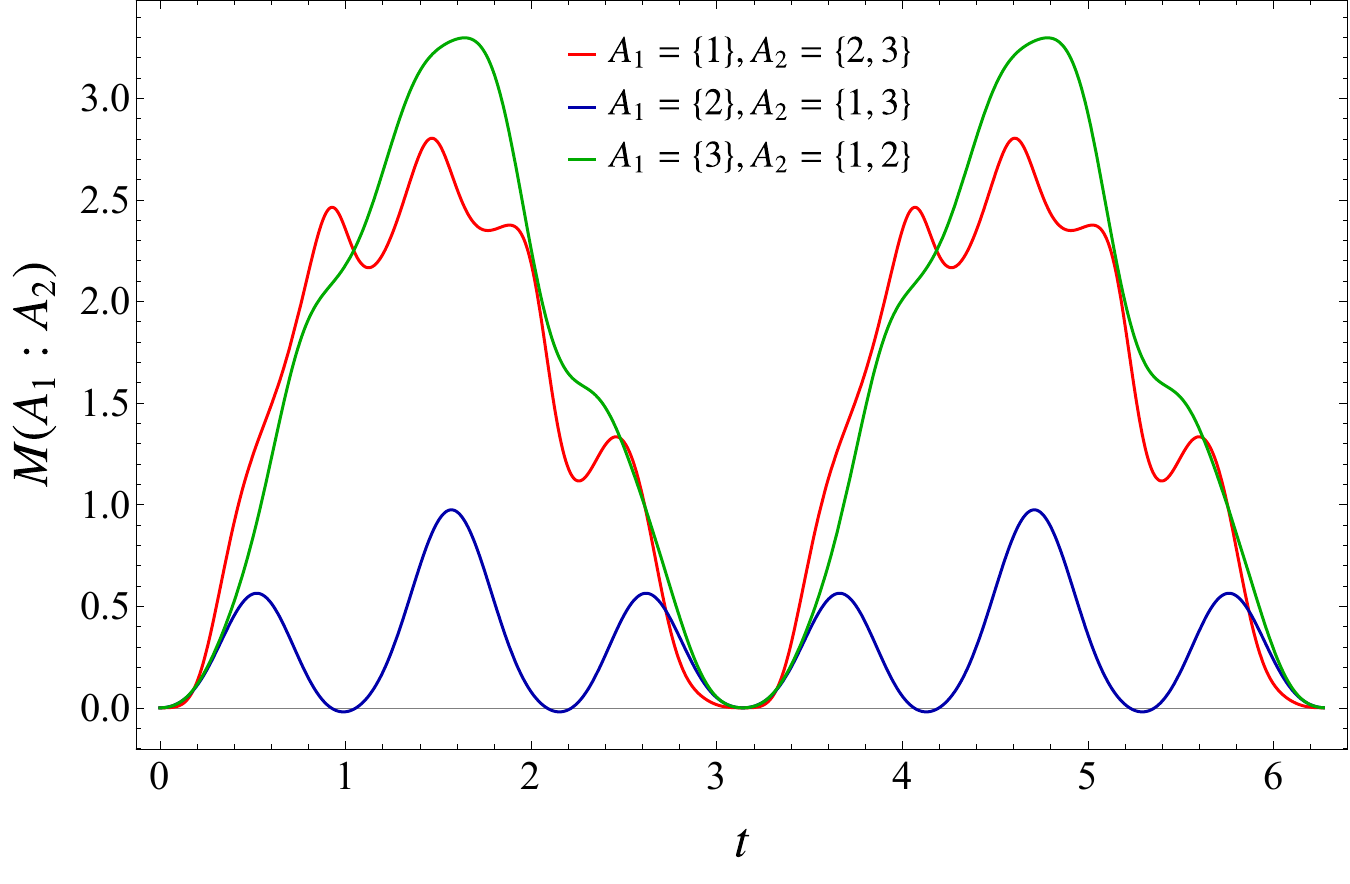}
\includegraphics[width=0.495\textwidth]{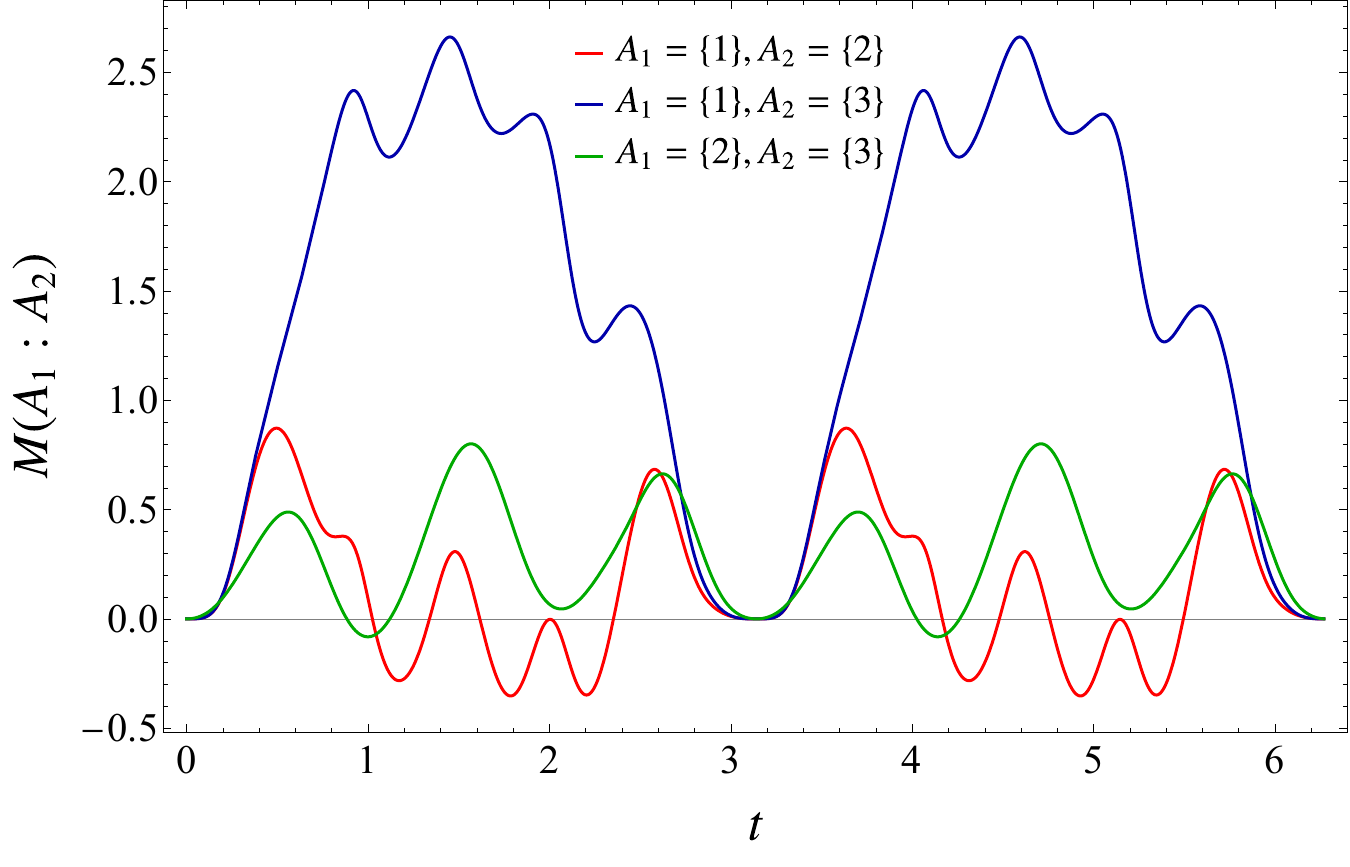}
\vspace{-.7cm}
\caption{
The initial state \eqref{eq:initst_3q} with parameters
$(\theta_1, \theta_2, \theta_3; \varphi_1, \varphi_2, \varphi_3)
= \left(\frac{\pi}{4}, 0, \frac{\pi}{2}; \frac{\pi}{2}, 0, 0\right)$ is evolved via the Hamiltonian \eqref{eq:Ham_3q}.
We plot the mutual complexity \eqref{eq:mutualcomplexity}, computed using the subsystem Krylov complexity (or the Krylov complexity of the density matrix when the state is pure). In the left panel, the subsystem $A$ is decomposed into $A_1$, consisting of a single qubit (specified in brackets), and $A_2$, containing the remaining two qubits. In the right panel, the curves are obtained by considering a subsystem $A$ made of two qubits, further partitioned into $A_1$ and $A_2$ as specified in the brackets.
}
\label{fig:6}
\end{figure}
\begin{figure}[b!]
\centering
\includegraphics[width=0.495\textwidth]{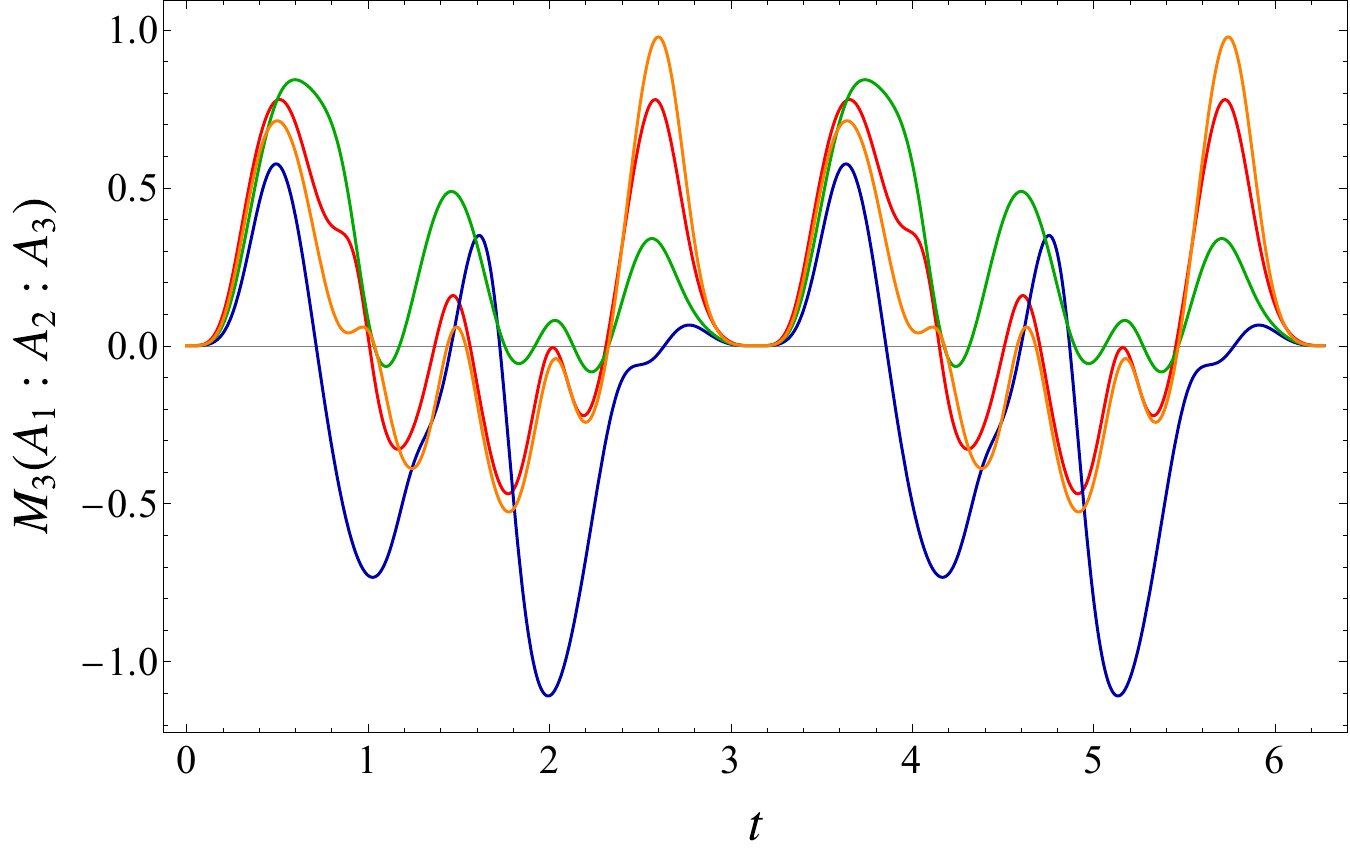}
\includegraphics[width=0.495\textwidth]{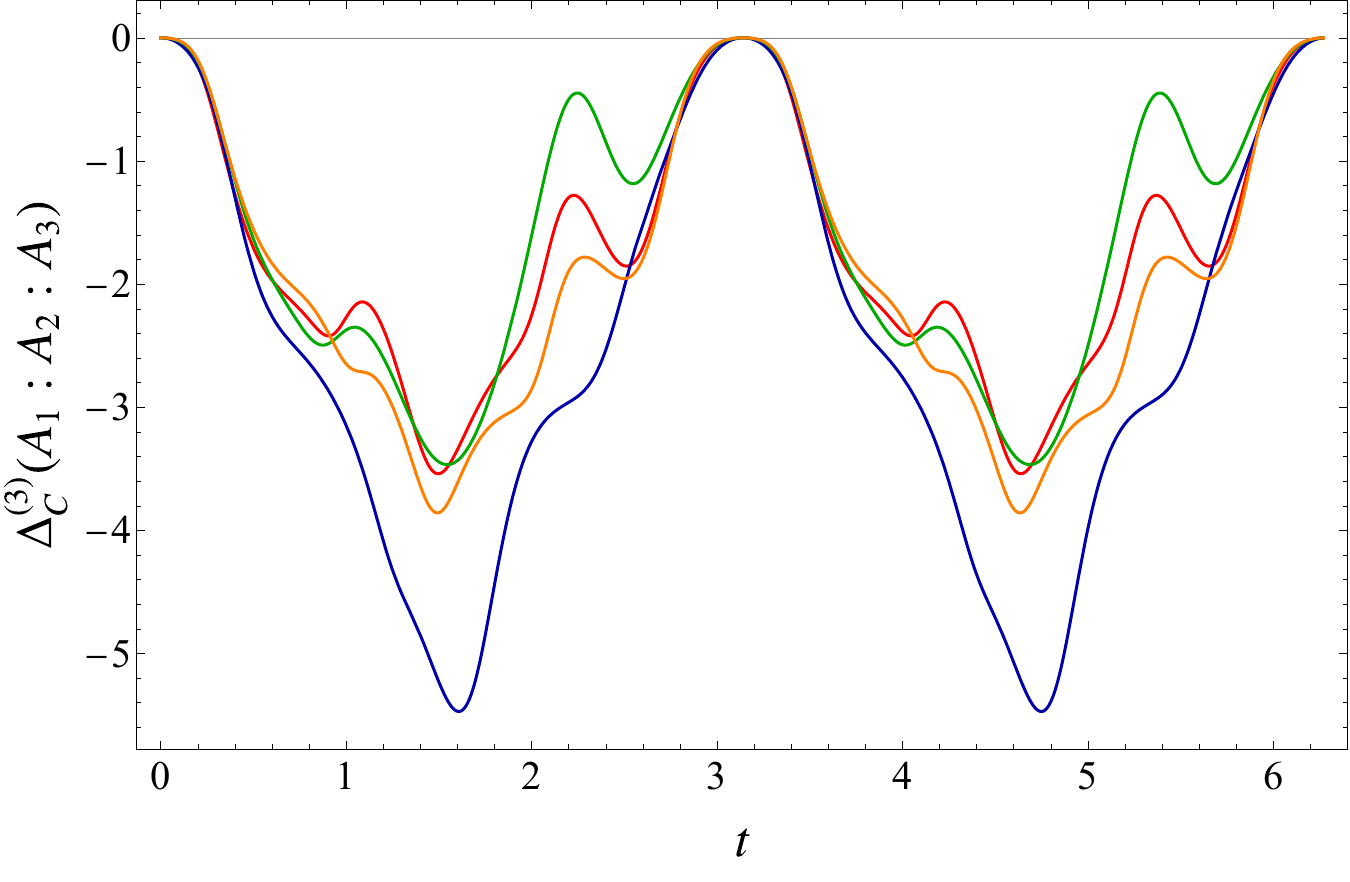}
\renewcommand{\arraystretch}{1.4}
\resizebox{0.3\textwidth}{!}{
\begin{tabular}{c|ccc|ccc}
\toprule
\textbf{Curve} & $\theta_{1}$ & $\theta_{2}$ & $\theta_{3}$ 
& $\varphi_{1}$ & $\varphi_{2}$ & $\varphi_{3}$ \\
\midrule

\tikz{\draw[thick,
color={rgb,255:red,255;green,0;blue,0}]
(0,0) -- (0.6,0);}
& $\frac{\pi}{4}$ & $0$ & $\frac{\pi}{2}$
& $\frac{\pi}{2}$ & $0$ & $0$ \\

\tikz{\draw[thick,
color={rgb,255:red,0;green,0;blue,170}]
(0,0) -- (0.6,0);}
& $\frac{\pi}{3}$ & $\frac{\pi}{4}$ & $\frac{2\pi}{3}$
& $\frac{\pi}{6}$ & $\frac{7\pi}{6}$ & $\pi$ \\

\tikz{\draw[thick,
color={rgb,255:red,0;green,170;blue,0}]
(0,0) -- (0.6,0);}
& $\frac{\pi}{2}$ & $\frac{\pi}{3}$ & $\frac{\pi}{5}$
& $\frac{7\pi}{4}$ & $\frac{5\pi}{4}$ & $\frac{3\pi}{2}$ \\

\tikz{\draw[thick,
color={rgb,255:red,255;green,128;blue,0}]
(0,0) -- (0.6,0);}
& $\frac{\pi}{7}$ & $\frac{4\pi}{7}$ & $\frac{5\pi}{7}$
& $\frac{\pi}{3}$ & $\frac{\pi}{2}$ & $\frac{\pi}{4}$ \\

\bottomrule
\end{tabular}}
\caption{Different initial states \eqref{eq:initst_3q}, with parameters specified in the table, are evolved under the Hamiltonian \eqref{eq:Ham_3q}. The three-qubit system is tri-partitioned into $A_1$, $A_2$, and $A_3$, each containing a single qubit. We plot the tripartite complexity \eqref{eq:tripartitecomplexity} (left panel) and the complexity gap in \eqref{eq:Mtilde} (right panel) as functions of time. These quantities are computed using the subsystem Krylov complexity, or the Krylov complexity of the density matrix whenever the corresponding state is pure.}
\label{fig:7}
\end{figure}

The presence of three qubits provides more possibilities for choosing bipartitions and, consequently, more configurations in which to study the mutual complexity \eqref{eq:mutualcomplexity} using the subsystem Krylov complexity. In Fig.\,\ref{fig:6}, we consider two representative cases.
In the left panel, the curves are obtained by choosing $A$ as a single qubit and $B$ as the subsystem formed by the remaining two qubits. In this case, the last term in \eqref{eq:mutualcomplexity} corresponds to the pure state Krylov complexity involving all three qubits.
In the right panel, instead, $A$ and $B$ are chosen to consist of two different qubits. Consequently, the last term in \eqref{eq:mutualcomplexity} is itself a subsystem Krylov complexity.
In both plots, the reported curves are oscillatory functions of time. However, it is interesting to observe that, while in the left panel all the curves remain positive (up to a short-time transient regime for the blue curve), the functions shown in the right panel can significantly deviate from zero on both the positive and negative sides.

In this model we can also consider a tripartition into three distinct qubits. Thus, quantities such as the tripartite complexity \eqref{eq:tripartitecomplexity} and complexity gap in \eqref{eq:Mtilde} can be computed within the subsystem Krylov complexity framework. The results for some choices of parameters are shown in Fig.\,\ref{fig:7}. Interestingly, while the tripartite complexity does not have a definite sign along the time evolution, the complexity gap remains always non-positive, and we have verified this for several additional sets of parameters.

We come back to the physical interpretation of these findings later in Sec.\,\ref{subsec:KrylovSubsyst_general}.

\subsection{Examples from coherent states}\label{subsec:coherentstates}

An important class of analytical results for spread and Krylov complexity has been obtained in cases where the evolution Hamiltonian can be expressed in terms of generators of a semi-simple Lie algebra \cite{Caputa:2021sib,Balasubramanian:2022tpr,Patramanis:2021lkx}. In such situations, the evolution can be formulated in terms of generalized coherent states \cite{Perelomov:1971bd}, and the corresponding Krylov space dynamics can be solved analytically.
In this section, we consider examples from this class in which a bipartition of the Hilbert space can be naturally identified, and the subsystem Krylov complexity \eqref{eq:subsystemKrylovcomplexity} can be computed analytically.

\subsubsection{SL(2,$\mathbb{R}$) dynamics}\label{subsc:SL2R}

We begin with the case in which the unitary dynamics of interest exhibits an emergent SL(2,$\mathbb{R}$) symmetry and can therefore be effectively described by a Hamiltonian built from the generators of the SL(2,$\mathbb{R}$) algebra. Since our focus is on subsystem Krylov complexity, which requires computing the reduced density matrix of a given subsystem through a partial trace over its complement, the two-mode representation of SL(2,$\mathbb{R}$) proves particularly convenient.
In what follows, we mainly focus on the subsystem dynamics, referring the reader to the standard literature \cite{Caputa:2021sib,Balasubramanian:2022tpr} for most details concerning the Krylov-space dynamics of the pure state.

Recall that the SL(2,$\mathbb{R}$) generators can be written as (see e.g. \cite{Caputa:2021sib})
\be
\label{eq:2moderep_SL2Rgenerators}
L_{-1}=c^\dagger d^\dagger\,,\qquad L_{1}=c d\,,\qquad L_0=\frac{1}{2}\left(c^\dagger c+d^\dagger d+1\right)\,,
\ee
in terms of two bosonic modes such that $[c,c^\dagger]=[d,d^\dagger]=1 $, with all the other commutators vanishing. These canonical commutation relations imply that the generators \eqref{eq:2moderep_SL2Rgenerators} satisfy the SL(2,$\mathbb{R}$) algebra $[L_n,L_m]=(n-m)L_{n+m}$ for $n,m\in\{-1,0,1\}$.

Now, consider a Hamiltonian of the form
\be
\label{eq:SL2RHamiltonian}
H=\alpha(L_1+L_{-1})
+\gamma L_0
\,,
\ee
with real parameters $\alpha$ and $\gamma$, and the time evolution
\be
\label{eq:2moderep_SL2Revolution}
\vert\psi(t)\rangle=e^{-{\rm i}H t}\vert k,0\rangle
\,,
\ee
where $\vert k,0\rangle$ is the tensor product of Fock states of the two modes $c$ and $d$ respectively. Crucially, the generators \eqref{eq:2moderep_SL2Rgenerators} act on $\vert k,0\rangle$ as on the SL(2,$\mathbb{R}$) highest weight state labeled by $h=\frac{k+1}{2}$.
Using the SL(2,$\mathbb{R}$) algebra, we can expand \eqref{eq:2moderep_SL2Revolution} as
\be
\label{eq:2moderep_SL2Revolution_gammanon0}
\vert\psi(t)\rangle=(1-|z|^2)^{\frac{k+1}{2}} e^{{\rm i}\frac{k+1}{2} F(z)}\sum^\infty_{n=0}z^n\sqrt{\frac{\Gamma(k+1+n)}{n!\Gamma(k+1)}}|n+k,n\rangle
\,,
\ee
where $z(t)$ is a known complex functions of time, $\alpha$ and $\gamma$, and $F(z)$ depends on the same parameters, as well as on $\arg(z(t))$. Explicit form these functions will not be important in our analysis, but we refer the interested reader to \cite{Caputa:2021sib,Balasubramanian:2022tpr} for details. 
From the action of the Hamiltonian on the Fock bases, we know that $|n+k,n\rangle$ is the Krylov basis. Importantly, it is also the Schmidt basis for the state $\vert\psi(t)\rangle $ with respect to the bipartition between the two modes.

From the Krylov space perspective, the dynamics of the state \eqref{eq:2moderep_SL2Revolution_gammanon0} is uniquely determined by the 
return amplitude
\be
\label{eq:returnamplitudeSl2R}
R_S(t)=e^{-{\rm i}\frac{k+1}{2}F(z)}(1-|z|^2)^{\frac{k+1}{2}}\,.
\ee
From this function, using standard techniques, we can analytically determine the spread complexity
\cite{Balasubramanian:2022tpr}
\be
\label{eq:complexitySl2R}
C_S(t)=(k+1)\frac{|z(t)|^2}{1-|z(t)|^2}\,.
\ee
We can also study the unitary dynamics of the full density matrix $\rho(t)=\vert\psi(t)\rangle\langle\psi(t)\vert $. In the Krylov space, this evolution is governed by the autocorrelation function. Using \eqref{eq:2moderep_SL2Revolution} in \eqref{eq:returnamplitudes}, we find that the autocorrelation function reads
\be
\label{eq:KreturnamplitudeSl2R}
R_K(t)={\rm Tr}(\rho(t)\rho(0))=(1-|z|^2)^{k+1}\,.
\ee
 The functions $R_K(t)$ and $R_S(t)$ differ not only by a power of two, but also by a phase factor with a non-trivial time dependence.
Unlike for $C_S(t)$ \cite{Caputa:2021sib}, due to this phase factor, the Krylov data of such operator dynamics cannot be simply identified with a single SL(2,$\mathbb{R}$) structure and solved analytically.
Still, the expression \eqref{eq:KreturnamplitudeSl2R} manifestly shows the relation $R_K(t)=\vert R_S(t)\vert^2$, which was used in \cite{Caputa:2024vrn} to e.g. show that the early time growth of $C_K(t)$ is two times faster than that of $C_S(t)$.

Most importantly, in this setup, we can analytically study the dynamics of the reduced density matrices of the two bosonic modes. If we refer to the first mode as the subsystem $A$ and to the second as $B$, from \eqref{eq:2moderep_SL2Revolution}, we can compute the partial traces obtaining
\be
\label{eq:SL2RRDM_gamma0}
\rho_A(t)=
(1-|z|^2)^{k+1}
\sum^\infty_{n=0}|z|^{2n}\frac{\Gamma(k+1+n)}{n!\Gamma(k+1)}|n+k\rangle\langle n+k|
\,,
\ee
and
\be
\label{eq:SL2RRDM_gamma0right}
\rho_B(t)=
(1-|z|^2)^{k+1}
\sum^\infty_{n=0}|z|^{2n}\frac{\Gamma(k+1+n)}{n!\Gamma(k+1)}|n\rangle\langle n|\,,
\ee
where in the two reduced density matrices only one of the two initial bosonic modes is left.
The corresponding subsystem return amplitudes \eqref{eq:generaldef_RL} read
\be
R_A(t)=R_B(t)=(1-|z|^2)^{k+1}=R_K(t)\,.
\ee
The identification with $R_K(t)$ can be traced back to the equivalence between the Krylov basis of the state evolution \eqref{eq:2moderep_SL2Revolution} and the Schmidt basis (see the discussion in Appendix \ref{app:SchmidtDecomp}).
This implies that the subsystem complexities are equal to Krylov complexity, namely
\be
\label{eq:complexityLRK_SL2R}
C_A(t)=C_B(t)=C_K(t)\,,
\ee
regardless of their explicit form that we are not able to derive in general.
Interestingly, in this case, all the Lanczos coefficients $a^{(A)}_n$ and $a^{(B)}_n$ are zero and the dynamics of the reduced density matrices is represented in the subsystem Krylov space by a unitary Schrödinger-like dynamics.
As commented before, all these complexities can be compared in general only at early times, where we observe that $C_S(t)\leq C_A(t)=C_B(t)=C_K(t)$.

It is also insightful to comment on the entanglement of this particular bipartition. From the reduced density matrices \eqref{eq:SL2RRDM_gamma0} and \eqref{eq:SL2RRDM_gamma0right}, we observe that the Schmidt eigenvalues are the same as the Krylov probabilities of the state \eqref{eq:2moderep_SL2Revolution}. As a consequence, the entanglement entropy $S_A$ of $\rho_A$ (and obviously that of $\rho_B$) is equal to the K-entropy of \eqref{eq:2moderep_SL2Revolution}. This is another consequence of the identification between Schmidt basis and Krylov basis. In the limit $\vert{z}\vert^2\to 1$, the leading behaviour of the entanglement entropy is
\cite{Balasubramanian:2022tpr}
\be
\label{eq:EEleft_SL(2R)}
S_A=-(k+1)\frac{(1-\vert{z}\vert^2)\ln(1-\vert{z}\vert^2)+\vert{z}\vert^2\ln\vert{z}\vert^2}{1-\vert{z}\vert^2}+\dots\,.
\ee
Varying the parameters of the Hamiltonian \eqref{eq:SL2RHamiltonian},
the entanglement entropy (as well as the spread complexity of the state \eqref{eq:2moderep_SL2Revolution}) can have three distinct behaviors. If $\gamma^2>\alpha^2$, $S_A$ (and $C_S$) are oscillating in time, if $\gamma^2<\alpha^2$, $S_A$ grows linearly in time ($C_S$ exponentially), while, if $\gamma^2=\alpha^2$, $S_A$ grows logarithmically ($C_S$ quadratically). Interestingly, the monotonicity properties of $S_A$ and $C_S$ are the same.

An interesting case where the analytical computations can be pushed further is the case where we set $\gamma=0$ in \eqref{eq:SL2RHamiltonian}. If this happens, we can check that $F(z)=0$ and, therefore, $R_K(t)=R_S^2(t)$.
This implies that, also the dynamics of the full density matrix is effectively characterized by a single SL(2,$\mathbb{R}$) symmetry. One can repeat the computation for $C_S$ by replacing  $k+1\to 2(k+1)$. Thus, we find
\be
\label{eq:complexityLRK_SL2R_gamma0}
C_A(t)=C_B(t)=C_K(t)=2(k+1)\frac{|z|^2}{1-|z|^2}\geq C_S(t)\,.
\ee
Thus, the hierarchy observed at early times when $\gamma\neq 0 $ is valid at any time when $\gamma= 0 $.
We remark that \eqref{eq:complexityLRK_SL2R} was observed in \cite{Das:2024zuu} for $h=1/2$, i.e. $k=0$, in the context of purification Krylov complexity. 

\subsubsection{SU(2) dynamics}

In this subsection, we adapt the previous analysis to the study of reduced density matrix dynamics in cases where the underlying unitary evolution exhibits an emergent SU(2) symmetry.
 
The two-mode representation of the SU(2) generators is given by 
\be
J_3=\frac{1}{2}(c^\dagger c-d^\dagger d)\,,\qquad J_+=c^\dagger d\,,\qquad J_-=d^\dagger c\,,
\ee
where the modes $c$ and $d$ are the same bosonic operators as in the previous subsection. Expressing the SU(2) Casimir in terms of the bosonic modes, we find 
\be
J_3^2+\frac{1}{2}\left(J_+J_-+ J_- J_+\right)=\frac{c^\dagger c+d^\dagger d}{2}\left(\frac{c^\dagger c+d^\dagger d}{2}+1\right)\,,
\ee
which implies that the irreducible representations (irreps) of SU(2) are labeled by $j=\frac{n_c+n_d}{2}$, where $n_c$ and $n_d$ are the eigenvalues of $c^\dagger c $ and $d^\dagger d $ respectively. Combining the expressions of $j$ and the one of $m$ (eigenvalue of $J_3$) in terms of  $n_c$ and $n_d$, we find that the states in the irreps of SU(2) are labeled by
\be
\vert j,n-j\rangle=\vert n_c=n, n_d=2j-n\rangle\,,
\ee
or the counterpart where we exchange $n_c$ and $n_d$. 

We now consider $\vert j,-j\rangle$ as initial state and let it evolve as
\be
\label{eq:2moderep_SU2Revolution_gammanon0}
\vert \psi(t)\rangle=e^{-{\rm i}H t}\vert j,-j\rangle=(1+|y|^2)^{-j}e^{{-\rm i}jG(y)} \sum^{2j}_{n=0}y^n\sqrt{\frac{\Gamma(2j+1)}{n!\Gamma(2j-n+1)}}\ket{j,-j+n}\,,
\ee
where
\be
\label{eq:SU2Hamiltonian}
H=\alpha(J_++J_{-})
+\gamma J_3
\,,
\ee
with real $\alpha$ and $\gamma$. Again, we used the SU(2) Baker–Campbell–Hausdorff (BCH) formula
to expand the state in the final expression. Here, $y(t)$ is a known complex function (distinct from the $z$ introduced in the previous section) of time and the parameters of the model, while $G(y)$ is a real function depending on the same parameters and on $\arg(y(t))$. Also in this example, we omit the explicit expressions for $y(t)$ and $G(y(t))$, referring the interested reader to \cite{Caputa:2021sib,Balasubramanian:2022tpr}.

From \eqref{eq:2moderep_SU2Revolution_gammanon0}, we can compute the total density matrix
\begin{eqnarray}
\rho(t)&=&\vert\psi(t)\rangle\langle\psi(t)\vert
\\
\nonumber
&=&(1+|y|^2)^{-2j}\sum^{2j}_{n,m=0}\frac{y^n (y^*)^m\Gamma(2j+1)}{\sqrt{n!\Gamma(2j-n+1)m!\Gamma(2j-m+1)}}\ket{j,-j+n}\bra{j,-j+m}\,,
\end{eqnarray}
and consider the natural bipartition into the two bosonic modes; we call $A$ the first mode and $B$ the second. The reduced density matrices of these two subsystems read
\be
\rho_A(t)=
(1+|y|^2)^{-2j}\sum^{2j}_{n=0}\frac{|y|^{2n} \Gamma(2j+1)}{n!\Gamma(2j-n+1)}\ket{n_c=n}\bra{n_c=n}\,,
\ee
and $\rho_B(t)$ obtained by replacing $\ket{n_c=n} $ with $\ket{n_d=2j-n} $.

The (subsystem) return amplitudes and the autocorrelation function can be explicitly computed for all three evolutions above. For the state evolution \eqref{eq:2moderep_SU2Revolution_gammanon0}, the return amplitude reads
\be
\label{eq:returnamplitude_SU2}
R_S(t)=(1+|y|^2)^{-j} e^{{\rm i}jG(y)}\,,
\ee
which, due to the time dependence in $y(t)$, is a return amplitude of the SU(2) symmetry-class \cite{Balasubramanian:2022tpr}. This and the fact that $\ket{j,n-j} $ are the Krylov vectors leads to the following expression for the spread complexity
\be
\label{eq:SU2_spreadcomplexity}
C_S(t)=2j\frac{|y(t)|^2}{1+|y(t)|^2}\,.
\ee
Due to the finite dimension of the Krylov space, by writing $y(t)$ in terms of the parameters of the Hamiltonian, we find that the spread complexity is an oscillatory function of time \cite{Balasubramanian:2022tpr}.

On the other hand, the autocorrelation function of  $\rho(t)$ reads
\be
\label{eq:autocorr_SU2}
R_K(t)=(1+|y|^2)^{-2j}\,,
\ee

but, unlike $R_S(t)$, it does not fall into a single SU(2)-symmetric dynamics, and we cannot determine an analytical expression for the Krylov complexity from it \cite{Caputa:2021sib}. Nevertheless, comparing the first Lanczos coefficients of this dynamics with the one obtained from $R_S(t)$, we again confirm that $2C_S(t)=C_K(t)$ at order $t^2$ and, therefore, at early time the spread complexity is always smaller than the corresponding Krylov complexity.

The subsystem return amplitudes of $A$ and $B$ are equal and have the same expression as \eqref{eq:autocorr_SU2}. 
As proven in general in Appendix \ref{app:SchmidtDecomp}, this is related to the equivalence between the Krylov basis of the state evolution and the Schmidt basis corresponding to the bipartition into $A$ and $B$, as it appears in \eqref{eq:2moderep_SU2Revolution_gammanon0}. 
The fact that $R_K(t)=R_A(t)=R_B(t)$ implies
$C_K(t)=C_A(t)=C_B(t)$. Thus, at order $t^2$, the hierarchy $C_S(t)\leq C_A(t)=C_B(t)=C_K(t)$ holds also in this example.

For the same reason discussed for the SL(2,$\mathbb{R} $) case in Sec.\,\ref{subsc:SL2R}, the entanglement entropy of $\rho_A$ is given by the K-entropy of the evolving state \eqref{eq:SU2_spreadcomplexity}. Thus, using the results for the K-entropy from \cite{Balasubramanian:2022tpr}, we find that the entanglement entropy exhibits oscillatory behavior in time, similar to the spread complexity.

To be more quantitative,  we consider the regime $\gamma=0$ in \eqref{eq:SU2Hamiltonian}. In that case, $R_K$, $R_A$ and $R_B$ become the same as $R_S$ up to $j\to 2j$ (and an irrelevant sign). This change in the exponent of the autocorrelation function translates into a rescaling of the Krylov and subsystem complexities by a factor two, namely
\be
C_K(t)=C_L(t)=C_R(t)=2C_S(t)\,,
\ee
implying that the hierarchy $C_S(t)\leq C_A(t)=C_B(t)=C_K(t)$ holds at all times.
Interestingly, for both coherent-state examples considered in this manuscript, the mutual complexity \eqref{eq:mutualcomplexity} is non-negative at all times, irrespective of whether $\gamma \neq 0$ or $\gamma = 0$.

\subsection{Pure and mixed state Krylov complexities}\label{subsec:KrylovSubsyst_general}

In this subsection, we summarize and discuss the results obtained for subsystem Krylov complexity, with the aim of identifying its general physical properties.

In Figs.\,\ref{fig:3} and \ref{fig:5}, and in Sec.\,\ref{subsec:coherentstates}, the subsystem Krylov complexity $C_A(t)$ has been compared with the Krylov complexity $C_K(t)$ and the spread complexity $C_S(t)$ of the corresponding pure states. While $C_S(t)$ is smaller than $C_A(t)$ and $C_K(t)$ for most values of time, we do not observe any universal ordering between $C_K(t)$ and $C_A(t)$ (see also Fig.\,\ref{fig:entangled1}). The fact that both $C_K(t)$ and $C_A(t)$ tend to dominate over $C_S(t)$ can be understood from the observation that, for a given system, the Hilbert space of states is smaller than the corresponding operator Hilbert space.

To mitigate the influence of the size of the operator Hilbert space on our analysis, we compare reduced density matrices with pure-state density matrices of the same Hilbert-space dimension. For example, given a fixed number of qubits, say $n$, we compare the Krylov-space dynamics of reduced density matrices on a subsystem $A$ composed of $n$ qubits with those of pure-state density matrices for systems also consisting of $n$ qubits. We begin with $n=1$. While examples involving subsystems composed of a single qubit have been analyzed in Sections \,\ref{subsec:2spins} and \ref{subsec:3spins}, we have not yet considered the pure-state Krylov dynamics of a single qubit.

This can be easily achieved using the Hamiltonian
\begin{equation}
    H = \mu_x \sigma_x + \mu_y \sigma_y + \mu_z \sigma_z\,,
\end{equation}
to evolve the initial state
\begin{equation}
   \vert \psi(0)\rangle=
\binom{\cos \frac{\theta}{2}}{e^{\mathrm{i} \varphi} \sin \frac{\theta}{2}}\,.
\end{equation}
We do not present the details of the Krylov space analysis for this simple example, reporting only that the (largest) Krylov spaces associated with the state and density matrix dynamics have dimensions $\mathcal{K}_S=2$ and $\mathcal{K}_K=3$, respectively. This should be compared with the largest subsystem Krylov space dimensionality for the reduced density matrix of a single qubit discussed in Sec.\,\ref{subsec:3spins}, namely $\mathcal{K}_A=7$. Thus, $\mathcal{K}_A\geq \mathcal{K}_K\geq \mathcal{K}_S$.
We refer to this phenomenon as an enhancement of the dimensionality of the effective subsystem Krylov space, and we believe it to be a general feature of the present approach to describing the dynamics of reduced density matrices. We suspect that this originates from forcing the intrinsically non-Schrödinger dynamics of reduced density matrices into an effective Schrödinger-like evolution in Krylov space.

The enhancement of the dimensionality of the effective subsystem Krylov space is also observed in the case $n=2$, i.e. for the dynamics of two-qubit pure states compared to the dynamics of reduced density matrices of two qubits. In Secs.\,\ref{subsec:2spins} and \ref{subsec:3spins}, we find $\mathcal{K}_S=2$, $\mathcal{K}_K=3$, and $\mathcal{K}_A=7$, where the subsystem $A$ consists of two qubits obtained by tracing out a total three-qubit system. This confirms the hierarchy previously observed for $n=1$.
A similar analysis can be performed for $n=3$, showing that, also in this case, $\mathcal{K}_A \geq \mathcal{K}_K$, with $\mathcal{K}_A$ reaching values as large as $29$ for suitable choices of the parameters.

In the single qubit example, we can highlight the notion of enhancement of Krylov space dimensionality from another perspective. 
Given an operator acting on a Hilbert space $\mathcal{H}$, \cite{Rabinovici:2020ryf} predicts a bound on the dimensionality of the corresponding Krylov space given by $\mathcal{K}\leq\dim (\mathcal{H})^2-\dim\mathcal{H}+1$. This holds true for operators evolving unitarily, in a Heisenberg (or Schrödinger, after vectorization) way. For a  $2\times 2$ operator, this bound is $\dim (\mathcal{H})^2-\dim\mathcal{H}+1=3$. We conclude that the single-qubit reduced density matrix violates this bound due to $\mathcal{K}_A=7$, providing a further notion of Krylov space enlargement.

\begin{figure}[t!]
\centering
\includegraphics[width=0.495\textwidth]{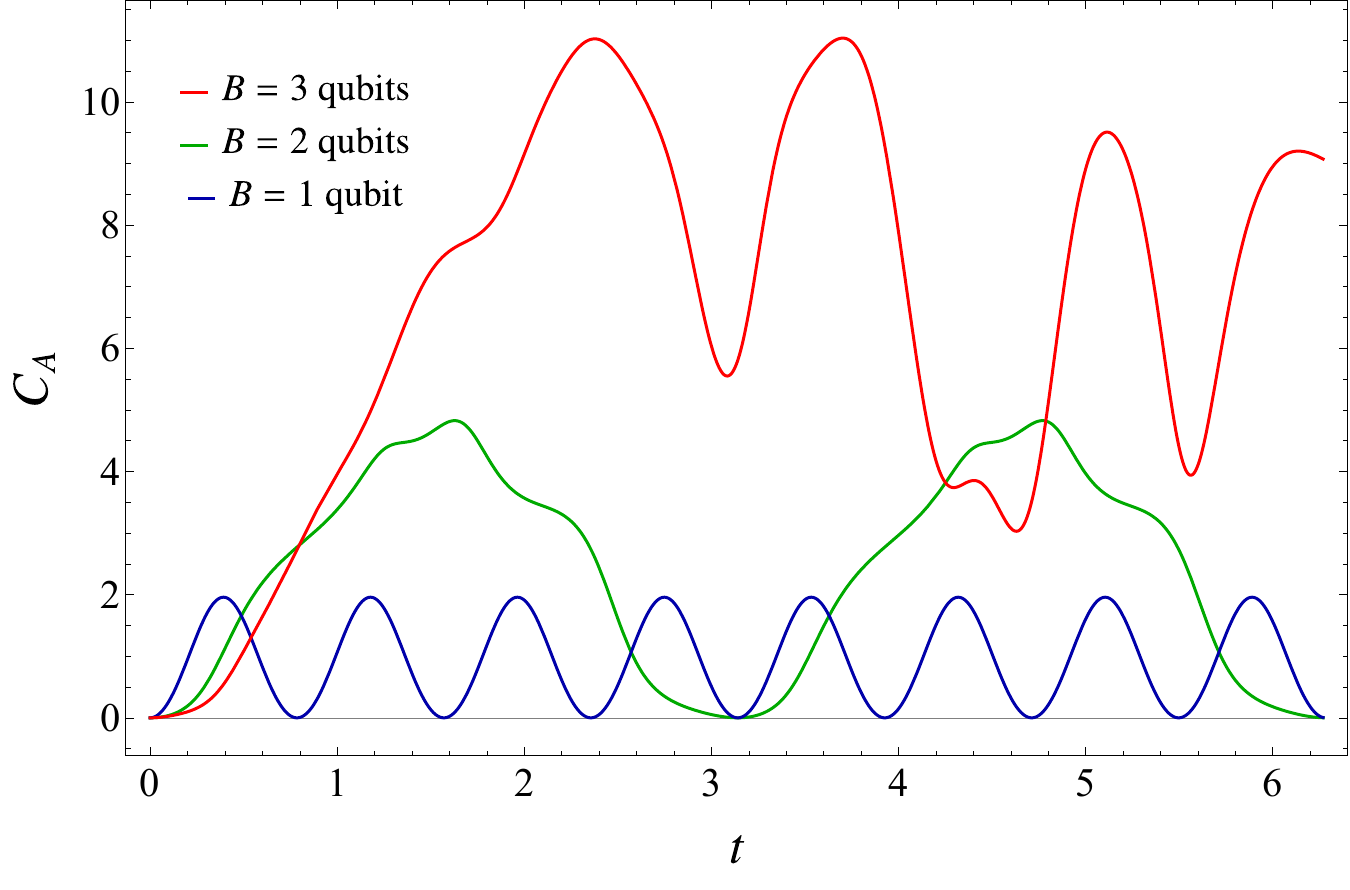}
\includegraphics[width=0.495\textwidth]{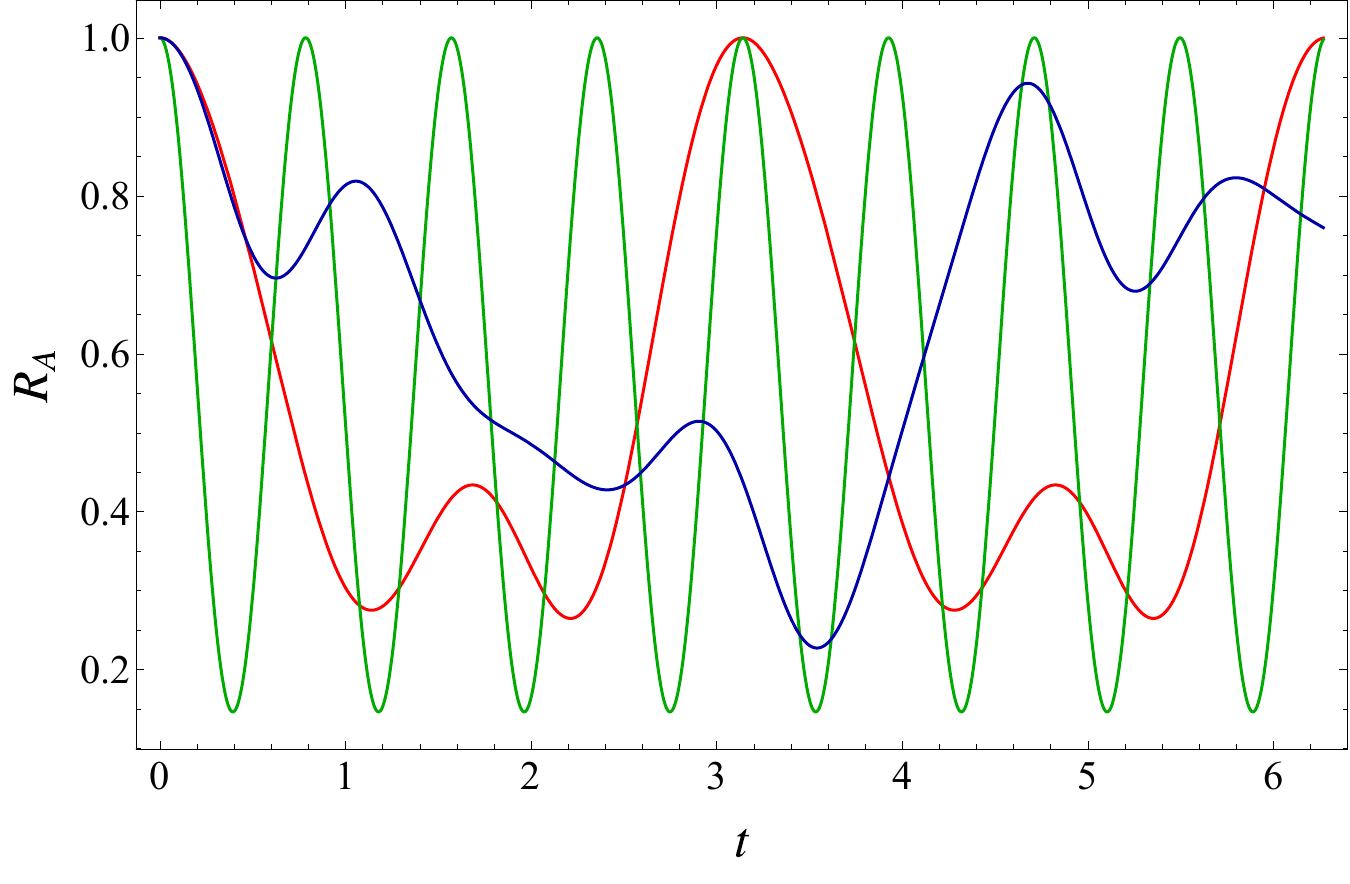}
\begin{tabular}{c|cccccccc}
\toprule
\textbf{Curve} 
& $\theta_{1}$ & $\theta_{2}$ & $\theta_{3}$ & $\theta_{4}$
& $\varphi_{1}$ & $\varphi_{2}$ & $\varphi_{3}$ & $\varphi_{4}$ \\
\midrule

\tikz{\draw[thick,
color={rgb,255:red,255;green,0;blue,0}]
(0,0) -- (0.6,0);}
& $\frac{\pi}{2}$ 
& $\frac{\pi}{3}$ 
& $\frac{\pi}{5}$ 
& $\frac{\pi}{4}$ 
& $\frac{7\pi}{4}$ 
& $\frac{5\pi}{4}$ 
& $\frac{3\pi}{2}$ 
& $\frac{\pi}{3}$ \\

\tikz{\draw[thick,
color={rgb,255:red,0;green,170;blue,0}]
(0,0) -- (0.6,0);}
& $\frac{\pi}{3}$ 
& $\frac{\pi}{4}$ 
& $\frac{2\pi}{3}$ 
& / 
& $\frac{\pi}{6}$ 
& $\frac{7\pi}{6}$ 
& $\pi$ 
& / \\

\tikz{\draw[thick,
color={rgb,255:red,0;green,0;blue,255}]
(0,0) -- (0.6,0);}
& $\frac{\pi}{8}$ 
& $\frac{5\pi}{8}$ 
& / 
& / 
& $0$ 
& $\pi$ 
& / 
& / \\

\bottomrule
\end{tabular}
\caption{
Subsystem Krylov complexity (left panel) and return amplitude (right panel) associated with the dynamics of the reduced density matrix of a single qubit, with the complementary subsystem $B$ consisting of one (blue), two (green), or three (red) qubits. The initial state is given by \eqref{eq:unentangled initialstate}, generalized to a tensor-product state involving up to four qubits, with the parameters specified in the table. The evolution Hamiltonian generalizes \eqref{eq:Ham_3q} and is given by $H_N=-\sum_{j=1}^N(X_jX_{j+1}+Y_jY_{j+1}+Z_jZ_{j+1})$, with $N=2,3,4$. 
}
\label{fig:1spin}
\end{figure}

To better understand the physical origin of the enhancement of the Krylov-space dimensionality for subsystem dynamics, we perform a further comparison.
We compare the subsystem Krylov complexities associated with a subsystem $A$ consisting of a single qubit, while varying the number of qubits in the traced-out complementary subsystem $B$. In the left panel of Fig.~\ref{fig:1spin}, we report these curves for a representative choice of parameters, showing that the larger the subsystem $B$, the larger the value around which the complexity oscillates. This corresponds to a larger dimensionality of the Krylov space associated with the reduced density matrix when the complementary subsystem becomes larger. The same analysis has been repeated for several other parameter choices, always leading to the same conclusion.
In the right panel of Fig.\,\ref{fig:1spin}, we also compare the behaviour of the subsystem return amplitudes corresponding to the complexities shown in the left panel. We observe that lower-frequency oscillations in the subsystem return amplitudes are associated with larger Krylov-space dimensionalities and larger values around which the subsystem complexity oscillates.
Our analysis relies on systems of few qubits. In the future, it would be insightful to understand whether, by further increasing the size of the complementary subsystem, the subsystem complexity continues to grow without bound or instead, due to the localization of correlations near the interfaces between the two regions, eventually saturates to a maximal value beyond a certain threshold in the size of the complementary subsystem.

Our analysis suggests that the enhancement of the Krylov space dimensionality for the reduced density matrix is driven by the complementary subsystem that is traced out. This is particularly interesting as it indicates that the subsystem Krylov complexity retains non-trivial information about the full evolving state. More precisely, we argue that, in the presence of a large number of degrees of freedom in the complement, the number of effective configurations available to the subsystem increases, leading to an increase in the subsystem Krylov space dimensionality.

Finally, we discuss the behaviour of $M(A_1:A_2)$, $M_3(A_1:A_2:A_3)$, and $\Delta^{(3)}_C(A_1:A_2:A_3)$ as functions of time analyzed in the previous section, and we argue for the implications of the enhancement of Krylov space dimensionality in subsystem dynamics on their signs.

When the subsystem $A_1\cup A_2$ coincides with the full system, the last term in \eqref{eq:mutualcomplexity} reduces to the pure state Krylov complexity. In this case, accounting for the enhancement of the Krylov space dimension in the mixed state construction, this contribution is expected to be suppressed compared to the first two terms. We therefore expect $M(A_1:A_2)$ to remain positive along the time evolution. This behaviour is indeed observed in Fig.\,\ref{fig:3}, in the left panel of Fig.\,\ref{fig:6} (up to very short negative transient regimes), and in the right panel of Fig.\,\ref{fig:entangled1} in Appendix \ref{app:nonvanishing_an}.
On the other hand, when $B$ is non-empty, $C_{A_1\cup A_2}$ is itself a subsystem Krylov complexity and its typical value is enhanced. As a consequence, the sign of $M(A_1:A_2)$ is no longer guaranteed to remain positive for all times. This is indeed what we observe in the right panel of Fig.\,\ref{fig:6}.

In the definition \eqref{eq:tripartitecomplexity} of $M_3(A_1:A_2:A_3)$, subsystem complexities appear with both positive and negative signs, making it difficult to expect a definite sign of the tripartite complexity along the dynamics. This is confirmed by the left panel of Fig.\,\ref{fig:7}, where $M_3(A_1:A_2:A_3)$ is computed for the three-qubit example discussed in Sec.\,\ref{subsec:3spins}, where the natural tripartition is given by assigning one qubit to each of $A_1$, $A_2$, and $A_3$.
To further support our picture, in the right panel of Fig.\,\ref{fig:7} we evaluate the complexity gap in \eqref{eq:Mtilde} for the same three-qubit system. Since this quantity involves a difference between a pure state Krylov complexity and subsystem Krylov complexities, a negative sign is expected. This is indeed observed in Fig.\,\ref{fig:7}.
This result further confirms our intuition regarding the enhancement of Krylov-space dimensionality in reduced density matrix dynamics.

\section{Inequalities for Fisher-Rao subsystem complexity}\label{sec:SRSpreadComplexity}

In the previous section, we proposed a method to represent the dynamics of $\rho_A(t)$ in an effective Krylov space with the aim of quantifying its complexity. Here, we discuss an alternative approach to this problem, applicable to the class of bosonic Gaussian states.
This method was originally developed in \cite{DiGiulio:2020hlz} and later studied in out-of-equilibrium settings in \cite{DiGiulio:2021oal,DiGiulio:2021noo}. In what follows, after a brief review, we apply it to investigate the out-of-equilibrium dynamics of the complexity-related quantities introduced in Sec.\,\ref{subsec:Inequalities}, involving subsystems composed of several disjoint components. We refer the interested reader to \cite{DiGiulio:2020hlz} for further details.

\subsection{Fisher-Rao subsystem complexity}\label{subsec:FR_review}

For bosonic systems, any density matrix, including reduced density matrices, can be described in terms of Wigner functions, namely quasi-probability distributions defined on a quantum phase space (see \cite{Serafini:2017rrn} for an extensive discussion). Therefore, given a reduced density matrix evolving in time, the associated Wigner function also acquires a non-trivial time dependence. A possible way to characterize the subsystem dynamics of bosonic systems is then to track the evolution of the Wigner function associated with the reduced density matrix.

The Fisher-Rao subsystem complexity provides a tool to characterize this dynamics, by interpreting the trajectory of the state in the space of Wigner functions as a quantum circuit implementing the quantum evolution. Even from this perspective, however, the problem is analytically very challenging. For this reason, it is convenient to restrict attention to the class of bosonic Gaussian states, which are more tractable while still playing a central role in several applications ranging from quantum information to quantum optics \cite{Weedbrook:2011wxo,Serafini:2017rrn}.

It is known that the states in this class can be described by their Wigner functions which are Gaussian functions of the quantum phase space variables. More precisely, we describe any $N$-mode bosonic Gaussian state $\hat\rho$ in terms of the Wigner function 
\begin{equation}
  W_{\hat{\rho}}(\boldsymbol{r})=\frac{\exp\left(-\frac{1}{2}(\boldsymbol{r}-\bar{\boldsymbol{r}})^\textrm{t}\gamma^{-1}(\boldsymbol{r}-\bar{\boldsymbol{r}})\right)}{(2\pi)^N\sqrt{\det\gamma}}  \,,
\end{equation}
where the vector $\boldsymbol{r}\in\mathbb{R}^{2N}$ can be written as $\boldsymbol{r}=(\boldsymbol{q},\boldsymbol{p})^{\textrm{t}}\equiv(q_1,\dots,q_N,p_1,\dots,p_N)^{\textrm{t}}$, with these $2N$ coordinates that parameterize the quantum phase space, and
\begin{equation}
   \bar{r}_j\equiv
{\rm Tr}\left[\hat{\rho}\,\hat{r}_j \right]\,,
\quad
\gamma_{ij}\equiv\frac{1}{2}
{\rm Tr}\left[ \hat{\rho}\,\{(\hat{r}_i-\bar{r}_i), (\hat{r}_j-\bar{r}_j)\}\right]
\,, 
\end{equation}
with $\{,\}$ denoting the anticommutator and $\boldsymbol{\hat{r}}\equiv \left(\hat{q}_1,\dots, \hat{q}_N ,\hat{p}_1,\dots, \hat{p}_N\right)^{\textrm{t}}$ containing the quadrature operators which satisfy the usual canonical commutation relations.
Since the average of the quadratures can always be set to zero via a unitary transformation on $\hat\rho$, we restrict our analysis to the case $\bar{r}_j=0 $. The space of  $N$-mode bosonic Gaussian state is now parameterized only by the $2N \times 2N$ covariance matrix $\gamma$, which being symmetric and positive definite, depends on $ N(2N+1)$ real parameters. 
To avoid confusion between operators and quantum phase space variables, in this section we denote the operators acting on the Hilbert space of bosonic states with a hat.

The physical idea behind the subsystem complexity proposal reviewed in this section is to study circuits made up by Gaussian gates, namely operations mapping bosonic Gaussian states into bosonic Gaussian states. Given a reference state $\gamma_R$ (with abuse of notation, we identify a bosonic Gaussian state with its covariance matrix), a target state $\gamma_T$ can be constructed by applying bosonic Gaussian gates to $\gamma_R$. In the Nielsen's picture \cite{Nielsen:2006cea}, if the gates are infinitesimal, this problem can be mapped to studying the curves connecting $\gamma_R$ and $\gamma_T$ in the geometry of bosonic Gaussian states, i.e. the geometry of Gaussian distributions, with the uncertainty principle implemented as a constraint on (the eigenvalues of) the covariance matrix as $\gamma+{\rm i }J/2\geq 0 $. 
In this picture, the complexity is the length of the geodesics connecting $\gamma_R$ and $\gamma_T$, interpreted as the optimal circuit built with Gaussian gates. 

A non-trivial ambiguity comes from the choice of the metric that, in Nielsen's picture, is related to the choice of allowed gates and the cost function. In \cite{DiGiulio:2020hlz} the Fisher-Rao metric was singled out based on physical considerations. Indeed, according to Chentsov’s theorem, it is the unique metric on the space of Gaussian states satisfying {\it information monotonicity} \cite{amaribook}. This means that, when distinguishability between Gaussian states is measured using the Fisher--Rao metric, pairs of states related by a coarse-graining transformation become less distinguishable. This property can be regarded as the counterpart, for Gaussian Wigner functions, of the contractivity properties satisfied by quantum distances \cite{Nielsen:2012yss}.

The geodesics length associated with this metric is known as the Fisher-Rao distance \cite{Bhatiabook}. We interpret it as the complexity of constructing  $\gamma_T$ from $\gamma_R$. More explicitly:
\begin{equation}
\label{eq:FRcomplexity}
    C_{\rm \tiny FR}(\gamma_R,\gamma_T)=\frac{1}{2\sqrt{2}
    }\sqrt{{\rm Tr}\left[\ln^2\left(\gamma_T\gamma_R^{-1}\right)\right]}\,.
\end{equation}
When $\gamma_T$ and $\gamma_R$ describe pure states, \eqref{eq:FRcomplexity} gives the Nielsen complexity calculated in \cite{Jefferson:2017sdb,Chapman:2018hou}. On the other hand, this formalism remains valid also for mixed states and, in particular, for covariance matrices associated with reduced density matrices (reduced covariance matrices). In those cases, we refer to  \eqref{eq:FRcomplexity} as subsystem Fisher-Rao (FR) complexity.

One of the main advantages of this proposal is its computability. Indeed, once the two-point functions in the reference and target states are known, the expression \eqref{eq:FRcomplexity} can be evaluated rather easily, both numerically and, in some cases, analytically. This is a direct consequence of the Gaussian nature of the states involved, which reduces an otherwise difficult problem to the determination of a relatively small number of two-point functions, namely a number that scales polynomially with the degrees of freedom of the (sub)system.

In the next subsection, we apply this definition of subsystem complexity to the study of out-of-equilibrium dynamics in chains of coupled quantum oscillators, namely harmonic chains.

In particular, we study the time evolution of the mutual complexity $M$, the tripartite complexity $M_3$ and the complexity gap $\Delta^{(3)}_C$ introduced in Sec.\,\ref{subsec:Inequalities}, where the various contributions entering these quantities are now computed using the subsystem FR complexity. Moreover, one may introduce different notions of mutual and tripartite complexity, and complexity gap that are specific to the FR subsystem complexity framework. Indeed, in \cite{DiGiulio:2020hlz,DiGiulio:2021oal}, it was shown that the FR subsystem complexity grows proportionally to $\sqrt{\ell}$, where $\ell$ denotes the subsystem size. In order to work with extensive quantities in the definitions \eqref{eq:mutualcomplexity}-\eqref{eq:Mtilde}, we define, for a given subregion decomposed as $A_1\cup A_2$,
\begin{equation}
\label{eq:mutualcompl_sq}
    M^{(2)}(A_1:A_2)\equiv C^2_{A_1}+C^2_{A_2}-C^2_{A_1\cup A_2}  \,,
\end{equation}
and, given a subsystem made by three parties $A_1\cup A_2\cup A_3$,
\begin{equation}
\label{eq:tripartitecompl_sq}
 M_3^{(2)}(A_1:A_2:A_3)
 \equiv C^2_{A_1}+C^2_{A_2}+C^2_{A_3}-C^2_{A_1\cup A_2}-C^2_{A_1\cup A_3}-C^2_{A_2\cup A_3}+C^2_{A_1\cup A_2\cup A_3}   \,,
\end{equation}
and
\begin{equation}
\label{eq:Mtilde_sq}
\Delta^{(3,2)}_C(A_1:A_2:A_3)
 \equiv 
 C^2_{A_1\cup A_2\cup A_3} -C^2_{A_1}-C^2_{A_2}-C^2_{A_3}  \,.
\end{equation}
In \cite{DiGiulio:2020hlz}, the quantity \eqref{eq:mutualcompl_sq} was studied in bosonic systems at equilibrium, where it was found not to possess a definite sign. In the next sections, we investigate this quantity, together with \eqref{eq:mutualcomplexity}, \eqref{eq:tripartitecomplexity},
\eqref{eq:Mtilde}, \eqref{eq:tripartitecompl_sq} and \eqref{eq:Mtilde_sq}, along out-of-equilibrium dynamics.

\subsection{Harmonic chains after a global quench of the mass}\label{subsec:HC_quench}

In this section, we study the subsystem complexity and related quantities such as mutual and tripartite complexity, and complexity gap in a harmonic chain, i.e. a one-dimensional system of coupled harmonic oscillators.

\subsubsection{The model}\label{subsec:model}

The harmonic chain is defined by the Hamiltonian
\begin{equation}
   \label{eq:HamHL}
\widehat{H}
=
\sum_{i=1}^L\left(
\frac{1}{2 m}\hat{p}_i^2+\frac{m \omega^2}{ 2}\hat{q}_i^2
\right)+\sum_{i=1}^L\frac{\kappa}{2}(\hat{q}_{i}-\hat{q}_{i-1})^2
\,,
\end{equation}
where $\hat{q}_{i}$ and $\hat{p}_{i}$ are the canonically commuting operators introduced before.

We assume that the harmonic chain lies on a segment with Dirichlet boundary conditions (DBC) at its boundaries, i.e. $\hat{q}_{0}=\hat{q}_{L}=0$ and $\hat{p}_{L}=0$. Without loss of generality, we can set $m=1$ and $\kappa=1$, which amounts to a rescaling of the canonically commuting variables.
We refer to the remaining frequency parameter $\omega$ as the mass. Indeed, $\omega$ sets the gap of the excitations of the model and, in the limit $\omega\to 0$, the harmonic chain becomes gapless and critical, with a continuum limit described by a free bosonic 1+1-dimensional CFT.

Since the Hamiltonian is quadratic, the ground state of this model is a Gaussian state labeled by the mass parameter $\omega$. After a bipartition and the partial trace, the reduced ground state becomes a mixed state but remains a Gaussian state.
In this work, we are interested in the out-of-equilibrium dynamics induced by a global mass quench. Such protocol is defined by initializing the system in the ground state of \eqref{eq:HamHL} with mass $\omega_0$ and letting it evolve via \eqref{eq:HamHL} with mass $\omega\neq\omega_0$. Given that the Hamiltonians with two different mass parameters do not commute, the described dynamics is highly non-trivial. Still, the Gaussian nature of the initial state, together with the quadratic form of the evolution Hamiltonian, ensures that the state evolved at generic times remains Gaussian.

To study the evolution of the subsystem FR complexity after this quench, we denote by $A$ the subsystem under consideration, and by $\vert A\vert$ the number of sites it contains.
We assume that the reference state is the initial state and the target state is the state evolved at time $t$. The choice of DBC is convenient because these boundary conditions avoid the occurrence of the zero mode \cite{DiGiulio:2021oal}. 

In practice, we have to construct the covariance matrices of the reference and target states. For $s\in {R,T}$, we can decompose the covariance matrices into  $\vert A\vert \times \vert A\vert$ blocks
\begin{equation}
    \gamma_s=
    \begin{pmatrix}
        Q_s & M_s
        \\
        M_s^{\rm t} & P_s
    \end{pmatrix}
    \,,
\end{equation}
where
\begin{eqnarray}
(Q_s)_{ij}&=&\Tr\left(\hat\rho_{s,A}\hat{q}_i\hat{q}_j\right)\,,
    \qquad
(P_s)_{ij}=\Tr\left(\hat\rho_{s,A}\hat{p}_i\hat{p}_j\right) \,,
    \\
    (M_s)_{ij}&=&{\rm Re}\left[\Tr\left(\hat\rho_{s,A}\hat{q}_i\hat{p}_j\right)\right]\,,
    \qquad
   i,j=1,\dots,\vert A\vert \,,
\end{eqnarray}
 and $\hat\rho_{s,A}$ is the reduced density matrix of either the reference or the target state. The indices of the blocks run over all lattice sites in the subsystem $A$.

These covariance matrices are usually referred to as {\it reduced covariance matrices}, since they contain two-point functions restricted to the subsystem $A$. We do not report here the explicit expressions for the correlators as functions of time and parameters of the model, as they have been extensively studied and employed throughout the literature. We refer the interested reader to \cite{DiGiulio:2021oal}, Sec.\,2.1.
As anticipated, in the remainder of this manuscript we use the two-point functions for the harmonic chain on a finite segment with DBC. The same analysis can also be carried out for a periodic harmonic chain, upon appropriately modifying the correlators (see \cite{DiGiulio:2021oal}).

The FR subsystem complexity after this quench has been studied in \cite{DiGiulio:2021oal} for a subsystem given by a single block of consecutive oscillators. Interesting qualitative similarities with results from holography \cite{Chen:2018mcc,Auzzi:2019mah} have been found, which, due to the recent progress in \cite{Haah:2025hyf,Fan:2025moc} are also features of subsystem complexity in random circuits.
In this work, we mainly consider subsystems made up of two or three blocks of consecutive oscillators, and we construct mutual and tripartite FR complexities, and the corresponding complexity gap, associated with these bipartitions.

\subsubsection{Analysis for two subsystems}\label{subsec:twointervals}

\begin{figure}[t!]
\hspace{-1.5cm}
\includegraphics[width=1.18\textwidth]{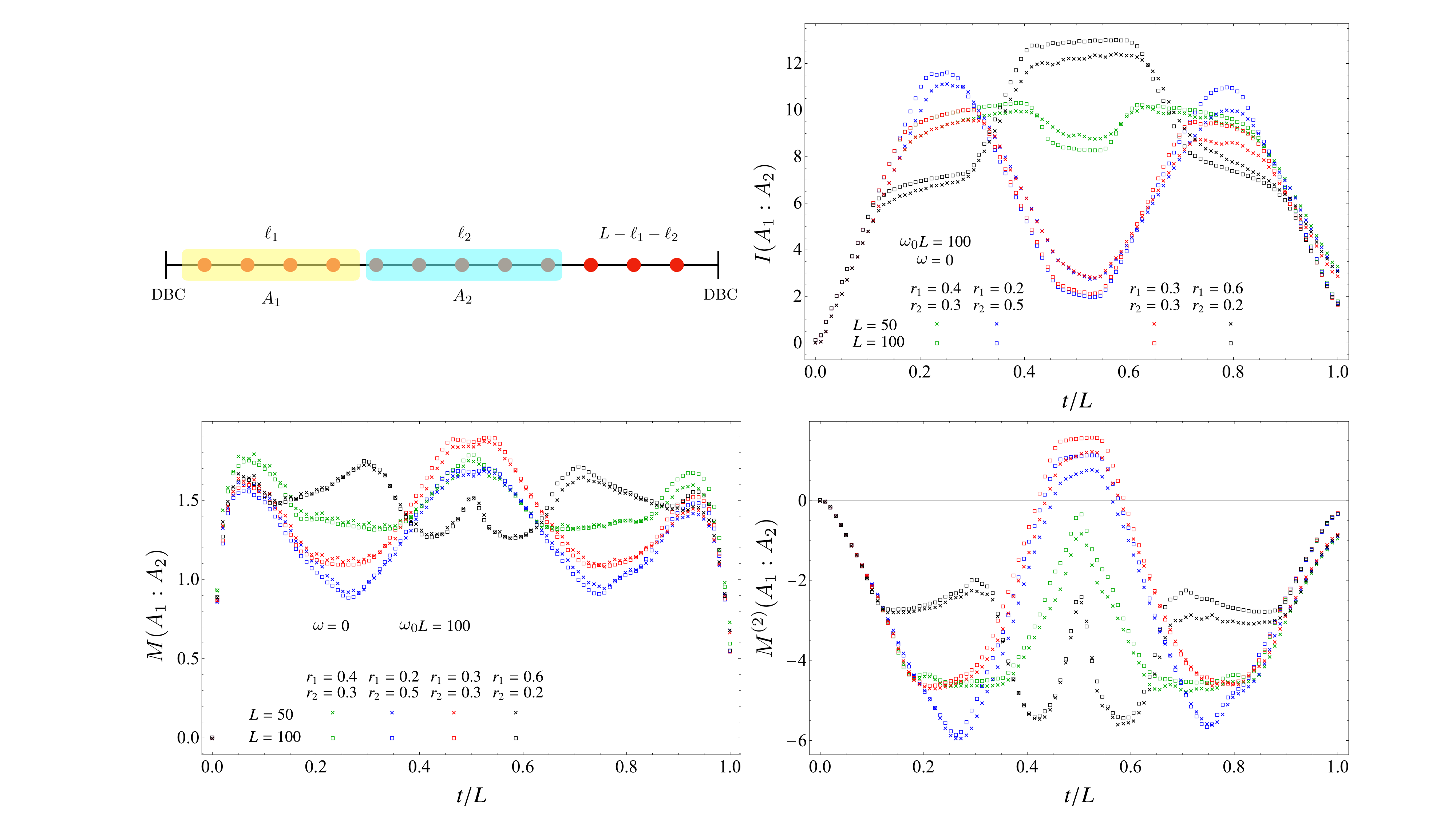}
\vspace{-0.8cm}
\caption{
We consider a partition of a harmonic chain consisting of $L$ oscillators with DBC at its boundaries into $A_1$, made of $\ell_1$ sites, $A_2$, made of $\ell_2$ sites, and $B$, made of the remaining $L-\ell_1-\ell_2$ sites, as represented in the top-left panel. We study the time evolution following a global quench of the mass parameter, from $\omega_0\neq 0$ to $\omega=0$, for various choices of subsystems parametrized by $r_i=\ell_i/L$.
In the top-right panel, we plot the mutual information \eqref{eq:mutualinfo} as a function of time. In the bottom panels, we show the mutual complexities \eqref{eq:mutualcomplexity} (bottom-left panel) and \eqref{eq:mutualcompl_sq} (bottom-right panel), computed using the FR subsystem complexity \eqref{eq:FRcomplexity}.
}
\label{fig:tripartition_time_DBC_nonemptyB}
\end{figure}

We begin our numerical analysis by considering a bipartition of the harmonic chain with Hamiltonian \eqref{eq:HamHL} (with DBC) into subsystems $A=A_1\cup A_2$ and $B$, where $A_1$ and $A_2$ are blocks of $\ell_1$ and $\ell_2$ consecutive oscillators, respectively. This bipartition is schematically represented in the top-left panel of Fig.\,\ref{fig:tripartition_time_DBC_nonemptyB}.

In the other panels of the same figure, we report the time evolution of the mutual information \eqref{eq:mutualinfo} (top-right panel) and the mutual complexities \eqref{eq:mutualcomplexity} (bottom-left panel) and \eqref{eq:mutualcompl_sq} (bottom-right panel) for this bipartition. The out-of-equilibrium protocol applied to the harmonic chain is the one discussed in Sec.\,\ref{subsec:model}.
The mutual information is computed using standard techniques for entanglement entropy in bosonic Gaussian states, reviewed in \cite{Weedbrook:2011wxo}. Since this is not the main focus of the present work, we do not provide further details and instead use the mutual information as a benchmark for our analysis of mutual complexity.
On the other hand, the mutual complexities are computed using the FR subsystem complexity \eqref{eq:FRcomplexity} in their definition. To this end, we construct the reduced covariance matrices of the reference state, i.e. the initial state, and of the target state given by the evolved state at time $t$, using the two-point correlation functions from \cite{Coser:2014gsa, DiGiulio:2021oal}.

The initial state is chosen with a mass parameter such that $\omega_0 L=100$, while the evolution Hamiltonian has $\omega=0$. The data points are obtained for two values of $L$ and strongly suggest the existence of limiting curves in the continuum limit, approached for $L\to\infty$.
Several choices of subsystem sizes, $\ell_1=r_1 L$ and $\ell_2=r_2 L$, are considered. All the curves in the three panels are periodic in time, with revivals induced by the finite size of the system \cite{Cardy:2014rqa}. Therefore, we focus on the time regime within the first period.

The time evolution of the mutual information is shown in the top-right panel of Fig.\,\ref{fig:tripartition_time_DBC_nonemptyB}. As expected, $I(A_1:A_2)$ remains positive throughout the evolution.
The entanglement dynamics of the harmonic chain after a global quench can be understood in terms of the quasi-particle picture \cite{Calabrese:2005in,Alba:2017ekd}. In this framework, pairs of excitations are emitted from every point in space after the quench and propagate in opposite directions with unit velocity. Quasi-particles emitted from the same point are entangled, and they begin to contribute to the entanglement entropy when one of them is inside the subsystem and the other in its complement. The entanglement entropy is therefore proportional to the number of entangled quasi-particle pairs shared between the subsystem and its complement.
For a subsystem consisting of a single interval in an infinite line, this picture leads to the well-known initial linear growth of the entanglement entropy, followed by a sudden saturation when the time $t$ reaches half the interval length, $t=\ell/2$.

This picture was originally introduced for critical dynamics described by CFTs \cite{Calabrese:2005in}, such as the one considered here for the harmonic chain. Combining these behaviors for the entanglement entropy, we obtain the quasi-particle description of the mutual information. This is reviewed in Appendix \ref{app:QuasiParticle} using the CFT results \cite{Coser:2014gsa} (see Fig.\,\ref{fig:I2CFT}).
The quasi-particle picture has been refined for free and integrable lattice models by including excitations with a non-trivial velocity spectrum arising from a non-linear dispersion relation \cite{Alba:2017ekd}\footnote{In holography, this effective picture breaks down already for subsystems given by two intervals \cite{Asplund:2015eha} and, in general chaotic systems, should be replaced by the membrane picture \cite{Jonay:2018yei,Mezei:2018jco}.}. This refinement leads to a smoother behaviour at the crossover between different time regimes in the entanglement evolution.

Indeed, in the top-right panel of Fig.\,\ref{fig:tripartition_time_DBC_nonemptyB}, the transitions between different regimes occur smoothly. However, the curves obtained from the numerical analysis differ from those shown in Fig.\,\ref{fig:I2CFT}.
The reason is that, unlike the CFT analysis in Appendix \ref{app:QuasiParticle}, these results are obtained for finite-size systems. At early times, the quasi-particles in the harmonic chain do not yet feel the presence of the boundaries and therefore behave as if they were in an infinite system, reproducing the characteristic CFT behavior. However, after a certain time, finite-size effects become relevant and modify the dynamics predicted by ballistic spreading in the infinite-volume limit. In particular, in the presence of boundaries, quasi-particles are reflected, leading to a different evolution from that described by the formulas in Appendix \ref{app:QuasiParticle}.
Consistently with these expectations, we observe that $I(A_1:A_2)$ grows linearly at early times up to $t\simeq \ell_2/2$ (strictly speaking, this holds only if $\ell_2/2 \leq \ell_1$; otherwise, finite-size effects set in before the end of the initial linear growth).

In the bottom panels of Fig.\,\ref{fig:tripartition_time_DBC_nonemptyB}, we show the evolution of the mutual complexities $M(A_1:A_2)$ and $M^{(2)}(A_1:A_2)$ after the global quench considered in this section. The sign of $M(A_1:A_2)$ is positive for all times. This property has been verified for several other choices of parameters, not reported here.
On the other hand, $M^{(2)}(A_1:A_2)$ does not have a definite sign and can take both positive and negative values during the time evolution. Therefore, our analysis suggests that subadditivity of the FR subsystem complexity $C_A$ is plausible, whereas neither superadditivity nor subadditivity of $C_A^2$ holds in general. This latter observation is consistent with the results found in \cite{DiGiulio:2020hlz} for systems at equilibrium.

The same considerations on the signs of $M(A_1:A_2)$ and $M^{(2)}(A_1:A_2)$ also apply in the presence of periodic boundary conditions. For the sake of conciseness, we do not report the results of this analysis. For the same choice of parameters, the curves are very similar to those presented in this section. The main differences are that, with DBC, the period of revivals is twice that of the periodic case, while, in the latter, one additionally observes an overall logarithmic growth in time due to the zero mode \cite{DiGiulio:2021oal}.

\subsubsection{Analysis for three subsystems}\label{subsec:threeintervals}

To conclude our analysis, we study the same mass quench dynamics of finite harmonic chains with $L$ sites and DBC, but we consider a bipartition into $A$ and $B$, where  the subsystem $A$ divided into $A_1$, $A_2$, and $A_3$, taken as adjacent blocks containing $\ell_1 \equiv r_1 L$, $\ell_2 \equiv r_2 L$, and $\ell_3 \equiv r_3 L$ oscillators, respectively.
In all cases investigated, we study the tripartite complexities \eqref{eq:tripartitecomplexity} and \eqref{eq:tripartitecompl_sq}, and the complexity gaps \eqref{eq:Mtilde} and \eqref{eq:Mtilde_sq}, computed using the FR subsystem complexity \eqref{eq:FRcomplexity} in their definitions. The FR subsystem complexity is evaluated as reviewed in Secs.\,\ref{subsec:model} and \ref{subsec:twointervals}. Additionally, we study the tripartite information \eqref{eq:tripartiteinfo}, computed numerically using bosonic Gaussian-state techniques \cite{Weedbrook:2011wxo}.

All the curves presented in this subsection have been obtained by setting the mass parameter of the evolution Hamiltonian to its critical value, $\omega=0$. As for the initial state, we choose the mass parameter $\omega_0$ such that $\omega_0 L=100$, with the total system size taken to be either $L=50$ or $L=100$.
As $L$ increases, we observe a collapse of the data points onto a limiting curve, presumably related to the underlying CFT description. We postpone a detailed CFT analysis of these behaviours to future work. To better explore the dynamics, we report data obtained for different choices of $r_i=\ell_i/L$.
Also for this bipartition, we observe revivals with period $\Delta t \simeq L$. Therefore, we report the data points only within the first period.

\begin{figure}[t!]
\hspace{-1.5cm}
\includegraphics[width=1.18\textwidth]{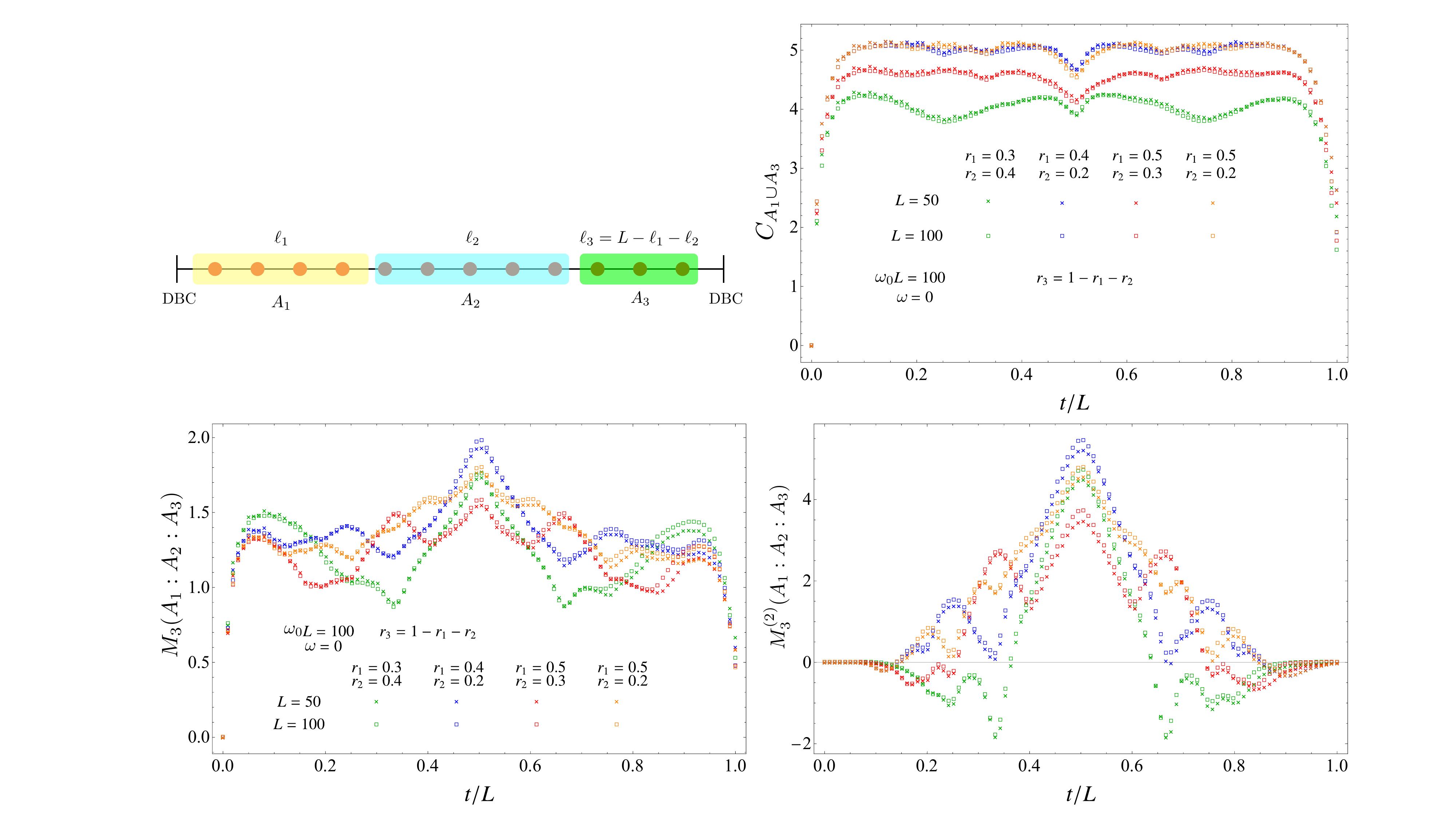}
\vspace{-0.8cm}
\caption{
We consider a partition of a harmonic chain consisting of $L$ oscillators with DBC at its boundaries into $A_1$, made of $\ell_1$ sites, $A_2$, made of $\ell_2$ sites, $A_3$, made of $\ell_3$ sites, as represented in the top-left panel. We study the time evolution following a global quench of the mass parameter, from $\omega_0\neq 0$ to $\omega=0$, for various choices of subsystems parametrized by $r_i=\ell_i/L$.
In the top-right panel, we plot the FR subsystem complexity $C_{A_1\cup A_3}$ obtained using \eqref{eq:FRcomplexity} as a function of time. In the bottom panels, we show the tripartite complexities \eqref{eq:tripartitecomplexity} (bottom-left panel) and \eqref{eq:tripartitecompl_sq} (bottom-right panel), computed using the FR subsystem complexity \eqref{eq:FRcomplexity}.
}
\label{fig:tripartitemutual_time_DBC}
\end{figure}

In Fig.\,\ref{fig:tripartitemutual_time_DBC}, we analyze the case in which the subsystem $B$ is empty, namely $L=\ell_1+\ell_2+\ell_3$. This configuration is schematically represented in the top-left panel of the figure. In the top-right panel, we report the time evolution of the FR subsystem complexity \eqref{eq:FRcomplexity} associated with the two disjoint intervals $A_1\cup A_3$.
To the best of our knowledge, the dynamics of FR subsystem complexity for two disjoint intervals has not been studied previously. Its evolution displays the initial growth, peak, and decay typical of the FR subsystem complexity for single intervals \cite{DiGiulio:2021oal}. However, after this regime, instead of saturating, the complexity exhibits a second rise and fall, until finite-size effects become relevant around $t/L\simeq 1/2$.

In the bottom panels, we plot the numerical results for the tripartite complexities \eqref{eq:tripartitecomplexity} and \eqref{eq:tripartitecompl_sq} as functions of time. Both quantities exhibit a rich dynamics; however, we emphasize some significant differences between them.
While $M_3(A_1:A_2:A_3)$ remains non-negative for all times, $M^{(2)}_3(A_1:A_2:A_3)$ displays an initial regime in which it vanishes and, after this transient, undergoes a non-trivial time evolution, taking both positive and negative values at different times.

\begin{figure}[t!]
\hspace{-1.5cm}
\includegraphics[width=1.18\textwidth]{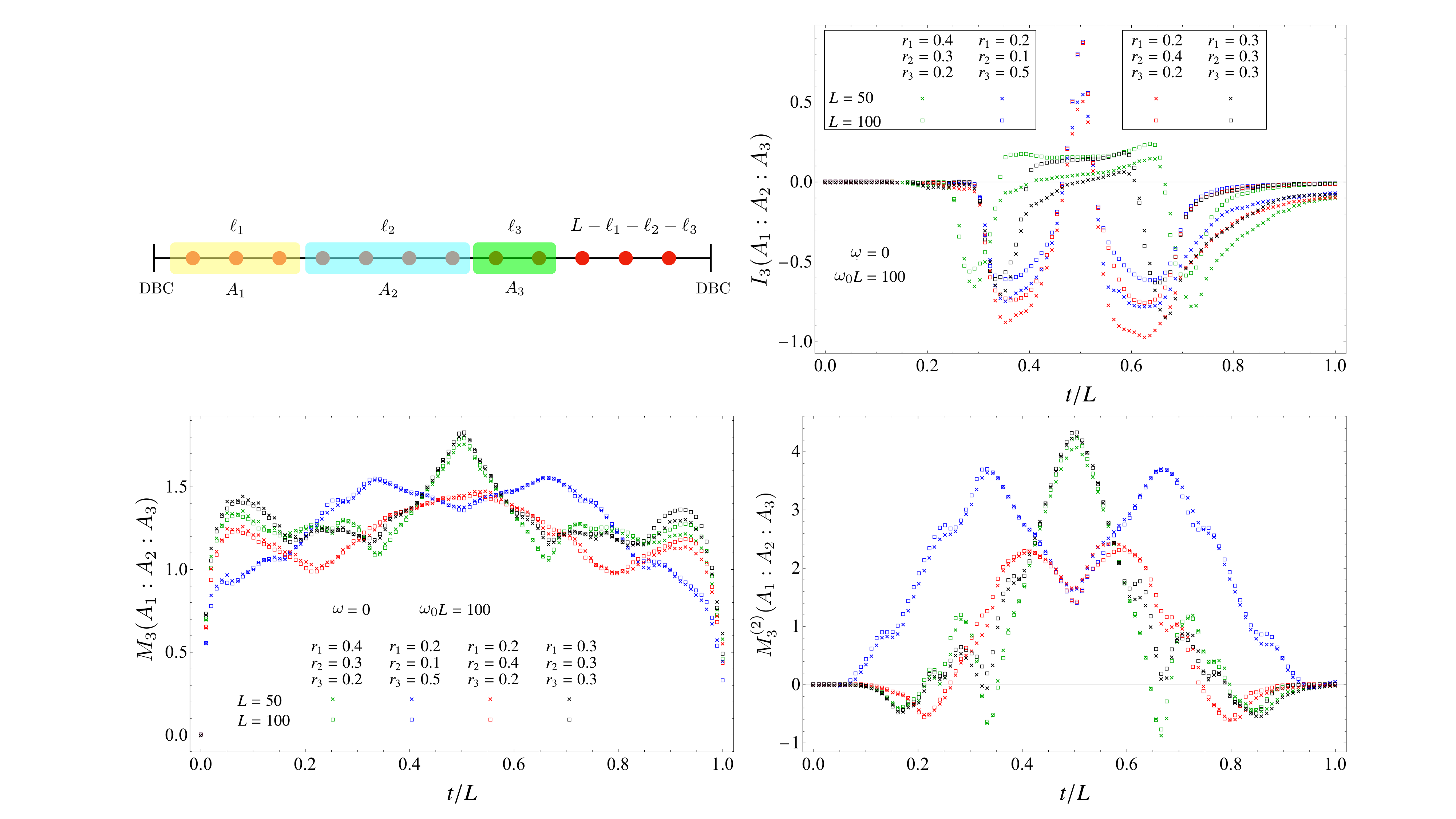}
\vspace{-0.8cm}
\caption{
We consider a partition of a harmonic chain consisting of $L$ oscillators with DBC at its boundaries into $A_1$, made of $\ell_1$ sites, $A_2$, made of $\ell_2$ sites, $A_3$, made of $\ell_3$ sites, and $B$, made of the remaining $L-\ell_1-\ell_2-\ell_3$ sites, as represented in the top-left panel. We study the time evolution following a global quench of the mass parameter, from $\omega_0\neq 0$ to $\omega=0$, for various choices of subsystems parametrized by $r_i=\ell_i/L$.
In the top-right panel, we plot the tripartite information \eqref{eq:tripartiteinfo} as a function of time. In the bottom panels, we show the tripartite complexities \eqref{eq:tripartitecomplexity} (bottom-left panel) and \eqref{eq:tripartitecompl_sq} (bottom-right panel), computed using the FR subsystem complexity \eqref{eq:FRcomplexity}.
}
\label{fig:quandripartition_time_DBC}
\end{figure}

In Fig.\,\ref{fig:quandripartition_time_DBC}, we investigate the bipartition in which $B$ is non-empty and contains $L-\ell_1-\ell_2-\ell_3\neq 0$ oscillators. This configuration is schematically represented in the top-left panel. For this choice of bipartition, the tripartite information \eqref{eq:tripartiteinfo} does not vanish, so in the top-right panel, we display the time evolution of $I_3(A_1:A_2:A_3)$.
According to the quasi-particle predictions reviewed in Sec.\,\ref{subsec:twointervals} and the CFT discussion in Appendix \ref{app:QuasiParticle}, the tripartite information should vanish identically during the quench dynamics in an infinite system. This is indeed what is observed in the early-time regime of the numerical curves shown in the figure.
The interpretation is similar to the one discussed for the mutual information in Fig.\,\ref{fig:tripartition_time_DBC_nonemptyB}: our results are consistent with the quasi-particle picture in infinite volume only at early times, namely until the excitations propagating through the system begin to feel the presence of the boundaries. At that stage, finite-size effects become relevant and lead to a dynamics different from the one predicted in the infinite volume limit. In particular, here the sign of $I_3(A_1:A_2:A_3)$ changes during the time evolution. To the best of our knowledge, this feature has not been observed previously for tripartite information in out-of-equilibrium many-body systems \cite{Maric:2020dpw,Maric:2022rsc,Caceffo:2023hns}, and is likely due to the presence of boundaries and to the reflection of quasi-particles at the endpoints of the system.

In the bottom panels, we display $M_3(A_1:A_2:A_3)$ and $M^{(2)}_3(A_1:A_2:A_3)$ as functions of time. Similar considerations to those discussed for Fig.\,\ref{fig:tripartitemutual_time_DBC} apply here as well. The tripartite complexity \eqref{eq:tripartitecomplexity}, shown in the bottom-left panel, remains positive throughout the evolution, while \eqref{eq:tripartitecompl_sq}, shown in the bottom-right panel, can assume both positive and negative values at different times, after an early time regime where it is zero.
The results reported here show that, at least for the quench dynamics considered in this work, $M^{(2)}_3(A_1:A_2:A_3)$ does not have a definite sign, while we cannot, a priori, rule out the possibility that $M_3(A_1:A_2:A_3)$, evaluated using the FR subsystem complexity, is a non-negative quantity.

\begin{figure}[t!]
\centering
\includegraphics[width=0.495\textwidth]{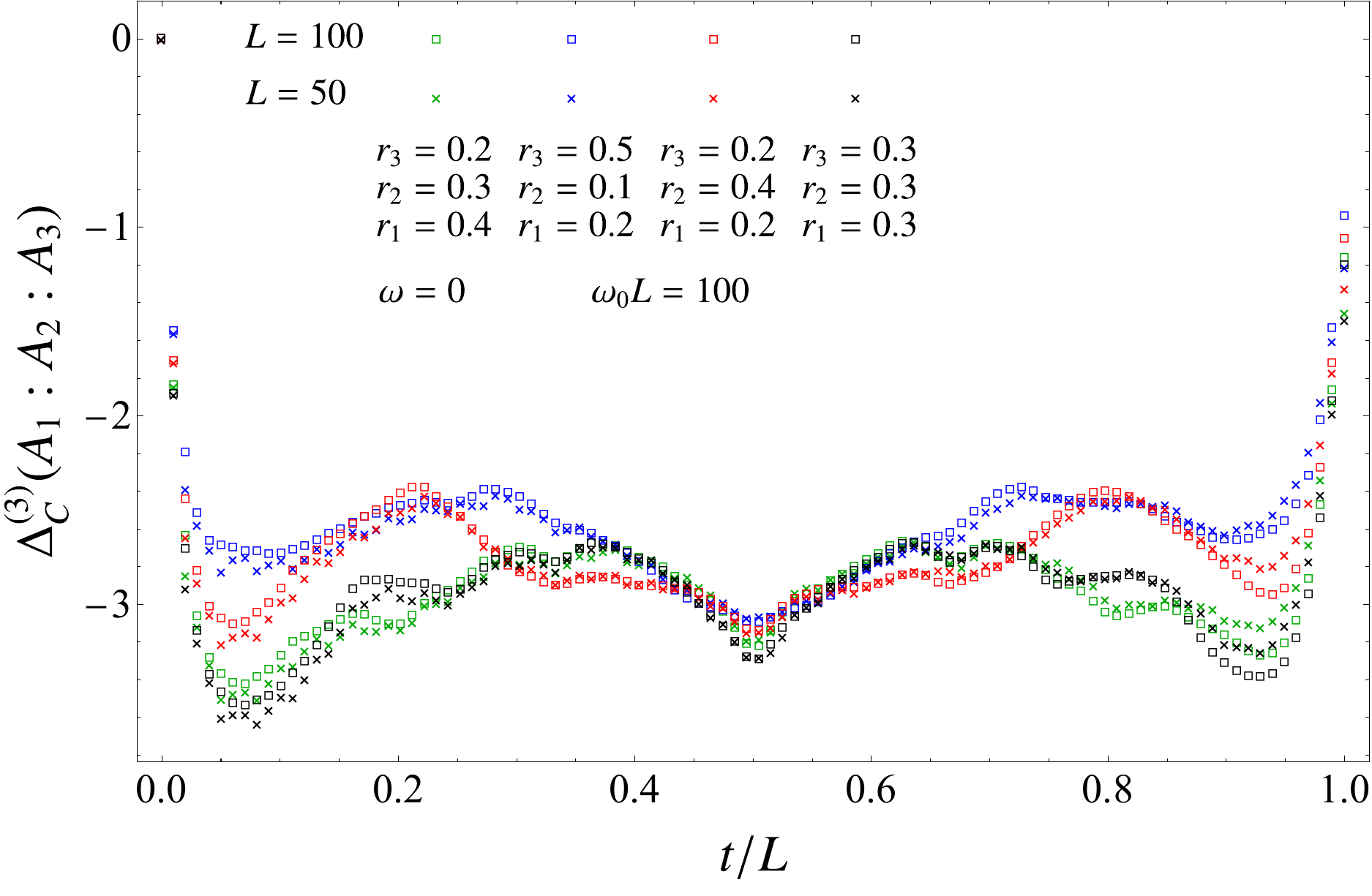}
\includegraphics[width=0.495\textwidth]{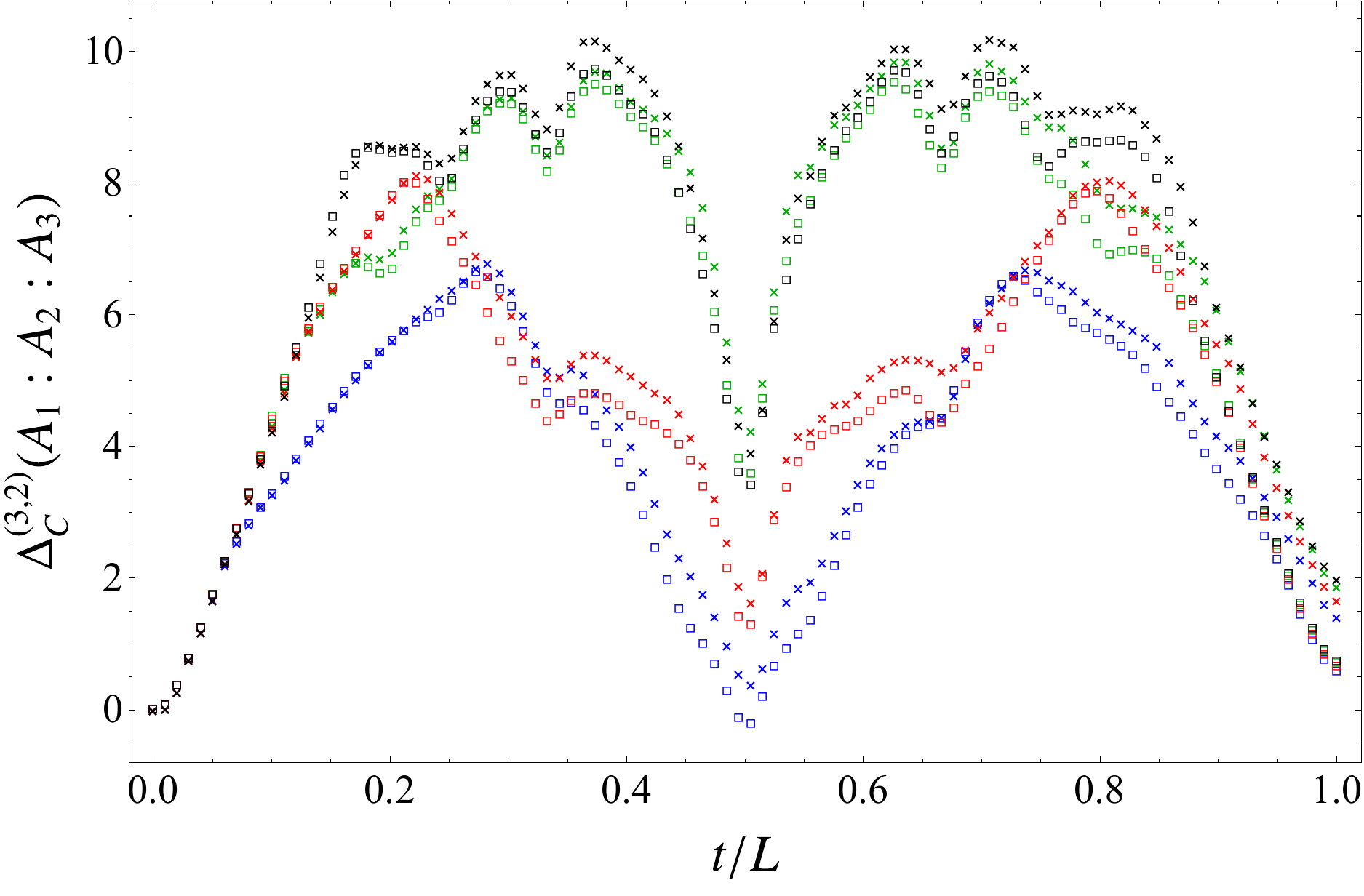}
\vspace{-0.5cm}
\caption{Complexity gaps \eqref{eq:Mtilde} (left panel) and \eqref{eq:Mtilde_sq} (right panel), computed using the FR subsystem complexity \eqref{eq:FRcomplexity}, as functions of time following a mass quench in a harmonic chain with $L$ oscillators and DBC. The quench is implemented by evolving an initial state with mass $\omega_0 \neq 0$ under a critical Hamiltonian with $\omega = 0$. The bipartition considered is the same as that shown in the top-left panel of Fig.~\ref{fig:quandripartition_time_DBC}. The data points are displayed for different values of the subsystem sizes, parametrized by $r_i = \ell_i/L$.}
\label{fig:Mtilde_FR}
\end{figure}

Finally, we discuss the time evolution of the complexity gaps \eqref{eq:Mtilde} and \eqref{eq:Mtilde_sq} computed using the FR subsystem complexity \eqref{eq:FRcomplexity}. The data curves are shown in Fig.\,\ref{fig:Mtilde_FR} for the bipartition represented in the top-left panel of Fig.~\ref{fig:quandripartition_time_DBC}.
Different values of subsystem sizes $\ell_i=r_i L$ and $L$ are considered and the curves suggest a scaling behavior in the regime of large enough values of $L$. We observe that $\Delta^{(3)}_C(A_1:A_2:A_3)$ is negative throughout the post-quench evolution, while $\Delta^{(3,2)}_C(A_1:A_2:A_3)$ is positive for any value of time. Interestingly, $\Delta^{(3,2)}_C(A_1:A_2:A_3)$ is the only quantity investigated here defined in terms of the squares of the FR subsystem complexity that, within our analyses, exhibit a definite sign along the entire dynamics. 

All the features discussed in this subsection, for Figs.\,\ref{fig:tripartitemutual_time_DBC}, \ref{fig:quandripartition_time_DBC}, and \ref{fig:Mtilde_FR}, have been tested and observed for several other choices of parameters that we do not analyze here, and appear to be very robust for this model and the corresponding dynamics. Moreover, the same phenomenology for $I_3(A_1:A_2:A_3)$, $M_3(A_1:A_2:A_3)$, $M^{(2)}_3(A_1:A_2:A_3)$, $\Delta^{(3)}_C(A_1:A_2:A_3)$, and $\Delta^{(3,2)}_C(A_1:A_2:A_3)$ has also been observed for harmonic chains with periodic boundary conditions.

\section{Conclusions and outlook}\label{sec:Conclusions}

In this work, we examined three distinct proposals for quantifying the complexity of reduced density matrices and their time evolution. Two of these are existing constructions: the holographic CV subsystem complexity, formulated in terms of bulk volumes in AdS, and the FR subsystem complexity, which relies on the dynamics of the Wigner function of a Gaussian reduced density matrix. We also proposed a new approach that generalizes Krylov-space methods, originally developed for the unitary evolution of pure states, to the study of subsystem dynamics.

In holography, one can derive families of inequalities involving entanglement entropies associated with partitions of the system made by multiple subregions. The resulting entropy cone provides a powerful framework for characterizing the correlation structure of holographic CFTs and their geometric dual descriptions. Motivated by this idea, we ask whether analogous combinations of subsystem complexities can be identified, potentially leading to a complexity cone analogue. From the holographic perspective, such a structure would provide an invaluable guide in constraining the search for a consistent definition of subsystem complexity in CFTs and many-body systems through inequalities satisfied by bulk complexity proposals.

Towards this goal, we analyze the mutual complexity \eqref{eq:mutualcomplexity} and tripartite complexity \eqref{eq:tripartitecomplexity}, together with the complexity gap \eqref{eq:Mtilde}, using different proposals for subsystem complexity in their definitions. In particular, we focused on their sign structure, with the aim of understanding whether these quantities could provide suitable candidates for complexity inequalities. The results of our analysis are summarized in Table~\ref{tab:summary}.

\begin{table}[b!]
\renewcommand{\arraystretch}{1.5}
\begin{tabular}{|l!{\vrule width 1.2pt}c|c|c|}
\hline
& $M(A_1:A_2)$
& $M_3(A_1:A_2:A_3)$
& $\Delta^{(3)}_C(A_1:A_2:A_3)$
\\
\noalign{\hrule height 1.2pt}
\text{CV holographic subsystem}
& $\le 0$
& ND
& $\ge 0$
\\
\hline
\text{Krylov subsystem}
& ND
& ND
& $\le 0$
\\
\hline
\text{Fisher-Rao subsystem}
& $\ge 0$
& $\ge 0$
& $\le 0$
\\
\hline
\text{(Fisher-Rao subsystem)$^2$}
& ND
& ND
& $\ge 0$
\\
\hline
\end{tabular}
\caption{Summary of the results obtained in this manuscript for the sign of the quantities \eqref{eq:mutualcomplexity}, \eqref{eq:tripartitecomplexity}, and \eqref{eq:Mtilde}. We report the sign (or ND for non-definite) restricted to the states and analyses performed in the previous sections. In the last row, we also include the corresponding results obtained using the square of the FR subsystem complexity, namely \eqref{eq:mutualcompl_sq}, \eqref{eq:tripartitecompl_sq}, and \eqref{eq:Mtilde_sq}.}
\label{tab:summary}
\end{table}

We remark that these findings refer to the various complexities computed in specific scenarios and models and, except for the negativity of $M$ arising in the holographic CV proposal, are not proven to hold in general. Nevertheless, our results can, on the one hand, rule out certain quantities from serving as building blocks for complexity inequalities, and, on the other hand, suggest expressions that merit further investigation in this context.

For example, the tripartite complexity \eqref{eq:tripartitecomplexity}, whose definition is modeled on that of the tripartite information \eqref{eq:tripartiteinfo}, does not exhibit a definite sign in several cases, in particular in the holographic setting of Sec.~\ref{subsec:static}. Thus, we do not expect $M_3(A_1:A_2:A_3)$ to play a prominent role in the definition of a complexity cone. On the other hand, the complexity gap $\Delta^{(3)}_C(A_1:A_2:A_3)$ has a definite sign in all the cases analyzed in this manuscript and might be a promising candidate for a general complexity inequality. We postpone a general proof and a more detailed analysis of $\Delta^{(3)}_C(A_1:A_2:A_3)$ to future work.

The positivity of the complexity gap, together with its interpretation as a bulk volume discussed in Sec.~\ref{subsec:emergence}, suggests that this quantity may serve as a useful benchmark for candidate holographic CFT duals of the CV subsystem complexity proposal. Indeed, given a quantum state with a known classical gravity dual, if the complexity gap computed using a given candidate proposal turns out to be negative, then that proposal is unlikely to represent the correct holographic dual of CV subsystem complexity. We hope to carry out this analysis in explicit CFT examples in future work.

In Sec.\,\ref{sec:subsystemKrylov}, we introduced a method for recasting the general and complicated NZ dynamics governing the non-unitary evolution of reduced density matrices, into an effective Schrödinger equation on a Krylov chain. This equation is constructed via a generalization of the moments recursion method used commonly for Krylov space analysis of unitary dynamics.
 
We show that the subsystem Krylov complexity \eqref{eq:subsystemKrylovcomplexity} defined within this framework can probe the number of effective subsystem configurations encoded in the associated Krylov space. In the few-qubit example discussed in Sec.~\ref{subsec:fewspins}, this manifests as a strong sensitivity to the complementary region: the larger the number of qubits in the complement, the greater the resulting subsystem complexity. This Krylov space-based approach to subsystem complexity has the advantage of being applicable to arbitrary dynamics and, as in the standard Lanczos algorithm, admits a recursive structure that makes it numerically implementable.

In the case of unitary dynamics, the number of Lanczos coefficients, and hence the dimension of the Krylov space, is bounded by the Hilbert space dimensionality. This bound no longer holds in our generalization, as we have shown examples in which the Krylov-space dimension exceeds that of the Hilbert space. It is therefore not yet clear after how many steps the moment recursion algorithm applied to the subsystem return amplitude terminates, nor whether this can be bounded in terms of parameters of the underlying model. We leave these questions, along with a more formal analysis of the method, for future work as well.

While the Krylov-space approach to the dynamics of reduced density matrices is based on an algebraic construction, the method reviewed and applied in Sec.~\ref{sec:SRSpreadComplexity} is instead more geometric in nature. Indeed, the FR subsystem complexity is defined as the geodesic length in the space of Wigner functions associated with bosonic Gaussian reduced density matrices.

Although this method is restricted to the class of bosonic Gaussian states, its geometric nature and the simplicity of the expression \eqref{eq:FRcomplexity} for subsystem complexity make it insightful for understanding dynamical questions involving reduced density matrices. Indeed, the dependence of \eqref{eq:FRcomplexity} only on the reference and target covariance matrices allows $C_A$ to be expressed in terms of a finite set of observables, namely two-point functions, which scale polynomially with the number of degrees of freedom. This favorable scaling has been exploited here and in previous works \cite{DiGiulio:2021oal,DiGiulio:2021noo} to access sublattices of harmonic oscillator systems large enough to probe continuum-limit scaling regimes, which at criticality are described by CFTs.

Finally, we stress that the two subsystem complexity proposals are intrinsically expected to access more dynamical information than entanglement entropies. Indeed, while entanglement entropies depend only on the Schmidt eigenvalues of the reduced density matrix, the Krylov subsystem complexity depends, through the corresponding subsystem return amplitude \eqref{eq:generaldef_RL}, also on the overlap between the Schmidt eigenvectors of the initial and the evolving state (see Appendix~\ref{app:SchmidtDecomp} for more details on this point).

On a similar note, the FR subsystem complexity does not depend solely on the symplectic spectrum of the evolving state, as entanglement entropies do, but instead encodes information about the relative change of basis between the reference (initial) and target (evolved) states \cite{DiGiulio:2020hlz}.

Our analysis opens several interesting directions for future research and we list some of them below.

We have focused on examples with deterministic dynamics. However, in order to capture universal features of chaotic systems, it is often useful to consider dynamics generated by random Hamiltonians and ensembles of late-time random states \cite{Dyson:1962es, Bohigas:1983er}. An interesting question is whether subsystem complexity can be exploited as a diagnostic or proxy for quantum chaos. From this perspective, it would be worthwhile to apply the proposal introduced in Sec.~\ref{sec:subsystemKrylov} to reduced density matrices obtained by tracing out subsystems of randomly evolving pure state density matrices.

As for the FR approach discussed in Sec.~\ref{sec:SRSpreadComplexity}, the restriction to bosonic Gaussian states prevents this subsystem complexity proposal from accessing fully random dynamics. Nevertheless, it would still be interesting to study random Gaussian states \cite{Serafini:2007fqx,Serafini:2007ohw,Fukuda:2019ycv} from the perspective of FR subsystem complexity. This ensemble is known to reproduce the late-time dynamics of the SYK model in the case of two-fermion interactions \cite{Tiutiakina:2023ilu} and is therefore of interest across several communities.

Another interesting scenario to explore is the regime in which a Lindblad dynamics for the reduced density matrix emerges effectively from the NZ evolution. Under this approximation, we expect that several explicit examples could be worked out analytically (see, for example, \cite{Hashimoto:2026kjy}), potentially providing further insight into subsystem dynamics.
In addition, a map evolving reduced density matrices in time can, in some instances, be studied using tensor network techniques \cite{Coppola:2025qxk}. An analytical characterization of this map would provide better control over the moment recursion method for reduced density matrices developed in Sec.\,\ref{sec:subsystemKrylov}.

Of course we only used three selected, computable approaches to subsystems complexity. It would be interesting to explore our constraints further, for instance in tensor network inspired approaches such as path integral optimization or related definitions \cite{Caputa:2017yrh,Boruch:2021hqs,Abt:2017pmf,Chen:2020nlj}, and other geometric methods \cite{Balasubramanian:2018hsu,Caceres:2019pgf,Ruan:2020vze,Bhattacharyya:2020iic,Jian:2023mdh,Pedraza:2021mkh,Pedraza:2021fgp,Gerbershagen:2024qlz}.

From the holographic perspective, recent attempts to constrain subsystem complexity proposals have been pursued using focusing conjectures \cite{Concepcion:2026fhv}. For a more complete characterization of the complexity gap introduced in this manuscript, it would be insightful to investigate this quantity from that perspective as well.

\bigskip
\noindent {\bf \large Acknowledgments}
\\

We are grateful to Mari Carmen Bañuls, Alice Bernamonti, Bowen Chen, Federico Galli, Victor Godet, Matthew Headrick, Juan Hernandez, Sara Murciano, Dimitrios Patramanis, Tadashi Takayanagi, Erik Tonni, Evita Verheijden, Nicolò Zenoni for useful discussions and comments on the draft. This work is supported by the ERC Consolidator grant (number: 101125449/acronym: QComplexity).  Views and opinions expressed are however those of the authors only and do not necessarily reflect those of the European Union or the European Research Council. Neither the European Union nor the granting authority can be held responsible for them. P.C. is supported by the NCN Sonata Bis 9 2019/34/E/ST2/00123 grant.
P.C. is supported by the  Swedish Research Council (VR) under Grant No. 2025-04154.

\appendix

\section{Holographic tripartite complexity and complexity gap in global AdS$_3$}

\label{app:globalAdS3}
In this appendix we compute the holographic CV subsystem complexity for a boundary 1+1-dimensional CFT defined on a circle of circumference $L$, using global coordinates for the AdS$_3$ space in the bulk. We show that the finite contributions corresponding to $w_A$, $w_{12}$, $w_{123}$, $\hat{w}$, and $\tilde{w}$ defined in Sec.~\ref{subsec:static} take the same integer values as in the case of Poincaré coordinates discussed in the main text. The only difference compared to the analysis in Poincaré coordinates is that the occurrence of different phases by tuning the separation between the boundary subregions is modified by the compactness of the circle.

\subsection{Setup}

The metric on a constant-time slice of global AdS$_3$ with radius $R$ is given by
\be
\label{eq:global_metric}
ds^2 = R^2\left(d\rho^2 + \sinh^2\rho\, d\phi^2\right)\,,
\ee
where $\phi \in [0, 2\pi)$ is the angular coordinate along the boundary circle and $\rho \geq 0$ is the radial coordinate, with the boundary located at $\rho \to \infty$.

To regularize geodesic lengths and bulk volumes, we introduce a UV cutoff at $\rho = \rho_{\max}$, defined as
\be
\label{eq:rhomax_bulk}
\rho_{\max} \equiv \ln\left(\frac{R}{\delta}\right)\,,
\ee
so that in the limit $\delta\to 0$, one has $\rho_{\max}\to\infty$, corresponding to the asymptotic boundary. Here, both $\delta$ and $R$ are bulk length scales. To relate them to the boundary circle length $L$ where the CFT is defined and the corresponding UV cutoff $\epsilon$, we proceed as follows.

Consider the infinitesimal line element on the boundary at fixed $\rho\gg 1$,
\begin{equation}
\label{eq:boundaryelength}
d s=\frac{2\pi}{L} R\sinh\rho d \sigma\,,
\end{equation}
where $\phi=2\pi\sigma/L$ defines a dimensionful boundary coordinate $\sigma$. Evaluating this expression at $\rho=\rho_{\max}$, we identify $R\frac{d \sigma}{ds}$ with the smallest spatial distance resolvable on the cutoff surface, namely the boundary UV cutoff $\epsilon$, i.e. $R\frac{d \sigma}{ds}=\epsilon$. Substituting this identification into \eqref{eq:boundaryelength} and using $R\gg \delta$, we obtain
\begin{equation}
\label{eq:boundaryelength_epsilon}
\frac{L}{\pi\epsilon} =2\sinh\rho_{\max}\simeq
\frac{R}{\delta}\,.
\end{equation}
Therefore, the cutoff radius \eqref{eq:rhomax_bulk} can be equivalently expressed as
\be
\label{eq:rhomax_bdry}
\rho_{\max} = \ln\left(\frac{L}{\pi\epsilon}\right)\,,
\ee
which will be used throughout this appendix to regularize geodesic lengths and bulk volumes.

\subsection{Single interval}

Consider a boundary subregion $A$ consisting of a single arc of length $\ell$, subtending an angle $\Delta\phi = 2\pi \ell / L$ at the boundary. The RT surface is the bulk geodesic anchored at the two endpoints of $A$. In the coordinates \eqref{eq:global_metric}, geodesics satisfy
\be
\tanh\rho\cos\phi = \tanh\rho_*\,,
\ee
where $\rho_*$ is the turning point, determined by the boundary condition $\cos(\Delta\phi/2)\tanh\rho_{\max} = \tanh\rho_*$. The length of this geodesic gives the area of the RT surface, namely
\be
\label{eq:global_geodesic}
\text{Area}(\gamma_A) = 2R\ln\!\left(\frac{L}{\pi\epsilon}\sin\frac{\pi \ell}{L}\right)\,,
\ee
which, as expected, reduces to $2R\ln(\ell/\epsilon)$ as $L\to\infty$, i.e. the result obtained in Poincaré coordinates. Using \eqref{eq:RTformula}, this leads to the expression of the entanglement entropy of a single interval in a 1+1-dimensional CFT on a circle \cite{Calabrese:2004eu}.

To compute the volume enclosed by the RT surface, we use the area element $dV = R^2\sinh\rho\, d\rho\, d\phi$ on the time slice \eqref{eq:global_metric}. Integrating over the bulk region bounded by the RT geodesic and the regulated boundary, we find
\be
\label{eq:global_volume}
V(\gamma_A) = \frac{\ell\, R^2}{\epsilon} - \pi R^2 + O(\epsilon)\,.
\ee
Substituting into \eqref{eq:holo_subsystemC} and neglecting the terms of $O(\epsilon)$, we obtain
\be
\label{eq:global_single_complexity}
C_A = \frac{R}{8\pi G_N}\left(\frac{\ell}{\epsilon} + \pi w_A\right)\,,
\qquad w_A = -1\,,
\ee
which is identical to the result \eqref{eq:2dCFT1int_holocomp} obtained in Poincaré coordinates.

\subsection{Two disjoint intervals}

For a subsystem $A = A_1 \cup A_2$ consisting of two arcs of lengths $\ell_1$ and $\ell_2$ with separation $d$, the RT surface has two possible configurations, as in Poincaré coordinates (cf. the left panel of Fig.\,\ref{fig:RT}, where the intervals are embedded in the boundary circle). The transition between them occurs at the critical separation $d_c$ determined by
\be
\label{eq:global_critical_2int}
\sin\frac{\pi \ell_1}{L}\sin\frac{\pi \ell_2}{L} = \sin\frac{\pi(\ell_1 + d_c + \ell_2)}{L}\sin\frac{\pi d_c}{L}\,.
\ee
In the limit $L\to\infty$, \eqref{eq:global_critical_2int} reduces to \eqref{eq:critical distance}. Since the volume enclosed by the RT surface in each phase is still given by a sum of contributions of the form \eqref{eq:global_volume}, the subsystem complexity takes the same form as in \eqref{eq:subsyste_comple_2int},
\be
C_{A_1\cup A_2} = \frac{R}{8\pi G_N}\left(\frac{\ell_1+\ell_2}{\epsilon} + \pi\, w_{12}(d)\right)\,,
\qquad
w_{12}(d) = \begin{cases} 0\,, & d < d_c\,, \\ -2\,, & d > d_c\,, \end{cases}
\ee
with the same integer values of $w_{12}$ as in the Poincaré case, where, as anticipated, the only difference compared to the analysis in the main text is the expression of $d_c$.

\subsection{Three disjoint intervals}

Consider now a subsystem $A = A_1\cup A_2\cup A_3$ on the circle consisting of three arcs of lengths $\ell_1$, $\ell_2$, $\ell_3$ with separation $d_1$, $d_2$, and, due to the compactness of the circle, $d_3 = L - \ell_1-\ell_2-\ell_3-d_1-d_2$.  The RT surface can be one of the five allowed planar configurations described in Sec.~\ref{subsec:static} (see the right panel Fig.\,\ref{fig:RT} and imagine the boundary embedded in a circle). The comparison between these configurations is now governed by products of the form $\prod_i \sin(\pi x_i/L)$ rather than $\prod_i x_i$, but the number of configurations and the structure of the phase diagram remain the same. The separation curves in the phase diagram are determined by
\be
\label{eq:global_critical_3int_diag}
\sin\frac{\pi \ell_1}{L}\sin\frac{\pi \ell_3}{L} = \sin\frac{\pi(d_1 + \ell_2 + d_2)}{L}\sin\frac{\pi d_3}{L}\,,
\ee
which generalizes \eqref{eq:criticalline_3int},
and other conditions, including those generalizing the boundaries $d_1 = d_c(\ell_1,\ell_2)$ and $d_2 = d_c(\ell_2,\ell_3)$, where now $d_c$ is determined from \eqref{eq:global_critical_2int}.

\begin{figure}[t!]
    \centering
    \begin{tabular}{ccc}
\includegraphics[width=0.3\textwidth]{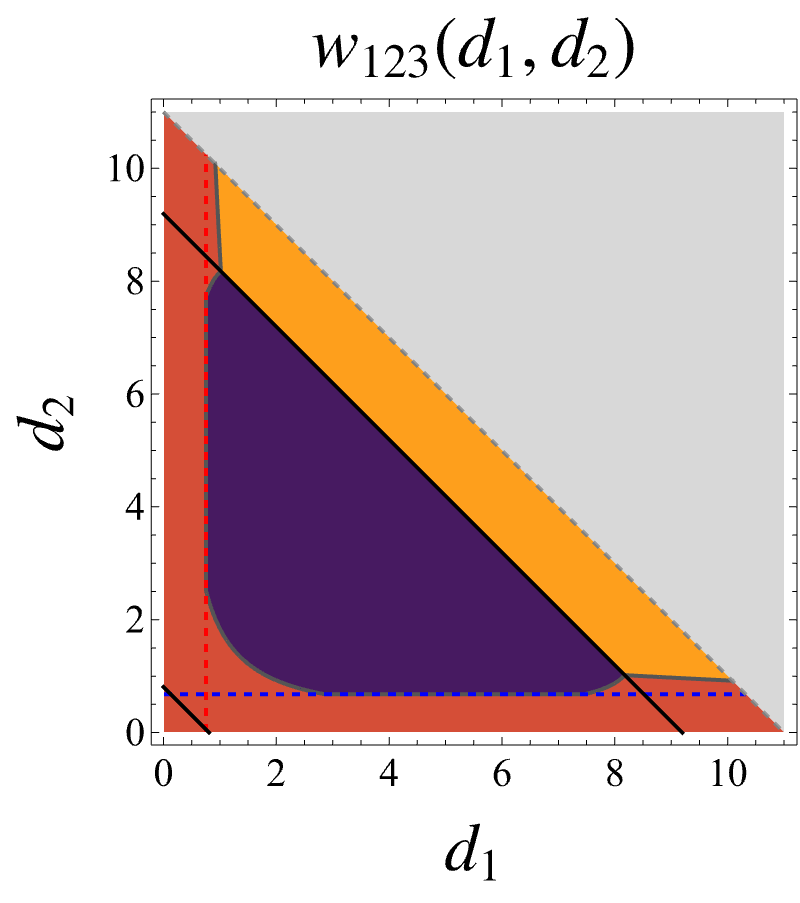} & \includegraphics[width=0.3\textwidth]{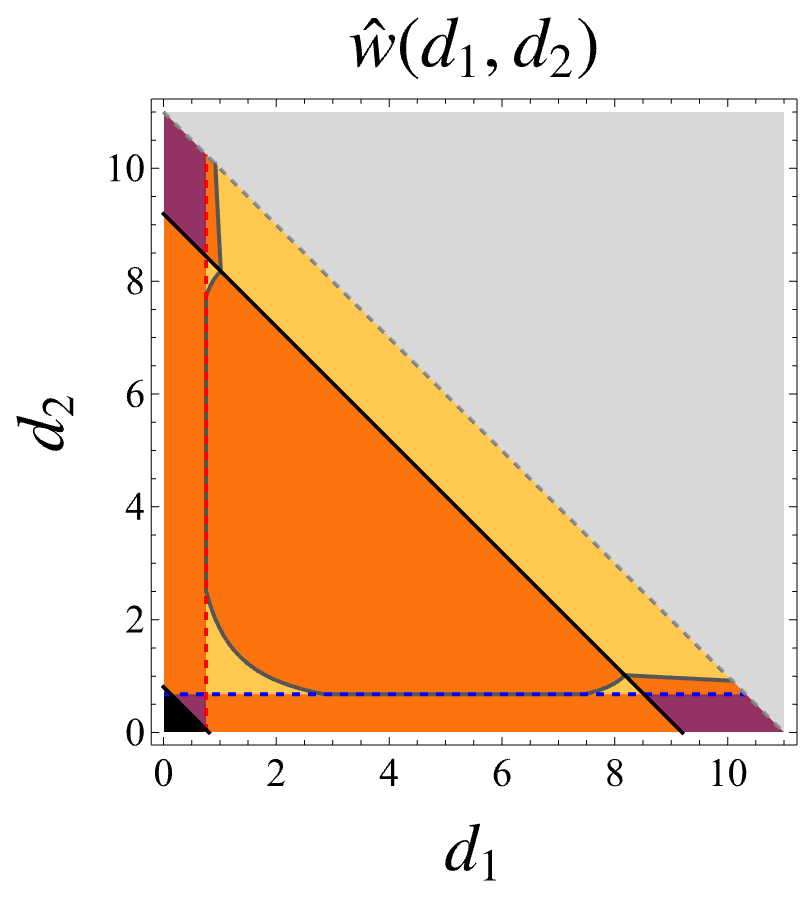} & \includegraphics[width=0.345\textwidth]{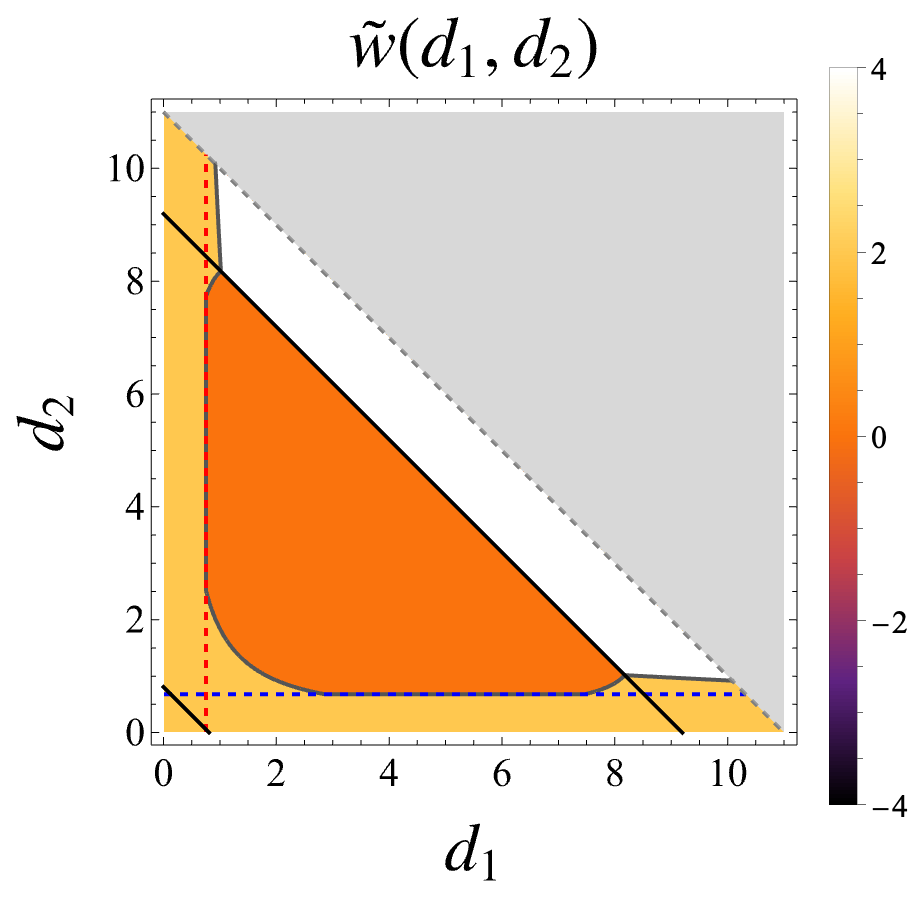}
    \end{tabular}
\caption{
Finite contributions to the holographic CV subsystem  complexity of three intervals (left panel), tripartite complexity (middle) and complexity gap (right).
The plots are obtained for the parameter choice $L=19$, $\ell_1 = 4$, $\ell_2 = 1$, and $\ell_3 = 3$. The left panel shows $w_{123}$ appearing in the subsystem complexity \eqref{eq:CVcompl_3int_global} of three disjoint intervals in global AdS$_3$ as a function of $d_1$ and $d_2$. The middle and right panels show $\hat{w}$ and $\tilde{w}$ in \eqref{eq:M3_global}, respectively. The shaded region corresponds to values of $(d_1,d_2)$ for which the remaining gap $
d_3=L-\ell_1-\ell_2-\ell_3-d_1-d_2
$ becomes negative and is therefore not physically allowed, reflecting the compact geometry of the boundary circle. Although the curves separating the phases are modified compared to the case of Poincaré coordinates, the finite quantities $w_{123}$, $\hat{w}$, and $\tilde{w}$ take the same integer values as those shown in Fig.~\ref{fig:mutualcomplexity}. In particular, $\tilde{w}$ remains non-negative throughout the entire allowed parameter space.
}
\label{fig:mutualcomplexityglobal}
\end{figure}

Since the volume enclosed by each RT configuration is still a sum of contributions of the form \eqref{eq:global_volume}, the subsystem complexity is
\be
\label{eq:CVcompl_3int_global}
C_{A_1\cup A_2\cup A_3} = \frac{R}{8\pi G_N}\left(\frac{\ell_1+\ell_2+\ell_3}{\epsilon} + \pi\, w_{123}(d_1,d_2)\right)\,,
\ee
where $w_{123}$ is piecewise constant and takes the same integer values as in the case of Poincaré coordinates. Combining this expression with the subsystem complexities of single interval and two disjoint intervals via the definitions \eqref{eq:tripartitecomplexity} and \eqref{eq:Mtilde}, the extensive contributions proportional to $1/\epsilon$ cancel exactly, leading to
\be
\label{eq:M3_global}
M_3(A_1:A_2:A_3) = \frac{R\,\hat{w}(d_1,d_2)}{8G_N}\,,
\qquad
\Delta^{(3)}_C(A_1:A_2:A_3) = \frac{R\,\tilde{w}(d_1,d_2)}{8G_N}\,,
\ee
where $\hat{w}$ and $\tilde{w}$ are defined as in \eqref{eq:what} and \eqref{eq:wtilde}, and take the same integer values as in Sec.~\ref{subsec:static}. In particular, $\tilde{w} \geq 0$ throughout the entire parameter space, so the complexity gap $\Delta^{(3)}_C(A_1:A_2:A_3)$ is non-negative also in the case of global coordinates. The corresponding density plots for $w_{123}$, $\hat{w}$, and $\tilde{w}$ are shown in Fig.~\ref{fig:mutualcomplexityglobal}. Because of the compactness of the boundary circle, only the region satisfying $d_3=L-\ell_1-\ell_2-\ell_3-d_1-d_2\ge0$ is physically allowed. We observe that the phases identified by these three quantities differ from those shown in Fig.~\ref{fig:mutualcomplexity}. This discrepancy arises from the different boundary geometries in global and Poincaré coordinates.

\section{Time evolution of a general reduced density matrix}\label{app:RDMevolution}

In this appendix, we discuss the time evolution of reduced density matrices obtained by partially tracing a unitarily evolving density matrix. Already in the simple case of a subsystem consisting of a single qubit extracted from a two-qubit system, the evolution of the reduced density matrix is governed by an integro-differential equation known as the Nakajima-Zwanzig equation \cite{Breuer:2007juk,Agarwalbook}.

\subsection{A simple example}\label{subapp:primeexample}

Consider a system with two qubits $A$ and $B$, and a Hilbert space given by $\mathcal{H}=\mathcal{H}_A\otimes\mathcal{H}_B=\textrm{Span}\left\lbrace\vert0,0\rangle,\vert 1,0\rangle,\vert0,1\rangle,\vert1,1\rangle\right\rbrace$. For simplicity, take the system in the initial state
\begin{equation}
   \vert \psi(0) \rangle= \vert 1,0\rangle\,,
\end{equation}
where the two spins are in opposite orientations and unentangled.
Written in the canonical basis of a four dimensional Hilbert space, the initial density matrix reads
\begin{equation}
   \rho(0)= \vert \psi(0) \rangle\langle\psi(0)\vert=
   \begin{pmatrix}
       0 &\; 0 &\;0 &\;0 \\
       0 &\; 1 &\;0 &\;0 \\
       0  &\; 0&\;0 &\;0 \\
       0 &\; 0 &\;0 &\;0 \\
   \end{pmatrix}\,.
\end{equation}
We also assume that the total system evolves unitarily via the Hamiltonian
\begin{equation}
\label{eq:evolutionXXX_app}
    H=-2J(X_A X_B+Y_AY_B+Z_AZ_B)\,,
\end{equation}
where $X_i$, $Y_i$, and $Z_i$ are the Pauli matrices associated with the two qubits.
This Hamiltonian entangles the initially decoupled qubits into a state described by the density matrix
\begin{equation}
   \rho(t)= e^{-{\rm i}H  t}\vert \psi(0) \rangle\langle\psi(0)\vert e^{{\rm i}H  t}=
   \begin{pmatrix}
       0 & 0 &0 &0 \\
       0 & \cos^2\left(4J t\right) &-\frac{\rm i}{2} \sin\left(8J t\right) &0 \\
       0 & \frac{\rm i}{2}\sin\left(8J t\right) &\sin^2\left(4J t\right) &0 \\
       0 & 0 &0 &0 \\
   \end{pmatrix}\,.
   \label{eq:2spin_evolving fullDM}
\end{equation}
By tracing out the qubit $B$, we obtain the reduced density matrix
\begin{equation}
\label{eq:evolving RDM}
     \rho_A(t)=  \begin{pmatrix}
        \cos^2\left(4J t\right) &0  \\
       0 & \sin^2\left(4J t\right) \\
   \end{pmatrix}
=\left(\frac{1}{2}-\sin^2\left(4J t\right)\right)Z_A+\frac{1}{2}\boldsymbol{1}_2
   \,,
\end{equation}
where, for convenience, in the last step we wrote $\rho_A(t)$ in terms of Pauli matrices and identity.

Next, we want to understand the operation  that implements the time evolution of $\rho_A(t)$ by acting on the initial reduced density matrix, formally written as
\begin{equation}
 \mathcal{E}_t   (\rho_A(0))=\rho_A(t)\,.
\end{equation}
Notice that $\mathcal{E}_t$ cannot be a unitary transformation. Indeed, if
\begin{equation}
\label{eq:unitaryev_rhoL}
\rho_A(t)=U_A(t)\rho_A(0)U^\dagger_A(t)\,,
\end{equation}
then we should have
\begin{equation}
\textrm{Tr}\left(\rho_A^2(t)\right)=\textrm{Tr}\left(\rho_A^2(0)\right)=1\,,
\end{equation}
which is not true for any value of $t$. Alternatively, we can write $U_A(t)$ as the most general $2\times 2$ unitary matrix, and verify that there is no dependence on time for its four real parameters so that \eqref{eq:unitaryev_rhoL} is satisfied.

It is well-known that due to their effective non-unitary evolution, reduced systems can be seen as open quantum systems, where the environment is represented by the complementary subsystem. Thus, the time evolution of the reduced density matrices can also be studied as an open quantum system dynamics.
In some circumstances, open quantum systems evolve according to the Lindblad master equation \cite{Breuer:2007juk}
\begin{equation}
   \dot \rho_A(t)=-{\rm i}[\tilde{H}_A,\rho_A(t)]+\sum_k\gamma_k\left(E_k\rho_A(t)E_k^\dagger-\frac{1}{2}\lbrace E_k^\dagger E_k,\rho_A(t)\rbrace\right)\,,
   \label{eq:Lindbladeq}
\end{equation}
where $\tilde{H}_A$ is an effective Hermitian Hamiltonian, $E_k$ are suitable jump operators, $\gamma_k$ are non-negative couplings, and $\{,\}$ denotes the anticommutator.
This way, we look for operators $E_k$ and $\tilde{H}_A$ and couplings $\gamma_k$ that implement the evolution \eqref{eq:evolving RDM}. We preliminarily observe that \eqref{eq:evolving RDM} implies that
\begin{equation}
     \dot \rho_A(t)=- 4J\sin\left(8J t\right)Z_A\,,
\end{equation}
namely the left-hand side of \eqref{eq:Lindbladeq}  is diagonal.
Since $\rho_A(t)$ is a $2\times 2$ matrix, a natural ansatz (although not strictly necessary) is to choose $E_k$ and $\tilde{H}_A$ from a basis of the space of $2\times 2$ matrices. In particular, we take this basis to consist of the Pauli matrices together with the identity matrix.
Given that $\rho_A(t)$ is a diagonal matrix, choosing $E_k$ and $\tilde{H}_A$ to be $Z_L$ or $\boldsymbol{1}_2$ would give a vanishing contribution to the right-hand side of \eqref{eq:Lindbladeq}. Thus, we have to use (linear combinations of) $X_A$ and $Y_A$, or, equivalently, the raising and lowering operators $X_A\pm {\rm i} Y_A$. By setting
\begin{equation}
\label{eq:choice1_Limbladops}
    E_1=\begin{pmatrix}
        0 &0  \\
       1 & 0 \\
   \end{pmatrix}\,,
   \qquad
\tilde{H}_A=\omega(E_1+E_1^\dagger)+{\rm i}\tilde{\omega}(E_1-E_1^\dagger)\,,
\end{equation}
and all the other $E_k$ to zero, we obtain
    \begin{equation}
   \dot \rho_A(t)=-{\rm i}[\tilde{H}_A,\rho_A(t)]+\gamma\left(E_1\rho_A(t)E_1^\dagger-\frac{1}{2}\lbrace E_1^\dagger E_1,\rho_A(t)\rbrace\right)\,,
   \label{eq:Lindbladeq_choice1}
\end{equation}
which is compatible with $\rho_A(t)$ in \eqref{eq:evolving RDM} if 
$\omega=\tilde{\omega}=0$ and $\gamma=8J \tan(4Jt)$.
We are not satisfied with the last condition, given that we want the couplings $\gamma_k$ to be non-negative and time-independent\footnote{We may interpret this result as yielding an effective time-dependent dissipator for the dynamics. Nevertheless, in the present work, we restrict our attention to Lindblad equations with time-independent, non-negative coefficients.}.
This means that the choice \eqref{eq:choice1_Limbladops} does not reproduce the dynamics we look for.
This is an example of the well-known fact that, in general, the Lindbladian dynamics does not reproduce the evolution of reduced density matrices. One of the main reasons is that Lindblad equation holds for Markovian dynamics, while the dynamics of reduced density matrices is generally non-Markovian \cite{Breuer:2007juk,Banerjee:2025mzc,Banerjee:2025uew}. Although within some approximations (weak coupling between subsystems, Markovian and Born approximations), Lindbladian dynamics can emerge, a general treatment of the time evolution for reduced density matrices requires even more complicated master equations, such as the Nakajima–Zwanzig equation that we now review.

\subsection{Nakajima–Zwanzig equation}\label{subapp:ZNequation}

Consider a system $A$ coupled to an environment $B$, with a total Hamiltonian given by
\begin{equation}
\label{eq:HamiltonianNZ}
H = H_0 + \alpha H_I,
\end{equation}
where $H_0$ generates the evolution of the two decoupled parties, $H_I$ describes the interaction, and $\alpha$ is a dimensionless coupling.
We denote by $\rho(t)$ the density matrix of the combined state of the system and the environment evolving via the Hamiltonian \eqref{eq:HamiltonianNZ}.
In the interaction picture, the Liouville–von Neumann equation reads
\begin{equation}
\label{eq:Liouville–von Neumann}
\frac{\partial}{\partial t}\rho(t) = -{\rm i}\alpha[H_I(t),\rho(t)] \equiv \alpha\hat{\mathcal{L}}(t)\rho(t).
\end{equation}
We introduce the projection operators $\hat{P}$ and $\hat{Q}$ such that
\begin{equation}
\hat{P}\rho = \operatorname{tr}_B\{\rho\} \otimes \rho_B \equiv \rho_A \otimes \rho_B, \qquad \hat{Q} = \boldsymbol{1} - \hat{P},
\end{equation}
where $\rho_B$ is a fixed, unit-normalized ($\operatorname{tr}_B(\rho_B )= 1$), time-independent reference state of the environment. 
In several instances, $\rho_B$ is taken to be the stationary Gibbs state \cite{Agarwalbook}, but in our discussion we assume $\rho_B\equiv\operatorname{tr}_B[\rho(0)]$, which justifies the notation. The operators $\hat{P}$ and $\hat{Q}$ satisfy the relations
\begin{align}
\hat{P}^2 &= \hat{P}, \qquad \hat{Q}^2 = \hat{Q}, \qquad \hat{P}\hat{Q} = \hat{Q}\hat{P} = 0.
\end{align}
Applying $\hat{P}$ and $\hat{Q}$ to the equation of motion \eqref{eq:Liouville–von Neumann} and inserting $\boldsymbol{1} = \hat{P}+\hat{Q}$ yields two coupled equations
\begin{align}
\label{eq:Pequation}
\partial_t \hat{P}\rho(t) &= \alpha\hat{P}\hat{\mathcal{L}}(t)\hat{P}\rho(t) + \alpha\hat{P}\hat{\mathcal{L}}(t)\hat{Q}\rho(t), \\
\partial_t \hat{Q}\rho(t) &= \alpha\hat{Q}\hat{\mathcal{L}}(t)\hat{P}\rho(t) + \alpha\hat{Q}\hat{\mathcal{L}}(t)\hat{Q}\rho(t).
\label{eq:Qequation}
\end{align}
The formal solution of \eqref{eq:Qequation} is
\begin{equation}
\hat{Q}\rho(t) = G(t,0)\hat{Q}\rho(0) + \alpha\int_{0}^t ds\, G(t,s)\hat{Q}\hat{\mathcal{L}}(s)\hat{P}\rho(s)\,,
\end{equation}
where the propagator is given by
\begin{equation}
G(t,s) \equiv \mathrm{T}_\leftarrow \exp\!\left[\alpha\int_s^t ds'\,\hat{Q}\hat{\mathcal{L}}(s')\right], \qquad \frac{\partial}{\partial t}G(t,s) = \alpha\hat{Q}\hat{\mathcal{L}}(t)G(t,s)\,,
\end{equation}
with the symbol $\mathrm{T}_\leftarrow$ indicating the time-ordered exponential, and the initial conditions imposed at time $t=0$. Substituting this into \eqref{eq:Pequation} gives the \textit{Nakajima--Zwanzig (NZ) equation} \cite{Breuer:2007juk,Agarwalbook}
\begin{equation}
\label{eq:NZequation}
\frac{\partial}{\partial t}\hat{P}\rho(t)
= \alpha\hat{P}\hat{\mathcal{L}}(t)G(t,0)\hat{Q}\rho(0)
+ \alpha\hat{P}\hat{\mathcal{L}}(t)\hat{P}\rho(t)
+ \int_{0}^t ds\,\mathcal{K}(t,s)\hat{P}\rho(s),
\end{equation}
with memory kernel
\begin{equation}
\mathcal{K}(t,s) = \alpha^2\hat{P}\hat{\mathcal{L}}(t)G(t,s)\hat{Q}\hat{\mathcal{L}}(s)\hat{P}\,.
\end{equation}
The first term on the right is an inhomogeneous contribution from initial correlations, while the integral in the last term encodes the full non-Markovian memory over the time interval $[0,t]$.
So far, we have not introduced any simplifying approximation; the NZ equation \eqref{eq:NZequation} is an exact integro-differential equation describing the dynamics of the reduced density matrix $\rho_A(t)$ appearing in the tensor-product structure obtained from $\hat{P}\rho(t)$. The price to pay, however, is that this equation is generally prohibitively difficult to solve exactly, and the dependence on $\rho_A(t)$ is implicitly encoded in the projection operator $\hat{P}$. 

To make further progress, we now introduce a number of simplifying assumptions.
The first one consists of requiring that there is no entanglement between the system and the environment in the initial state, i.e. $\rho(0)=\rho_A(0)\otimes\rho_B$. This implies that $\hat{Q}\rho(0)=0$, which cancels the first term on the right-hand side of \eqref{eq:NZequation}. An extra condition we impose is that the Liouvillian $\hat{\mathcal{L}}$ in \eqref{eq:Liouville–von Neumann} is time-independent.

We stress that, even with this simplification, the Nakajima–Zwanzig equation \eqref{eq:NZequation}, being an integro-differential equation, is more general than the Lindblad equation \eqref{eq:Lindbladeq}. As discussed at the end of the previous section, the NZ equation reduces to the Lindblad equation only under additional strong approximations. We verify later in this section that the apparently simple dynamics of the reduced density matrix of a single qubit in a two-qubit system, which does not follow \eqref{eq:Lindbladeq} as shown in Sec.~\ref{subapp:primeexample}, instead evolves according to \eqref{eq:NZequation}.

To proceed with the analysis of the NZ equation,  it is useful to decompose the projection operator $\hat{P}$ as
\begin{equation}
\label{eq:decompositionProjectors}
\hat{P} = \widehat{\ell}\,\widehat{r}\,,
\end{equation}
where $\widehat{r}\rho=\rho_A$ implements the partial trace  and $\widehat{\ell}\rho_A=\rho_A\otimes \rho_B$ brings back the operator to the space where the Liouvillian acts.
It is also helpful to introduce the auxiliary function
\begin{equation}
y(t) = \int_0^t ds\, e^{\alpha(t-s)\hat{Q}\hat{\mathcal{L}}}\hat{Q}\hat{\mathcal{L}}\widehat{\ell}\rho_A(s),
\end{equation}
which propagates in time the reduced density matrix of $A$.
By simple algebraic steps, the conditions we have imposed and the auxiliary function $y(t)$ allow to decompose the NZ into two coupled differential equations
\begin{align}
\label{eq:diffeq_rhoS}
\partial_t\rho_A(t) &=\widehat{r}\hat{\mathcal{L}}\widehat{\ell}\rho_A(t)+\widehat{r}\hat{\mathcal{L}}y(t), \\
\partial_t y(t) &= \hat{Q}\hat{\mathcal{L}}\widehat{\ell}\rho_A(t) + \hat{Q}\hat{\mathcal{L}}y(t)\,.
\end{align}
Notice that the integral nature of the NZ equation is now hidden in the definition of $y(t)$.
This system of equations can be conveniently rewritten as
\begin{equation}                \frac{d}{dt}\begin{pmatrix}{\rho}_A(t)\\y(t)\end{pmatrix} = \underbrace{
    \begin{pmatrix} \widehat{r}\hat{\mathcal{L}}\widehat{\ell}&& \widehat{r}\hat{\mathcal{L}} \\  \hat{Q}\hat{\mathcal{L}}\widehat{\ell}&& \hat{Q}\hat{\mathcal{L}}
    \end{pmatrix}}_{\hat\Lambda} \begin{pmatrix}{\rho}_A(t)\\y(t)\end{pmatrix},
    \label{eq:NZ-block}
    \end{equation} 
    which leads to the formal solution
\begin{equation}\begin{pmatrix}\rho_A(t)\\y(t)\end{pmatrix}
    = e^{\hat\Lambda t}\begin{pmatrix}\rho_{A}(0)\\0\end{pmatrix}\,,
    \label{eq:NZ-matexp}
    \end{equation}
    where we have used that, by definition, $y(0)=0$. We stress that the exponential operator is a superoperator that acts on the vectorized operators $\rho_A(t)$ and $y(t)$, conveniently collected in a unique vector.

In standard textbooks \cite{Breuer:2007juk,Agarwalbook}, an additional condition is often imposed, namely  that the odd moments of the action of $\hat{\mathcal{L}}$ vanish, so that
\begin{equation}
\hat{P}\hat{\mathcal{L}}(t_1)\cdots\hat{\mathcal{L}}(t_{2n+1})\hat{P} = 0, \qquad n = 0,1,2,\ldots,
\end{equation}
which in particular implies $\hat{P}\hat{\mathcal{L}}(t)\hat{P} = 0$. If we denote the identity superoperator by $\hat{I}$, the matrix $\hat\Lambda$ in \eqref{eq:NZ-block} reduces to
\begin{equation}
\label{eq:NZ-matexp_simpl}
 \hat\Lambda=  \begin{pmatrix}0 && \hat{I}\\ 
\hat{Q}\hat{\mathcal{L}}\widehat{\ell}&& \hat{Q}\hat{\mathcal{L}}
 \end{pmatrix} \,,
\end{equation}
simplifying the solution \eqref{eq:NZ-matexp}. 

\begin{figure}[t!]
\centering
\includegraphics[width=0.6\textwidth]{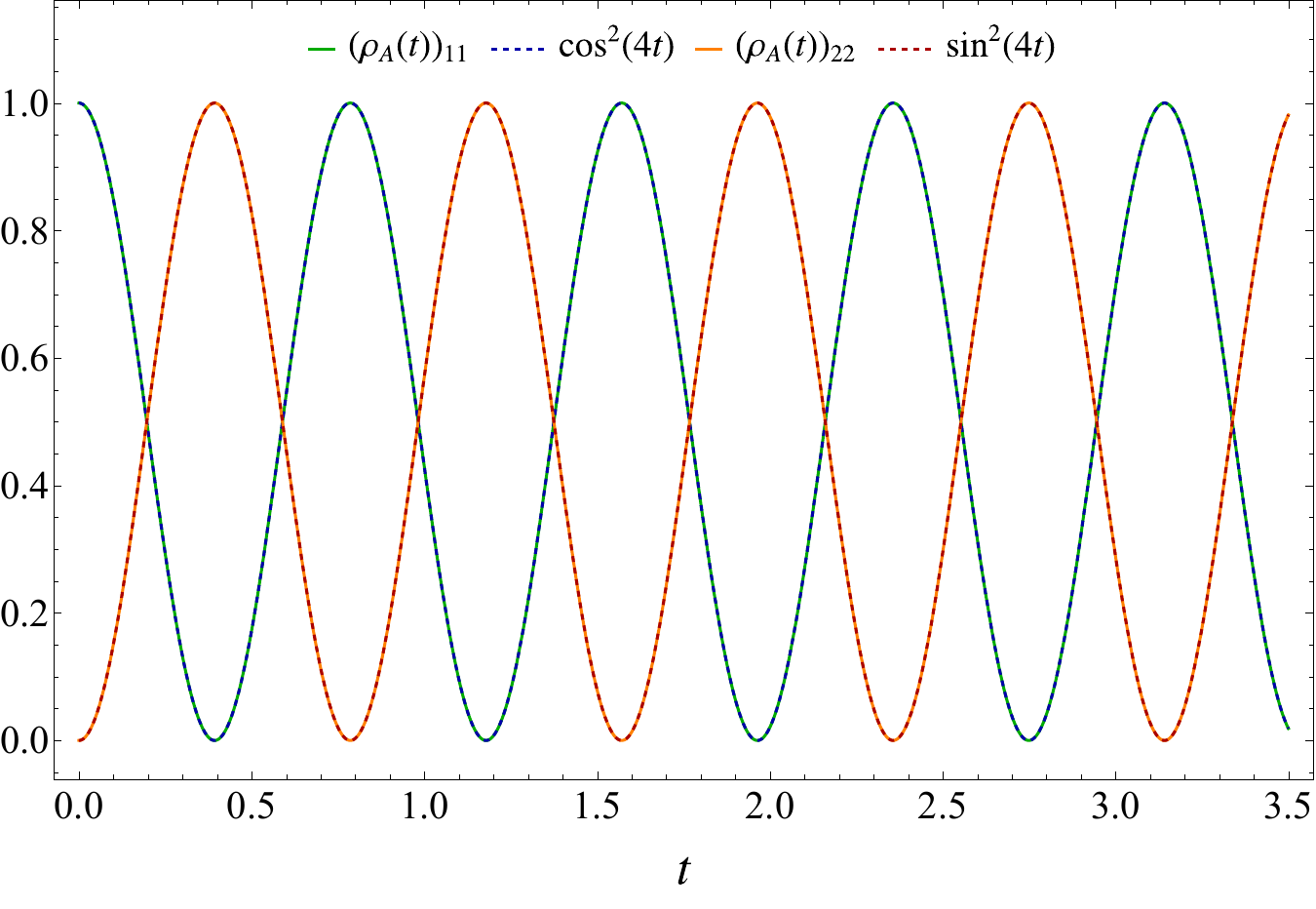}
\caption{The solid curves represent the non-vanishing entries of the reduced density matrix $\rho_A(t)$ obtained by solving the NZ equation \eqref{eq:NZ-block} for a subsystem consisting of a single qubit in the two-qubit system whose full time-evolving density matrix is given in \eqref{eq:2spin_evolving fullDM}. The results are benchmarked against the known analytical expressions in \eqref{eq:evolving RDM}, shown as dashed curves.}
\label{fig:1}
\end{figure}

As a consistency check of this approach, we can recast the example discussed in Sec.\,\ref{subapp:primeexample} in the formalism introduced in this section.
In that case, after vectorization, $\rho_A$ and $y$ have four and sixteen entries respectively. The superoperator $\hat{\mathcal{L}}$ can be constructed as
\begin{equation}
\hat{\mathcal{L}}=
-{\rm i}\left[\boldsymbol{1}_4\otimes H_I - H_I\otimes\boldsymbol{1}_4\right]
\,,   
\end{equation}
where $H_I$ is given by \eqref{eq:evolutionXXX_app}.
Since this Liouvillian is time-independent and the initial state is a tensor product of contributions from the subsystem and its complement, the NZ equation governing the dynamics of the reduced density matrix takes the form \eqref{eq:NZ-block}.
The projectors $\hat{Q}$ and $\hat{P}$ and the decomposition \eqref{eq:decompositionProjectors} are explicitly found from their action on the initial density matrix. The matrix $\hat\Lambda$ can be, therefore, numerically obtained. It has the form \eqref{eq:NZ-matexp_simpl} since, in this example, $\hat{P}\hat{\mathcal{L}}(t)\hat{P} = 0$.
The exponential operator can be applied to the initial density matrix as in \eqref{eq:NZ-matexp}, thus obtaining $\rho_A(t)$ and $y(t)$. Crucially, the time dependence of $\rho_A(t)$ is the same as \eqref{eq:evolving RDM}. 
The non-vanishing entries obtained from the solution \eqref{eq:NZ-matexp} are shown in Fig.\,\ref{fig:1}, where the analytical expressions \eqref{eq:evolving RDM} are represented by dashed curves. We observe perfect agreement between the solid and dashed curves.
In addition, we also obtain the entries of $y(t)$, which, once rearranged into a $4\times 4$ matrix, confirm that $y(t)$ is Hermitian, as expected from its definition.

The analysis presented in this appendix illustrates how intricate the time evolution of a reduced density matrix can be, even for a subsystem associated with a low-dimensional Hilbert space. The non-unitary and non-Markovian nature of the dynamics leads to the integro-differential NZ equation. Although, under certain simplifying assumptions, a formal solution such as \eqref{eq:NZ-block} can be obtained, its explicit evaluation quickly becomes prohibitive for systems with many degrees of freedom, namely for large Hilbert spaces, or when the aforementioned assumptions are relaxed.
For this reason, it is natural to ask whether such a complicated dynamics can be reformulated in terms of a simpler effective evolution that still captures all the relevant properties of the subsystem. In Sections \ref{sec:subsystemKrylov} and \ref{sec:SRSpreadComplexity}, we have discussed two approaches aimed at achieving this goal.

We should also mention that a possible and popular strategy is to first purify the density matrix to some auxiliary Hilbert space (e.g. by canonical purification) and study the unitary dynamics of this pure state. We find this approach very questionable when it comes to quantifying the complexity of $\rho_A(t)$. Indeed one often treats the purification itself as a free operation but in any physical system it would require the use of costly operations/gates. In principle one could try to define complexity by minimizing over all possible purifications but this would also make the definition not particularly computable, especially in many-body settings. For these reasons, in the main text we have worked with more direct approaches.

\section{Subsystem Krylov complexity in finite-dimensional Hilbert space}

\label{app:subsystemKrylovdetails}

\subsection{Computational details on the subsystem Lanczos coefficients}
\label{app:nonvanishing_an}

To keep the discussion in Sec.\,\ref{sec:subsystemKrylov} concise and focused on the main aspects of the construction and its physical implications, in this Appendix we collect several formulas, examples, and computations that complement and support the analysis presented in the main text.

\subsubsection{General case}
\label{app:Lanczos general}

In the main text, we discussed the difficulty of proving the sign of the squared subsystem coefficients $\left(b_n^{(A)}\right)^2$. Based on the explicit examples analyzed, all $\left(b_n^{(A)}\right)^2$ are found to be positive, implying that the corresponding coefficients $b_n^{(A)}$ are real. To support this conjecture from an analytical perspective, we focus on $\left(b_1^{(A)}\right)^2$ and show that, under certain assumptions, it is indeed positive.

 To progress in this direction, we can write an explicit expression of $\left(b_1^{(A)}\right)^2$ in \eqref{eq:b1_generalprocedure} as
 \begin{equation}
\label{eq:b1_generalprocedure_app}
 \left(b_1^{(A)}\right)^2
 =
\frac{\textrm{Tr}\left[\textrm{Tr}_B\left[\hat{\mathcal{L}}^2\rho(0)\right]\rho_A(0)\right]\textrm{Tr}\left(\rho^2_A(0)\right)-\left(\textrm{Tr}\left[\textrm{Tr}_B\left[\hat{\mathcal{L}}\rho(0)\right]\rho_A(0)\right]\right)^2}{\left[\textrm{Tr}\left(\rho^2_A(0)\right)\right]^2}\,,
\end{equation}
where $\hat{\mathcal{L}}$ is the Liouvillian superoperator, i.e. $\hat{\mathcal{L}}\,\cdot= {\rm i}\,[H,\cdot]$, and $\rho(0)$ is the pure density matrix of the full initial state. Notice the presence of the partial trace $\Tr_B$ over the Hilbert space of the subsystem $B$ appearing on the right-hand side. This is a crucial difference between these subsystem Lanczos coefficients and those arising in the standard Krylov-space approach for unitary operator dynamics.

Using the property $\textrm{Tr}_B[(X_A \otimes \mathds{1}_B) \cdot Y] = X_A \, \textrm{Tr}_B[Y]$, for any operator $Y$ acting on $\mathcal{H}_A\otimes \mathcal{H}_B$ and $X_A$ acting on $ \mathcal{H}_A$,
and defining $\widetilde{\mathcal{O}} \equiv \rho_A(0) \otimes \mathds{1}_B$, we obtain
\begin{eqnarray}
\textrm{Tr}[\rho_A^2(0)]
&=& \textrm{Tr}[\widetilde{\mathcal{O}}  \rho(0)] \,,
\quad
\textrm{Tr}_A \{ \textrm{Tr}_B[\hat{\mathcal{L}} \rho(0)]  \rho_A(0) \} 
= \textrm{Tr}[\widetilde{\mathcal{O}}\hat{\mathcal{L}} \rho(0)  ]
\,,
\\
&&
\textrm{Tr}_A \{ \rho_A(0)  \textrm{Tr}_B[\hat{\mathcal{L}}^2 \rho(0)] \} 
= \textrm{Tr}[\widetilde{\mathcal{O}}  \hat{\mathcal{L}}^2 \rho(0)]\,,
\end{eqnarray}
which allows us to rewrite
\eqref{eq:b1_generalprocedure_app} as
\begin{equation}
\label{eq:b1_generalprocedure_v2}
 \left(b_1^{(A)}\right)^2=
\frac{\textrm{Tr}[\widetilde{\mathcal{O}}  \hat{\mathcal{L}}^2 \rho(0)]\textrm{Tr}[\widetilde{\mathcal{O}}  \rho(0)] - \{\textrm{Tr}[\widetilde{\mathcal{O}}\hat{\mathcal{L}} \rho(0)  ] \}^2}{\left[\textrm{Tr}(\widetilde{\mathcal{O}}  \rho(0))\right]^2}\,.
\end{equation}
In what follows, we check that $\left(b_1^{(A)}\right)^2>0$ for the special case where $\tilde{O}$ and $\rho(0)$ commute. This is true, for example, where the initial state is a product state with factors in $\mathcal{H}_A$ and $\mathcal{H}_B$, as for the cases discussed in Secs.\,\ref{subsec:2spins} and \ref{subsec:3spins}.
In this case, the second term in the numerator of \eqref{eq:b1_generalprocedure_v2} vanishes. Indeed, if this is the case, both operators $\tilde{O}$ and $\rho(0)$ can be diagonalized in the same basis, namely 
\begin{equation}
\label{eq:common basis}
\rho(0)=\sum_\alpha p_\alpha |\alpha\rangle\langle\alpha|,
\qquad
\widetilde{\mathcal O}=\sum_\alpha o_\alpha |\alpha\rangle\langle\alpha|\,.
\end{equation}
In this basis, the diagonal entries of $[H, \rho(0)]$ read $([H, \rho(0)])_{\alpha \alpha}=\sum_\beta\left(H_{\alpha \beta} \rho_{\beta \alpha}-\rho_{\alpha \beta} H_{\beta \alpha}\right)=\sum_\beta H_{\alpha \beta} p_\beta \delta_{\beta \alpha}-p_\alpha \delta_{\alpha \beta} H_{\beta \alpha}=0$. Since the operator $\tilde{O}$ is diagonal in the same basis, when multiplied by $[H, \rho(0)]$ and evaluated in the trace, only the diagonal elements of the latter operator contribute, leading to
\begin{equation}
\mathrm{Tr}(\widetilde{\mathcal O}[H,\rho(0)]) =\mathrm{Tr}[\widetilde{\mathcal O}\hat{\mathcal L} \rho(0)]= 0.
\end{equation}
This fact, not only simplifies \eqref{eq:b1_generalprocedure_v2}, but also implies $\mu_1^{(A)}=0$. Notice that this is compatible with the findings in Secs.\,\ref{subsec:2spins} and \ref{subsec:3spins}, where the initial states are unentangled and, therefore, $[\widetilde{\mathcal O}, \rho(0)]=0$.

The remaining trace in the numerator of \eqref{eq:b1_generalprocedure_v2} can be evaluated explicitly, working in the basis introduced in \eqref{eq:common basis}. Given that
$([H,[H,\rho]])_{\alpha\alpha}
= 2\sum_\beta (p_\alpha-p_\beta)H_{\alpha\beta}H_{\beta\alpha}$ and using the Hermiticity of $H$, i.e.
$
H_{\alpha\beta}H_{\beta\alpha} = |H_{\alpha\beta}|^2 \ge 0
$,
we obtain
\begin{equation}
\left(b_1^{(A)}\right)^2 =
\frac{2\sum_{\alpha,\beta} o_\alpha (p_\alpha-p_\beta)|H_{\alpha\beta}|^2}
{\sum_\alpha o_\alpha p_\alpha}.
\end{equation}
Symmetrizing the numerator under $\alpha \leftrightarrow \beta$, we find
\begin{equation}
2\sum_{\alpha,\beta} o_\alpha (p_\alpha-p_\beta)|H_{\alpha\beta}|^2
=
\sum_{\alpha,\beta}
\left[
o_\alpha (p_\alpha-p_\beta)
+
o_\beta (p_\beta-p_\alpha)
\right]
|H_{\alpha\beta}|^2\,,
\end{equation}
which can be further simplified, leading to
\begin{equation}
\left(b_1^{(A)}\right)^2 =
\frac{
\sum_{\alpha,\beta}
(o_\alpha - o_\beta)(p_\alpha-p_\beta)|H_{\alpha\beta}|^2
}{
\sum_\alpha o_\alpha p_\alpha
}.
\end{equation}
Focusing on the interesting case where $\rho(0)=\rho_A(0)\otimes \rho_B(0)$ with $\rho_A$ and $\rho_B$ pure state density matrices,
 $\rho(0)$ has a single non-zero eigenvalue equal to $1$, while all other eigenvalues vanish.
 Given that $[\widetilde{O}, \rho(0)]=0$,
in the same basis, the operator $\widetilde{\mathcal O}$
has eigenvalues equal to one (with degeneracy $\dim\mathcal H_B$) in the subspace corresponding to the support of $\rho_A$ and vanishing elsewhere.
In particular, the eigenvectors corresponding to the non-zero eigenvalue of $\rho(0)$ also belong to the subspace where $\widetilde{\mathcal O}$ has eigenvalue $1$. As a consequence, for any pair of basis indices $\alpha$ and $\beta$, we have that
$(o_\alpha - o_\beta)(p_\alpha - p_\beta) \ge 0$.
Moreover, the denominator $\sum_\alpha o_\alpha p_\alpha$
is non-negative, since it is a sum of non-negative (zeros and ones) terms.
Putting everything together, we conclude that $\left(b_1^{(A)}\right)^2 \ge 0$.
Hence, $b_1^{(A)}$ is real.

Generalizing this proof to more generic initial states and to all higher Lanczos coefficients is a challenging task. We leave this analysis for future work, which will be more specifically devoted to the mathematical characterization of the subsystem Krylov complexity introduced in Sec.\,\ref{sec:subsystemKrylov}.

\subsubsection{Two-qubits system}
\label{app:2q}

In Sec.\,\ref{subsec:2spins}, we studied the subsystem Krylov complexity of a single qubit embedded in a two-qubit system, and compared it with the Krylov complexity of the full pure state density matrix and with the spread complexity of the corresponding evolving state. In this appendix, we report the autocorrelation function $R_K$ and the return amplitude $R_S$ used to compute these complexities.
In turn, these complexities are achieved by determining the Lanczos coefficients characterizing the two Krylov space dynamics. We denote by $a^{(K)}_n$ and $b^{(K)}_n$ the Lanczos coefficients associated with the growth of the density matrix, and by $a^{(S)}_n$ and $b^{(S)}_n$ those associated with the spreading of the state vector.

We start from the evolution of the full density matrix. The Lanczos coefficients and the Krylov complexity can be extracted from the autocorrelation function \eqref{eq:returnamplitudes}, which, in this case, reads
\begin{equation}
\label{eq:Krylovautocorr_2spin_unent}
\begin{aligned}
R_K(t)
=&\frac{1}{256}\Bigl(172 + 84\cos 8t + 4\cos 2\theta_1 + 4\cos 2\theta_2
    + 6\cos(2\theta_1-2\theta_2) \\
  &+ 6\cos 2(\theta_1+\theta_2)
+ 32\cos(\theta_1-\theta_2)  + 32\cos(\theta_1+\theta_2)
    - 2\cos(8t-2\theta_1) \\
  &- 2\cos(8t+2\theta_1)
   - 2\cos(8t+2\theta_2) - 3\cos(8t+2\theta_1-2\theta_2) \\
  &- 3\cos(8t-2\theta_1+2\theta_2)
   - 3\cos(8t-2(\theta_1+\theta_2)) - 16\cos(8t-\theta_1-\theta_2) \\
  &- 16\cos(8t+\theta_1-\theta_2)
  - 2\cos(8t-2\theta_2)
   -
    16\cos(8t-\theta_1+\theta_2) \\&- 16\cos(8t+\theta_1+\theta_2) - 3\cos(8t+2(\theta_1+\theta_2))\\&+ 32\sin^2 4t\bigl(\cos(2\varphi_1-2\varphi_2)\sin^2\theta_1\sin^2\theta_2
    \\&+ \cos(\varphi_1-\varphi_2)(4\sin\theta_1\sin\theta_2
    + \sin 2\theta_1\sin 2\theta_2)\bigr)\Bigr)\,.
\end{aligned}
\end{equation}
We show two examples of $R_K$ evaluated for different parameter choices as a function of time in Fig.~\ref{fig:2}.
The corresponding non-vanishing Lanczos coefficients are found to be
\begin{align}
b^{(K)}_{1}
= &
2\sqrt{2}\left(3 - 2\cos\theta_1\cos\theta_2 - \cos^2\theta_1\cos^2\theta_2 - 2\cos(\varphi_1 - \varphi_2)\sin\theta_1\sin\theta_2\right.
\nonumber\\&
\left.- 2\cos\theta_1\cos\theta_2\cos(\varphi_1 - \varphi_2)\sin\theta_1\sin\theta_2 - \cos^2(\varphi_1 - \varphi_2)\sin^2\theta_1\sin^2\theta_2\right)^{1/2}\,, 
\\
b^{(K)}_{2}
=&2\sqrt{2}\left(5 + 2\cos\theta_1\cos\theta_2 + \cos^2\theta_1\cos^2\theta_2 + 2\cos(\varphi_1 - \varphi_2)\sin\theta_1\sin\theta_2\right.
\nonumber\\&
\left.+ 2\cos\theta_1\cos\theta_2\cos(\varphi_1 - \varphi_2)\sin\theta_1\sin\theta_2 + \cos^2(\varphi_1 - \varphi_2)\sin^2\theta_1\sin^2\theta_2\right)^{1/2}\,,
\end{align}
while, due to the Hermiticity of the density matrix, the $a^{(K)}_n$ coefficients are vanishing.
We stress that the index ${K}$ is introduced here to make a clear distinction between these Lanczos coefficients, the ones reported in the main text encoding the subsystem dynamics ${(A)}$ and those, reported later, characterizing the spread of the state vector ${(S)}$.
The number of non-vanishing Lanczos coefficients implies that the dynamics of the full density matrix is occurring in a three-dimensional Krylov space, i.e. $\mathcal{K}_K=3$, as reported in the main text.

To compute the spread complexity, instead,  we start from the return amplitude in \eqref{eq:returnamplitudes}, which, for this specific example, is given by
\begin{equation}
\begin{aligned}
     R_S(t)
    &=[\cos(2t) + {\rm i}\sin(2t)]
    \Big[\cos(4t)
  - 2{\rm i}\cos(t)\cos(2t)\sin(t)
    \\ 
   &\times 
  \left(1 + \cos\theta_1\cos\theta_2
  + \cos(\varphi_1-\varphi_2)\sin\theta_1\sin\theta_2\right)\Big]\,.
\end{aligned}
\label{eq:spreadautocorr_2spin_unent}
\end{equation}
This expression is plotted in Fig.\,\ref{fig:2} as a function of time for a representative choice of parameters.
In this case, the only non-vanishing Lanczos coefficients are given by
\begin{align}
a^{(S)}_{0} &= -2 \left( \cos\theta_1 \cos\theta_2 
+ \cos(\varphi_1 - \varphi_2)\, \sin\theta_1 \sin\theta_2 \right)\,, \\
a^{(S)}_{1} &= 2 \left( 2 + \cos\theta_1 \cos\theta_2 
+ \cos(\varphi_1 - \varphi_2)\, \sin\theta_1 \sin\theta_2 \right)\,,
\end{align}
and
\begin{align}
    b^{(S)}_{1}&= 
2\left(3 - 2\cos\theta_1\cos\theta_2 - \cos^2\theta_1\cos^2\theta_2 - 2\cos(\varphi_1 - \varphi_2)\sin\theta_1\sin\theta_2\right.\\
&\left.- 2\cos\theta_1\cos\theta_2\cos(\varphi_1 - \varphi_2)\sin\theta_1\sin\theta_2 - \cos^2(\varphi_1 - \varphi_2)\sin^2\theta_1\sin^2\theta_2\right)^{1/2}\,.
\nonumber
\end{align}
As mentioned in the main text, the dimension of the Krylov space associated with the spread of the state is therefore $\mathcal{K}_S=2$.

The Lanczos coefficients $a^{(K)}_n$, $b^{(K)}_n$, $a^{(S)}_n$, and $b^{(S)}_n$, reported here for generic values of the parameters, are used to determine, via standard methods, the Krylov and spread complexities. These expressions are then used to obtain the curves shown in Figs.\,\ref{fig:3} and \ref{fig:4}.

\subsection{Insights from the Schmidt decomposition}
\label{app:SchmidtDecomp}

A privileged basis for representing vectors in a bipartite Hilbert space is the Schmidt basis. Since subsystem complexity requires, by definition, a bipartition and therefore a bipartite Hilbert space, in this appendix we analyze the subsystem return amplitude \eqref{eq:generaldef_RL}, together with other relevant quantities, from the perspective of the Schmidt decomposition.

Consider a system bipartite into $A$ and $B$ with Hilbert space given by $\mathcal H_A \otimes \mathcal H_B$. Given an initial state $|\psi(0)\rangle$, we can write its Schmidt decomposition as
\begin{equation}
\label{eq:initialbipartitestate}
|\psi(0)\rangle = \sum_{i=1}^{m}\sqrt{\lambda_i(0)}|i_A(0)\rangle \otimes |i_B(0)\rangle\,,
\end{equation}
where $\lambda_i(0) \geq 0$, $m\equiv\min(\dim\mathcal H_A,\dim\mathcal H_B)$, and $\sum_i \lambda_i(0) = 1$. In the Schmidt basis, the reduced density matrices of the two subsystems are
\begin{equation}
\rho_A(0) = \sum_{i=1}^{m} \lambda_i(0)|i_A(0)\rangle\langle i_A(0)|\,, \quad
\rho_B(0) = \sum_{i=1}^{m} \lambda_i(0)|i_B(0)\rangle\langle i_B(0)|\,.
\end{equation}
Recall that the Schmidt decomposition is the spectral decomposition of the reduced density matrices and guarantees that the nonzero eigenvalues of $\rho_A(0)$ and $\rho_B(0)$ are the same.
The state evolves unitarily via the operator $U = e^{-{\rm i} H t}$ such that
\begin{equation}
\label{eq:SchmidtDecomposition_timeevolution}
|\psi(t)\rangle = U |\psi(0)\rangle = \sum_{j=1}^{m} \sqrt{\lambda_j(t)}|j_A(t)\rangle \otimes |j_B(t)\rangle\,.
\end{equation}
Here, $\lambda_j(t)$ are the eigenvalues of the reduced density matrices at time $t$, and $|j_A(t)\rangle, |j_B(t)\rangle$ are the corresponding eigenvectors, which are not simply related to $|i_A(0)\rangle$ and $|i_B(0)\rangle$.
The reduced density matrices at time $t$ are
\begin{equation}
\begin{aligned}
\rho_A(t) & = \sum_{j=1}^{m} \lambda_j(t)|j_A(t)\rangle \langle j_A(t)|\,, \qquad
\rho_B(t) = \sum_{j=1}^{m} \lambda_j(t)|j_B(t)\rangle \langle j_B(t)|\,.
\end{aligned}
\end{equation}
Let us focus on the subsystem $A$ and the corresponding subsystem return amplitude \eqref{eq:generaldef_RL}. 
Plugging the Schmidt decompositions in the numerator of \eqref{eq:generaldef_RL}, we find
\begin{equation}
\begin{aligned}
\operatorname{Tr}\left[\rho_A(t) \rho_A(0)\right] & =\operatorname{Tr}\left[\left(\sum_{j=1}^m \lambda_j(t)\left|j_A(t)\right\rangle\left\langle j_A(t)\right|\right)\left(\sum_{i=1}^m \lambda_i(0)\left|i_A(0)\right\rangle\left\langle i_A(0)\right|\right)\right] \\
& =\sum_{i, j=1}^m \lambda_i(0) \lambda_j(t) \underbrace{\left\langle i_A \mid j_A(t)\right\rangle\left\langle j_A(t) \mid i_A\right\rangle}_{\left|\left\langle i_A(0) \mid j_A(t)\right\rangle\right|^2} \\
& =\sum_{i, j=1}^m \lambda_i(0) \lambda_j(t)\left|\left\langle i_A(0) \mid j_A(t)\right\rangle\right|^2 \,.
\end{aligned}
\end{equation}
Similarly, for the subsystem $B$, we find
\begin{equation}
\label{eq:RR_Schmidt}
\operatorname{Tr}[\rho_B(t) \rho_B(0)] = \sum_{i, j=1}^m \lambda_i(0) \lambda_j(t)|\langle i_B(0) | j_B(t)\rangle|^2\,.
\end{equation}
The denominators in $R_A(t)$ and $R_B(t)$ are independent of the subsystem and read $\operatorname{Tr}[\rho_A^2(0)]=\operatorname{Tr}[\rho_B^2(0)]=\sum_j \lambda_j^2(0)$. Thus, we conclude that
\begin{equation}
   R_A(t)=\frac{\sum_{i, j=1}^m \lambda_i(0) \lambda_j(t)|\langle i_A(0) | j_A(t)\rangle|^2}{\sum_j \lambda_j^2(0)} \,,
    \quad
  R_B(t)=\frac{\sum_{i, j=1}^m \lambda_i(0) \lambda_j(t)|\langle i_B(0) | j_B(t)\rangle|^2}{\sum_j \lambda_j^2(0)}   \,.
\end{equation}
We can use this representation of the subsystem return amplitudes to identify cases when $R_A(t)=R_B(t)$, and, as a consequence, $C_A(t)  =C_B(t)$. This happens, for instance, in the example discussed in Sec.\,\ref{subsec:2spins}. Comparing the above expressions for $R_A(t)$ and $R_B(t)$, we find that they are equal when 
\begin{equation}
\label{eq:modulusoverlap}
\left|\left\langle i_A(0) \mid j_A(t)\right\rangle\right|=\left|\left\langle i_B(0) \mid j_B(t)\right\rangle\right|\,,
\end{equation}
for any pair of elements in the Schmidt bases.
In Appendix \ref{app:RARB}, we prove a condition for \eqref{eq:modulusoverlap} to be true in the case of finite-dimensional Hilbert spaces.

We can use the Schmidt decomposition \eqref{eq:SchmidtDecomposition_timeevolution} to rewrite the return amplitude $R_S$ and the autocorrelation function of the full density matrix $R_K$. From \eqref{eq:returnamplitudes}, we find
\begin{equation} R_S(t)=\sum_{i,j=1}^m\sqrt{\lambda_i(t)\lambda_j(0)}\omega_{ij}^{(B)}(t)\omega_{ij}^{(A)}(t)\,,
\end{equation}
and
\begin{equation}
R_K(t)=\sum_{i,j,k,l=1}^m\sqrt{\lambda_i(t)\lambda_j(t)\lambda_k(0)\lambda_l(0)}\omega_{jk}^{(B)}(t)\omega_{jk}^{(A)}(t)\left(\omega_{il}^{(B)}(t)\right)^*\left(\omega_{il}^{(A)}(t)\right)^*\,,
\end{equation}
where, given $s=A,B$,
\begin{equation}
   \omega_{ij}^{(s)}(t) \equiv\left\langle i_s(t) \mid j_s(0)\right\rangle\,.
\end{equation}
These expressions make manifest the known fact that $R_K(t)=\vert R_S(t)\vert^2 $.
In this new notation, the two subsystem return amplitudes read
\begin{equation}
R_A(t)=\frac{\sum_{i, j=1}^m \lambda_i(0) \lambda_j(t)|\omega_{ij}^{(A)}(t)|^2}{\sum_{i=1}^m \lambda^2_i(0)}\,,
\qquad
R_B(t)=\frac{\sum_{i, j=1}^m \lambda_i(0) \lambda_j(t)|\omega_{ij}^{(B)}(t)|^2}{\sum_{i=1}^m \lambda^2_i(0)}\,.
\end{equation}
In the case where $\omega_{ij}^{(s)}(t)$ is independent of time (for instance, because the Schmidt vectors are time independent), we have
\begin{equation}
\label{eq:time indep_Schmidtvect}
    \omega_{ij}^{(s)}(t)=\omega_{ij}^{(s)}(0)=\delta_{ij}\,,
\end{equation}
which implies
\begin{equation}
R_s(t)=\frac{\sum_{i=1}^m \lambda_i(0) \lambda_i(t)}{\sum_{i, j=1}^m \lambda^2_i(0)}\,.
\end{equation}
If \eqref{eq:time indep_Schmidtvect} occurs for both $s=A$ and $s=B$,
then $R_A(t)  =R_B(t)$ and, correspondingly, $C_A(t)  =C_B(t)$.
In this case, we also find
\begin{equation}
  R_S(t)  = 
\sum_{i=1}^m\sqrt{\lambda_i(t)\lambda_i(0)}
  \,,
  \qquad
   R_K(t)  = 
\sum_{i,j=1}^m\sqrt{\lambda_i(t)\lambda_i(0)\lambda_j(t)\lambda_j(0)}\,,
\end{equation}
and,
since both $R_K(t) $ and $R_S(t)$ are real, their relation simplifies to $(R_S(t))^2=R_K(t)$.
Following the discussion of Sec.\,\ref{subsec:KrylovSubsyst_generalframework} and extracting the moments from the (subsystem) return amplitudes and the autocorrelation function, we could write the Lanczos coefficients in terms of the time derivatives of $\omega_{ij}^{(s)}(t)$ and $\lambda_i(t)$. However, these Schmidt data do not have a generally established behaviour in time, and using their time derivatives to understand the general properties of the Lanczos coefficients or potential hierarchies between spread, Krylov and subsystem complexities is not a promising strategy.

Another interesting scenario is when the initial state is unentangled. If
\begin{equation}
|\psi(0)\rangle=|\bar{j}_A\rangle\otimes|\bar{j}_B\rangle\,,
\end{equation}
we obtain
\begin{eqnarray}
R_S(t)&=&\sum_{i=1}^m\sqrt{\lambda_i(t)}\omega_{i\bar{j}}^{(B)}(t)\omega_{i\bar{j}}^{(A)}(t)\,,
\\
R_K(t)&=&\sum_{i,k=1}^m\sqrt{\lambda_i(t)\lambda_k(t)}\omega_{k\bar{j}}^{(B)}(t)\omega_{k\bar{j}}^{(A)}(t)\left(\omega_{i\bar{j}}^{(B)}(t)\right)^*\left(\omega_{i\bar{j}}^{(A)}(t)\right)^*
\,,
\end{eqnarray} 
and
\begin{equation}
R_A(t)=\sum_{i=1}^m \lambda_i(t)|\omega_{\bar{j}i}^{(A)}(t)|^2\,,
\qquad
R_B(t)=\sum_{i=1}^m \lambda_i(t)|\omega_{\bar{j}i}^{(B)}(t)|^2\,.
\end{equation}
If, in addition, \eqref{eq:time indep_Schmidtvect} is valid, we obtain
\begin{equation} 
\label{eq:retampl_timeindevec_unent}
R_S(t)=\sqrt{\lambda_{\bar{j}}(t)}
\,,\qquad
R_K(t)=R_A(t)=R_B(t)=\lambda_{\bar{j}}(t)
\,,
\end{equation}
namely, the Krylov, spread and subsystem complexities are governed by the time evolution of a unique Schmidt eigenvalue. The second equation in \eqref{eq:retampl_timeindevec_unent} implies the same relation among the complexities, i.e. $C_K(t)=C_A(t)=C_B(t)$.  Notice that \eqref{eq:retampl_timeindevec_unent} is valid when the Schmidt basis in \eqref{eq:SchmidtDecomposition_timeevolution} coincides with the Krylov basis. Indeed, for this circumstance to happen, the initial state has to be a product state, given that the Krylov amplitudes satisfy $\psi_n(0)=\delta_{n0}$, and the Schmidt eigenvectors are time independent (as the Krylov vectors should be), implying \eqref{eq:time indep_Schmidtvect}.

\subsection{A condition for $R_A=R_{B}$}
\label{app:RARB}

In the main text, we have discussed that, given a system bipartitioned into two complementary subregions $A$ and $B$, in general $R_A \neq R_B$. However, the cases in which $R_A = R_B$ are particularly interesting, as they imply $C_A = C_B$. In this appendix, we prove a sufficient condition under which $R_A = R_B$ holds. We stress that this condition does not encompass the most general class of dynamics for which this equality is satisfied.

Consider a bipartite Hilbert space $\mathcal{H}=\mathcal{H}_A\otimes\mathcal{H}_B $ where the two factors have the {\it same} finite dimensionality, an initial state of the form \eqref{eq:initialbipartitestate} and a Hamiltonian $H$ on $\mathcal{H}$.
Consider the operator $P_{AB}:\mathcal{H}_A\otimes\mathcal{H}_B\to\mathcal{H}_A\otimes\mathcal{H}_B$
such that $P_{AB}\bigl( |\phi_A\rangle \otimes |\chi_B\rangle \bigr)
= |\chi_B\rangle \otimes |\phi_A\rangle$, for any pair of basis vectors of $\mathcal{H}_A$ and $\mathcal{H}_B$.
Let us call $\ket{i_A}\in\mathcal{H}_A$ and $\ket{i_B}\in\mathcal{H}_B$ the elements of the Schmidt bases of the initial state. Since $\mathcal{H}_A$ and $\mathcal{H}_B$ are isomorphic, we can think of the Schmidt vectors $\ket{i_A}$ and $\ket{i_B}$ as belonging to the same Hilbert space $\tilde{\mathcal{H}}\simeq\mathcal{H}_A\simeq\mathcal{H}_B$. In $\tilde{\mathcal{H}}$, the two sets $\ket{i_A}$ and $\ket{i_B}$ are not mutually orthogonal and linearly independent. We now define a Hermitian operator $W:\tilde{\mathcal{H}}\to\tilde{\mathcal{H}}$ such that 
\begin{equation}
W |i_A\rangle = |i_B\rangle,
\qquad
W |i_B\rangle = |i_A\rangle .
\end{equation}
In our case, as will become clear below, a redefinition of $\ket{i_B}$ by an overall phase does not affect the result. We can therefore fix this phase freedom by $
\ket{i_B} \to e^{-{\rm i}\arg(\langle{i_A} \ket{i_B})} \ket{i_B} ,$
so that the overlap $\langle{i_A}|{i_B}\rangle$ becomes real and non-negative. With this choice of phase convention, we can then construct the operator $W$ as 
\begin{equation}
W
= \frac{(|i_A \rangle+| i_B\rangle) (\langle i_A| + \langle i_B|)}
{(\langle i_A| + \langle i_B|)(|i_A \rangle+| i_B\rangle)}
- \frac{(|i_A \rangle-| i_B\rangle) (\langle i_A| - \langle i_B|)}
{(\langle i_A| - \langle i_B|)(|i_A \rangle-| i_B\rangle)}.
\end{equation}

\noindent
\textbf{Statement:} 
If $[H, P_{AB}]=[H, W \otimes W]=0$, then \eqref{eq:modulusoverlap} holds, which implies $R_A(t)=R_B(t)$. 
\\

\noindent
\textbf{Proof:}
We begin by observing that, $P_{AB}^2 = \boldsymbol{1}$, and that, since the Hamiltonian commutes with $P_{AB}$, also the evolution operator does, i.e.
\begin{equation}
[H, P_{AB}] = 0 
\;\Longrightarrow\; 
[U(t), P_{AB}]=0.
\end{equation}
If the Hamiltonian commutes with $W \otimes W$, then we also have
\begin{equation}
[H, W \otimes W] = 0
\;\Longrightarrow\;
[U(t), W \otimes W] = 0.
\end{equation}
The strategy is to rewrite $P_{AB}|\psi(t)\rangle$ in two equivalent ways.
By definition of $P_{AB}$, we have
\begin{equation}
\label{PLRpsi(t)_1}
P_{AB} |\psi(t)\rangle
=\sum_j\sqrt{\lambda_j(t)} P_{AB}|j_A(t)\rangle \otimes |j_B(t)\rangle=\sum_i\sqrt{\lambda_j(t)} |j_B(t)\rangle \otimes |j_A(t)\rangle .
\end{equation}
On the other hand, using that $P_{AB}$ and $U(t)$ commute, we have
\begin{equation}
P_{AB} |\psi(t)\rangle
= U(t)\sum_i\sqrt{\lambda_i(0)} 
P_{AB}|i_A\rangle \otimes |i_B\rangle = U(t)\sum_i\sqrt{\lambda_i(0)} 
|i_B\rangle \otimes |i_A\rangle\,. 
\end{equation}
Exploiting the action of $W$ on the Schmidt basis of the initial state and the commutation of $W \otimes W$ with the evolution operator, we finally obtain
\begin{equation}
\begin{split}
P_{AB} |\psi(t)\rangle
  &= U(t)\sum_i \sqrt{\lambda_i(0)} 
|i_B\rangle \otimes |i_A\rangle
= (W \otimes W)\sum_j \sqrt{\lambda_j(t)} 
|j_A(t)\rangle \otimes |j_B(t)\rangle 
\\
&=
\label{PLRpsi(t)_2}
\sum_j \sqrt{\lambda_j(t)} 
W|j_A(t)\rangle \otimes W |j_B(t)\rangle
.
\end{split}
\end{equation}
Comparing \eqref{PLRpsi(t)_1} with \eqref{PLRpsi(t)_2}, we find that the left and right set of vectors in the tensor product can be identified up to a phase factor, namely
\begin{equation}
|j_B(t)\rangle
= e^{{\rm i}\phi_j} W |j_A(t)\rangle .
\end{equation}
Computing the overlap for the Schmidt basis vectors of the subsystem $B$, we find 
\begin{equation}
\langle i_B \mid j_B(t)\rangle
= \langle i_B |
 e^{{\rm i}\phi_j} W |j_A(t)\rangle,
\end{equation}
and, using the Hermitian action of $W$, we obtain
\begin{equation}
\langle i_B \mid j_B(t)\rangle
= e^{{\rm i}\phi_j}
\langle i_A \mid j_A(t)\rangle .
\end{equation}
Hence, their absolute values are equal
\begin{equation}
\left|
\langle i_B \mid j_B(t)\rangle
\right|
=
\left|
\langle i_A \mid j_A(t)\rangle
\right|\,,
\end{equation}
and, therefore, \eqref{eq:modulusoverlap} is valid. Since \eqref{eq:modulusoverlap} is satisfied for any pairing between the Schmidt eigenvalues of the initial and time-evolved states, it follows that $R_A(t)=R_B(t)$.

\hfill\qedsymbol

\subsection{Two-qubits example with non-vanishing $a_n$ coefficients}
\label{app:nonvanishing_an}

The example of two-qubit dynamics analyzed in Sec.\,\ref{subsec:2spins} led to subsystem Lanczos coefficients $a_n^{(A)} = 0$. In this appendix, we show that this is not a generic feature of two-qubit time evolution.

We consider an initial state which entangles two qubits
\begin{equation}
\label{eq:entangled initial state}
    \ket{\Psi(0)} = \sin\frac{\theta}{2}\ket{1,0} + e^{{\rm i}\varphi}\cos\frac{\theta}{2}\ket{0,1},
\end{equation}
with $\theta \in [0,\pi)$ and $\varphi \in [0,2\pi)$, and the corresponding initial
density matrix $\rho(0) = \ket{\Psi(0)}\bra{\Psi(0)}$. When $\theta\neq 0, \pi/2$, $\ket{\Psi(0)}$ is an entangled state. We let the system evolve with \eqref{eq:evolutionXXX}, and we perform the partial trace on the second qubit. 
The autocorrelation function of the resulting reduced density matrix reads
\begin{equation}
\label{eq:returnamplitude_subsystem_2sp_ent}
R_A(t)=\frac{2 + 2\cos(8t)\cos^2\theta + \sin \varphi\sin(8t)\sin(2\theta)}{3 + \cos(2\theta)}\,.
\end{equation}
Notice that, also for this two-qubit example, $R_A(t)=R_B(t)$.
We apply the moment recursion method from Sec.\,\ref{subsec:KrylovSubsyst_generalframework} to \eqref{eq:returnamplitude_subsystem_2sp_ent} and we obtain five non-vanishing subsystem Lanczos coefficients 
\begin{align}
a_0^{(A)} &= -\frac{8{\rm i}\sin\varphi\sin 2\theta}{3+\cos 2\theta}\,,\\
b_1^{(A)} &= \frac{8\sqrt{2\cos^2\theta(3+\cos 2\theta)+\sin^2\varphi\sin^2 2\theta}}{3+\cos 2\theta}\,,\\
a_1^{(A)} &=- \frac{8{\rm i}\cos\theta(10+2\cos 2\varphi+\cos(2\varphi-2\theta)+2\cos 2\theta+\cos 2(\varphi+\theta))\sin\varphi\sin^3\theta}{(3+\cos 2\theta)(2\cos^2\theta(3+\cos 2\theta)+\sin^2\varphi\sin^2 2\theta)}\,,\\b_2^{(A)} &= \frac{16\sqrt{2}\abs{\sec\theta}\sqrt{(3+\cos 2\theta)(3+\cos 2\theta-2\cos 2\varphi\sin^2\theta)}}{14-2\cos 2\varphi+\cos(2\varphi-2\theta)+2\cos 2\theta+\cos 2(\varphi+\theta)}\,,\\a_2^{(A)} &=\frac{32{\rm i}\sin\varphi\tan\theta}{7+\cos 2\theta-2\cos 2\varphi\sin ^2\theta}\,.
\end{align}
Compared with the previous example, here $R_A$ is not an even function of time and, therefore, leads to non-vanishing $a_n^{(A)}$ coefficients. Crucially, these are purely imaginary, making the effective Schrödinger-like Krylov dynamics of the reduced density matrix non-unitary. We stress again that this is not surprising, since it is well-known that reduced density matrices of unitarily evolving systems behave similarly to open quantum systems. The dimension of this effective Krylov space for the reduced density matrix evolution is $\mathcal{K}_A=\mathcal{K}_B=3$. We omit the cumbersome expression of the subsystem Krylov complexity $C_A(t)$, which can be obtained from the subsystem Lanczos coefficients above. We discuss the properties of $C_A(t)$ via the figures shown in this appendix. Notice that, due to $R_A(t)=R_B(t)$, we have $C_A(t)=C_B(t)$.

On the other hand, as expected, the dynamics of the full density matrix is unitary and its Krylov complexity can be determined by considering the autocorrelation function in \eqref{eq:returnamplitudes}, which, for this example, reads 
\begin{eqnarray}
\label{eq:Krylovautocorr_2spin_ent}
R_K(t)&=&\frac{1}{16}\Big(10+2\cos 2\varphi-\cos(2\varphi-8t)+6\cos 8t
-\cos(2\varphi+8t)\nonumber
\\
&-&8\cos^2\varphi\cos 2\theta\sin^2(4t)\Big)\,.
\end{eqnarray}
Via the standard procedure, this expression leads to the non-vanishing Lanczos coefficients
\begin{align}
b^{(K)}_1 &= 2\sqrt{2}\sqrt{3+\cos 2\theta-2\cos 2\varphi\sin^2\theta}\,,\\
b_2^{(K)} &= \frac{\sqrt{12\cos 2\varphi+3\cos 4\varphi+8\cos^4\varphi(\cos 4\theta-4\cos 2\theta)-55}}{\sqrt{2\cos^2\varphi\sin^2\theta-2}}\,,
\end{align}
corresponding to the dimension $\mathcal{K}_K=3$ for the Krylov space where the dynamics of the full density matrix is happening. From these coefficients, we straightforwardly determine the Krylov complexity $C_K(t)$. We do not report its explicit expression here; however, later in this appendix, we plot it for selected parameter values. 

\begin{figure}[t!]
\centering
\includegraphics[width=0.495\textwidth]{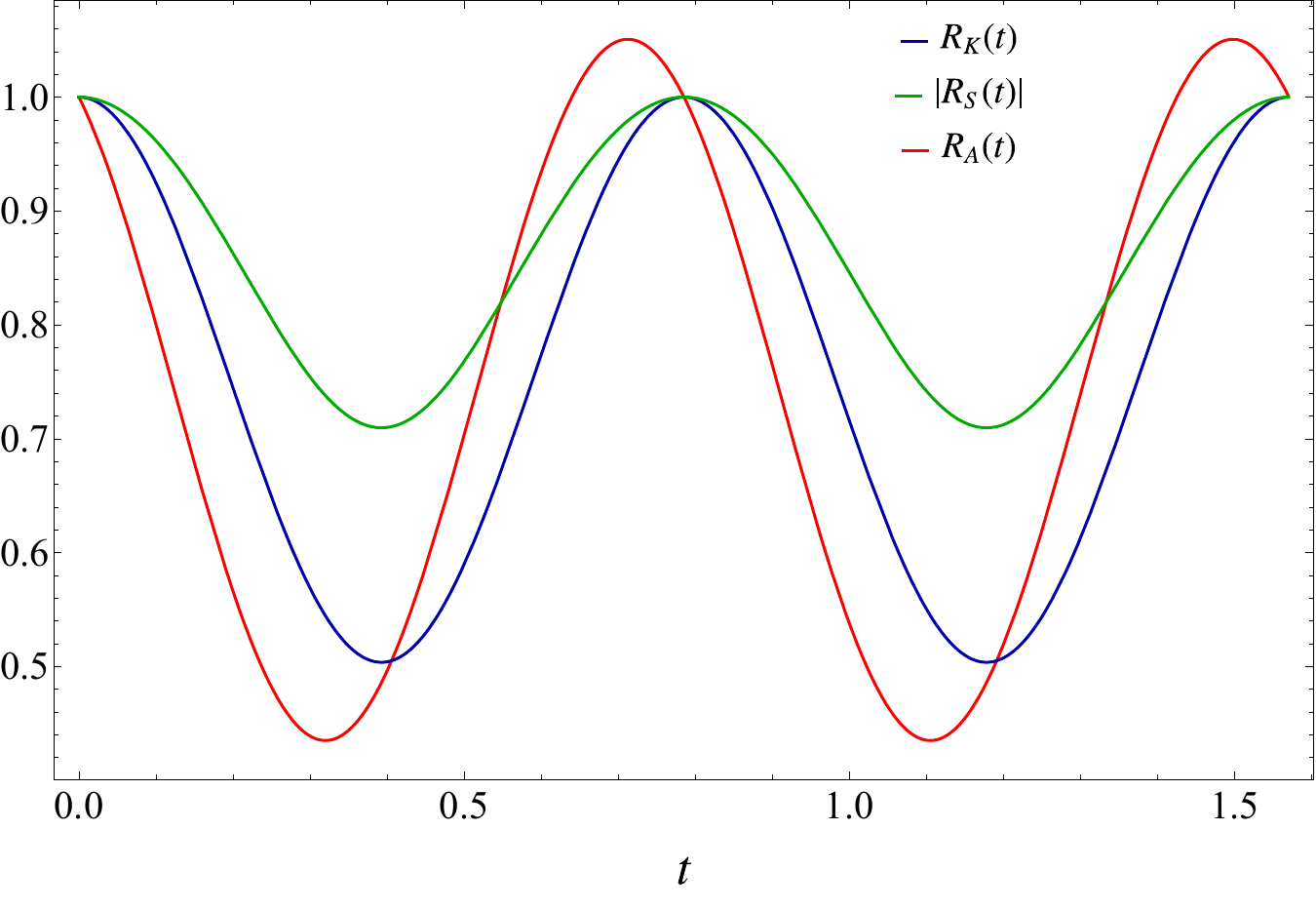}
\includegraphics[width=0.495\textwidth]{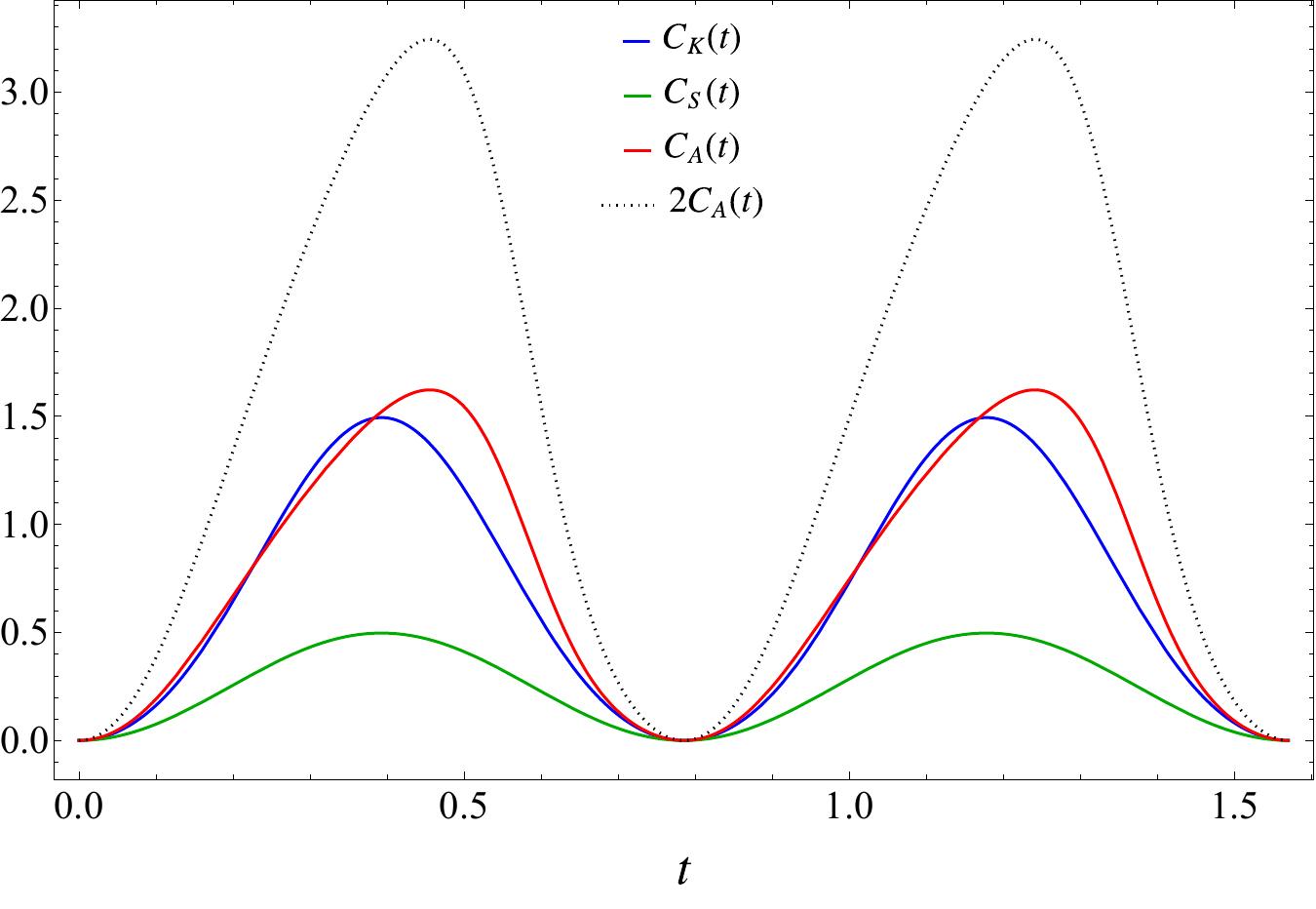}

\caption{
The initial state \eqref{eq:entangled initial state}, with parameters $\theta = 2.2$ and $\varphi = 0.5$, is evolved via the Hamiltonian \eqref{eq:evolutionXXX}. The system is bipartitioned into a subsystem $A$ consisting of the first qubit and the complementary subsystem $B$ containing the remaining one. In the left panel, we plot the return amplitude \eqref{eq:spreadautocorr_2spin_ent} and the autocorrelation function \eqref{eq:Krylovautocorr_2spin_ent} of the pure state density matrix, together with the subsystem return amplitudes \eqref{eq:returnamplitude_subsystem_2sp_ent} associated with $A$, as functions of time. In the right panel, we show the corresponding complexities.}
\label{fig:entangled1}
\end{figure}
\begin{figure}[t!]
\centering
\includegraphics[width=0.495\textwidth]{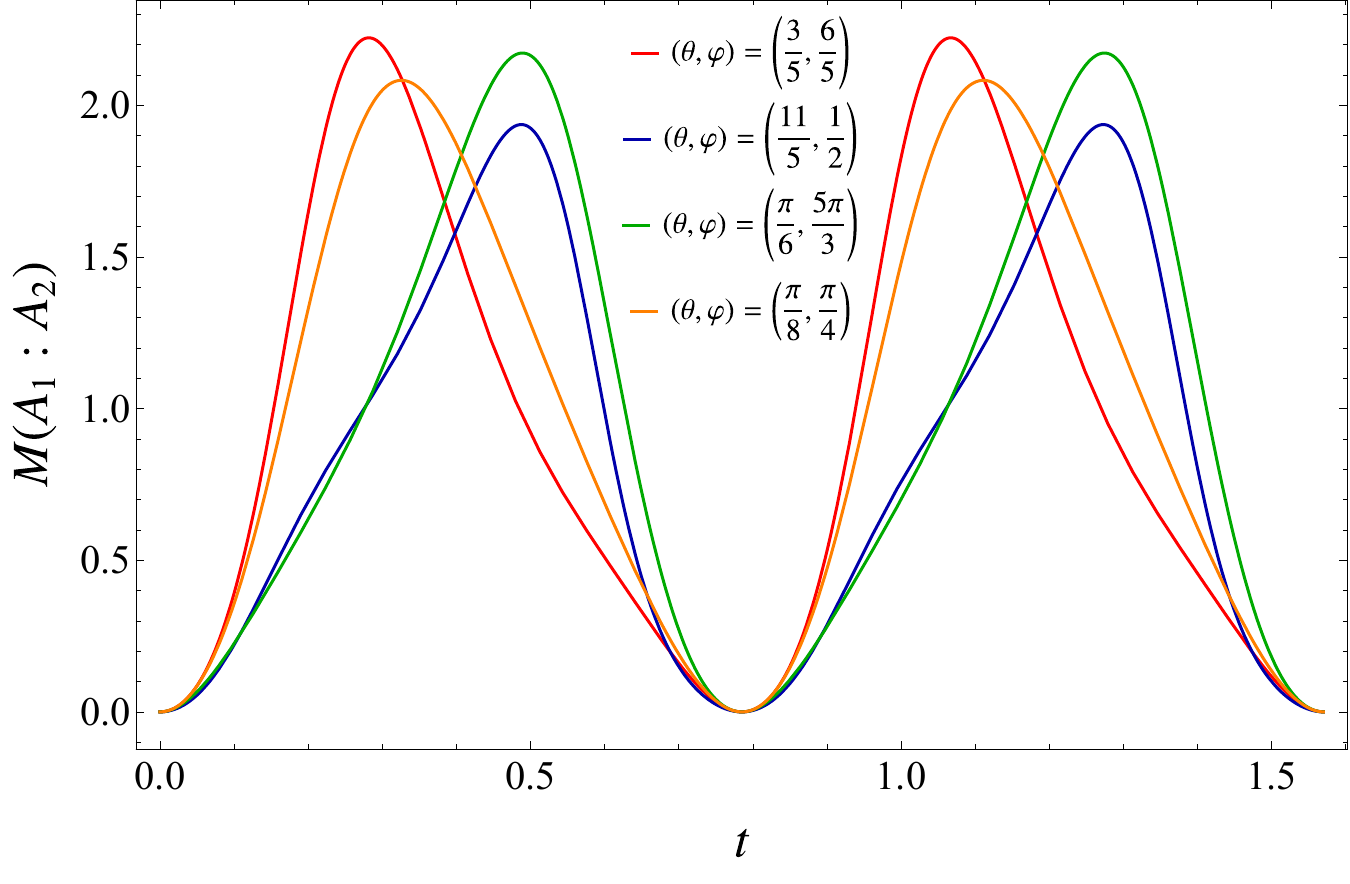}
\includegraphics[width=0.495\textwidth]{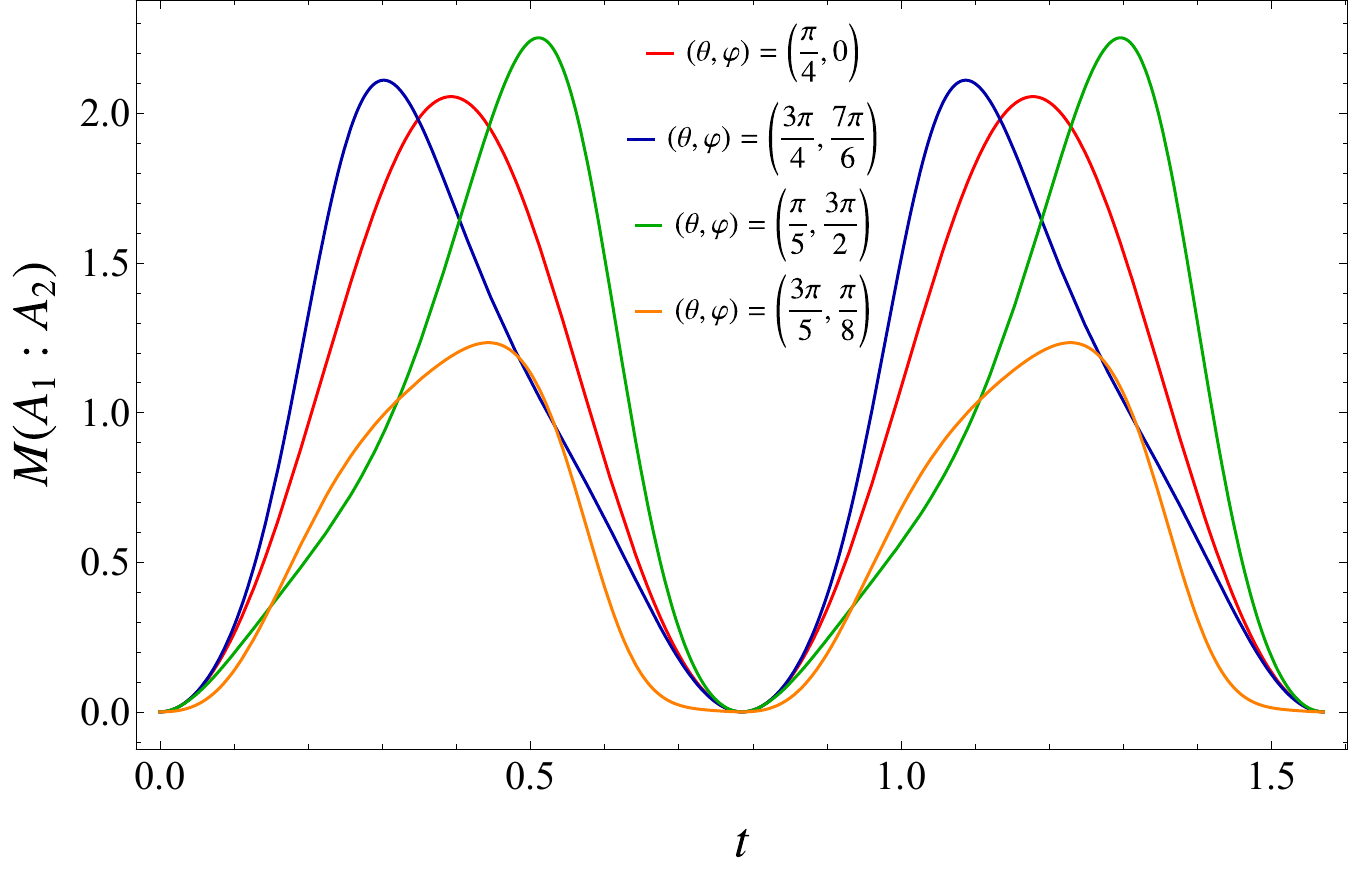}

\caption{Mutual complexity \eqref{eq:mutualcomplexity} for a bipartition of a two-qubit system into its individual qubits. The quantity is computed using the subsystem Krylov complexity and the Krylov complexity of the corresponding pure state density matrix. The initial state is \eqref{eq:entangled initial state} with the parameters reported in the panels, while the evolution Hamiltonian is given by \eqref{eq:evolutionXXX}.}
\label{fig:entangled2}
\end{figure}

Finally, to compute the spread complexity of the evolving state vector, we start from the return amplitude in \eqref{eq:returnamplitudes}
\begin{equation}
\label{eq:spreadautocorr_2spin_ent}
R_S(t)=(\cos 2t + {\rm i}\sin 2t)(\cos 4t - {\rm i}\cos\varphi\sin 4t\sin\theta)\,,
\end{equation}
and we extract three non-vanishing Lanczos coefficients 
\begin{equation}
a^{(S)}_{1} = 2 - 4\cos\varphi\sin\theta\,,\quad
b^{(S)}_{1} = 4\sqrt{1 - \cos^2\varphi\sin^2\theta}\,,\quad
a^{(S)}_{2} = 2 + 4\cos\varphi\sin\theta\,,
\end{equation}
corresponding to a Krylov space for the spread dynamics of dimension $\mathcal{K}_S=2$. The spread complexity $C_S(t)$ is finally computed from the Lanczos coefficients. Its expression, not reported here, is used in the figures of this appendix.

Although we have considered a different initial state, the values $\mathcal{K}_S=2<\mathcal{K}_K=\mathcal{K}_A=3$ are the same as those found in Sec.\,\ref{subsec:2spins} for the initial unentangled state \eqref{eq:unentangled initialstate}. We therefore expect a qualitatively similar comparison between subsystem, Krylov, and spread complexities, which is shown in the right panel of Fig.\,\ref{fig:entangled1}, together with the corresponding return amplitudes for the same choice of parameters (left panel).
However, differently from what is observed in Fig.\,\ref{fig:3}, in this case, while the spread complexity is always the smallest quantity, there is no fixed ordering between $C_K$ and $C_A$ throughout the time evolution.

Finally, in Fig.\,\ref{fig:entangled2}, we show the mutual complexity \eqref{eq:mutualcomplexity} evaluated using the subsystem Krylov proposal as a function of time. The curves displayed are obtained for various choice of parameters and all of them exhibit a positive sign for any value of time.

\section{Mutual and tripartite information from CFT and quasi-particle picture}
\label{app:QuasiParticle}

The out-of-equilibrium dynamics of entanglement measures in $1+1$-dimensional CFTs can be understood in terms of the spreading of counter-propagating quasi-particles emitted from every point in space and moving ballistically. If, at a given time, the emitted quasi-particles belong to two complementary regions, they contribute to the entanglement between them. As briefly reviewed in the main text, this constitutes the basis of the quasi-particle picture of entanglement dynamics, originally introduced for CFT evolutions.
In this appendix, we report the CFT formulas describing the evolution of the entanglement entropies of disjoint intervals. Using these expressions, we can construct the corresponding CFT predictions for the mutual and tripartite information, \eqref{eq:mutualinfo} and \eqref{eq:tripartiteinfo} respectively, which help us understand the early-time behavior shown in Figs.\,\ref{fig:tripartition_time_DBC_nonemptyB} and \ref{fig:quandripartition_time_DBC}.

\begin{figure}[t!]
\includegraphics[width=0.5\textwidth]{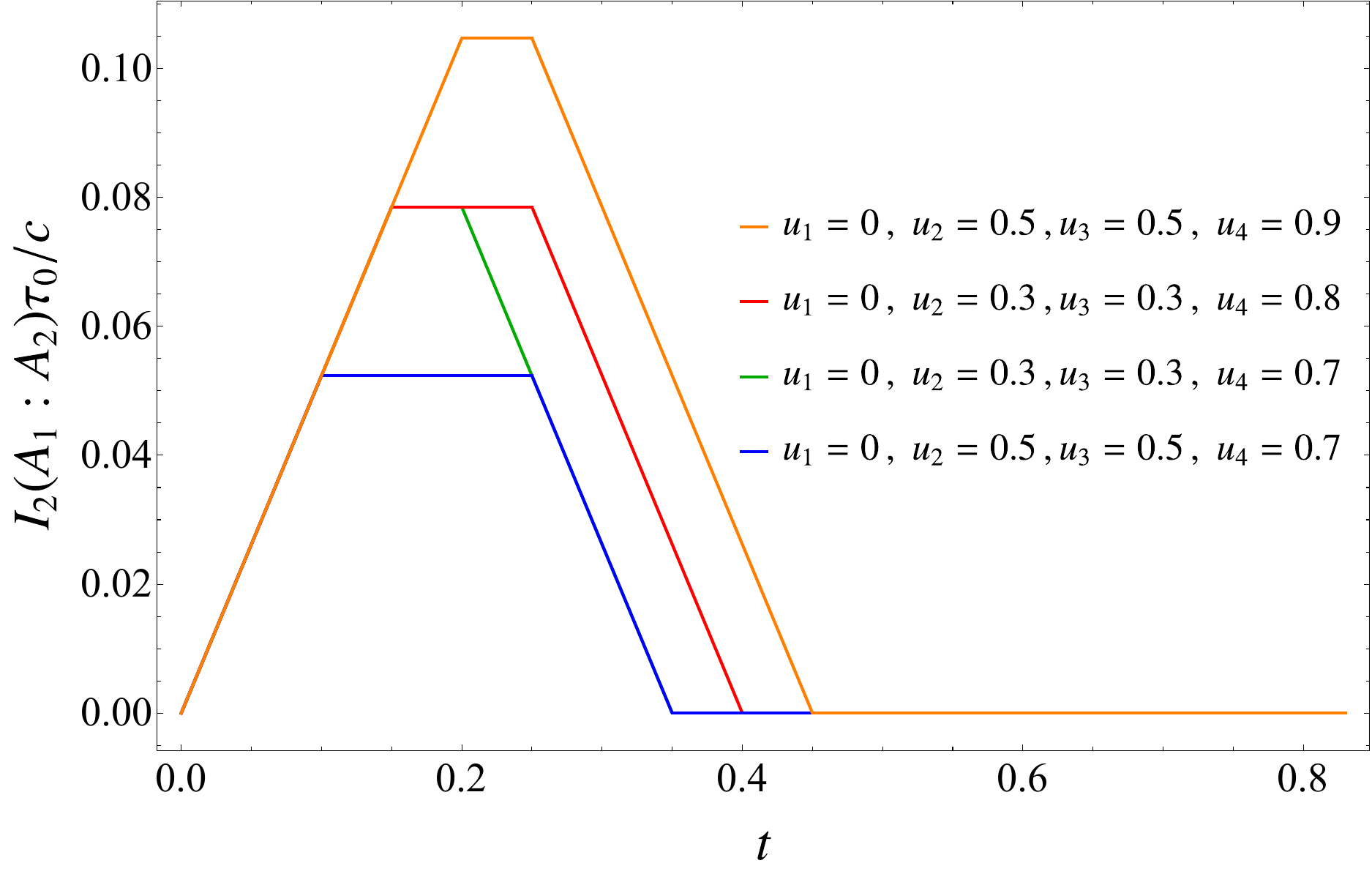}
\includegraphics[width=0.5\textwidth]{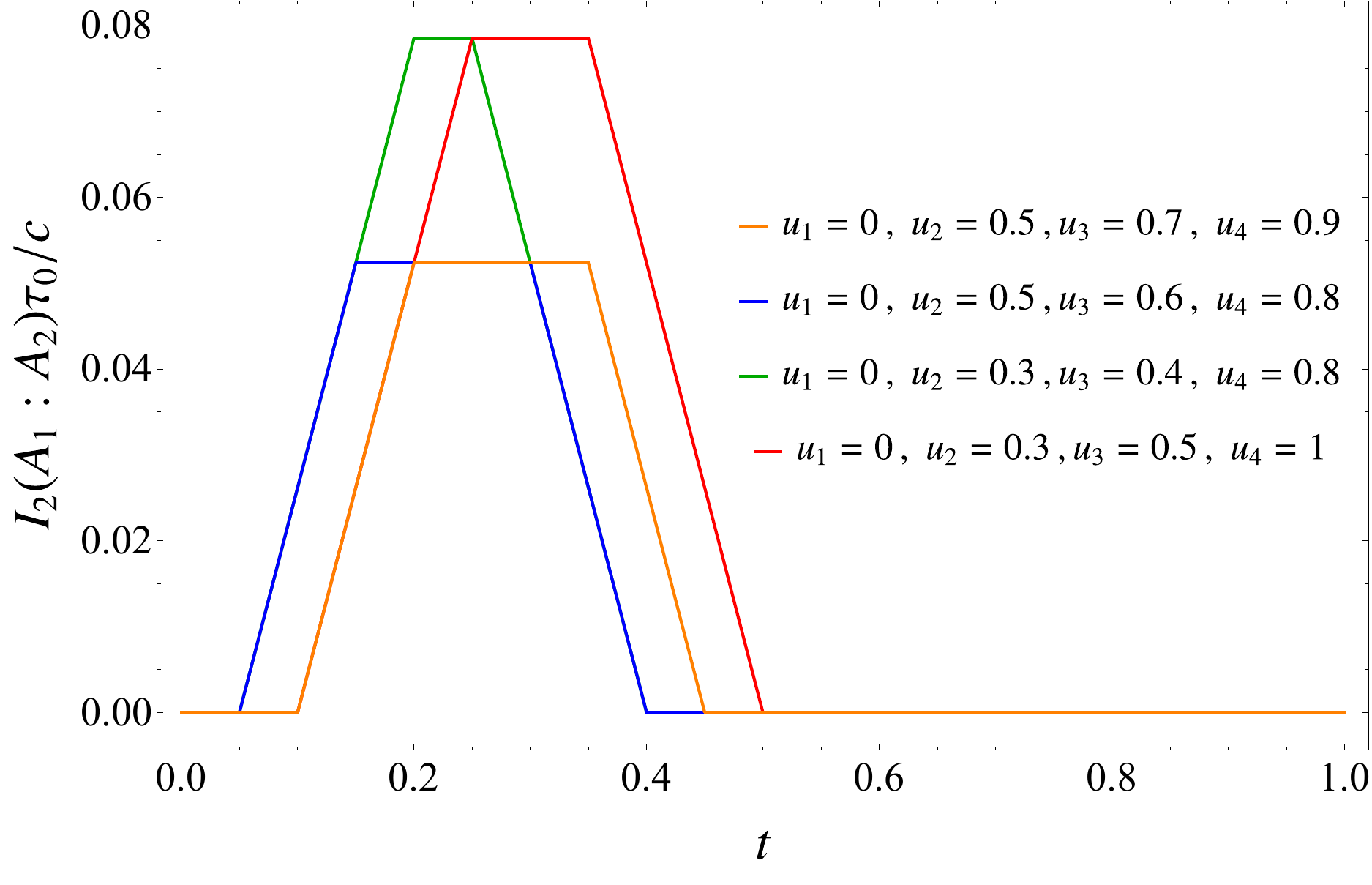}
\vspace{-.7cm}
\caption{Mutual information \eqref{eq:I2CFT} as a function of time for $A_1$ and $A_2$ disjoint intervals with endpoints $u_1$, $u_2$ and $u_3$, $u_4$, respectively. In the left panel, we consider cases where the two intervals are adjacent, while in the right panel we assume a finite separation between them.}
\label{fig:I2CFT}
\end{figure}

Consider an initial state of the form \eqref{eq:CFT quench initial state}.
The initial state is evolved via a CFT Hamiltonian and we want to study the evolution of its entanglement. 
We consider the general bipartition into $ A=\bigcup_{j=1}^N A_j$, made by $N$ disjoint intervals $A_j=[u_{2j-1},u_{2j}]$, and its complement, and we explore the regime where
$t\gg \tau_0$ and $\vert u_{i}-u_{j}\vert\gg \tau_0$. The results for the Rényi entropies  in this limit can be written for a general number of intervals \cite{Coser:2014gsa}. Let us focus on the case we are interested in, namely $n=1$ (entanglement entropy), $N=2$ and $N=3$, where we have (up to a negligible amount of entanglement in the initial state) 
\begin{eqnarray}
\label{eq:CFT_EE_2int}
\nonumber
   S_{A_1\cup A_2}&=&  
   \frac{c\pi}{6\tau_0}
\Big[
2t
+ q(t,|u_1-u_2|)
+ q(t,|u_1-u_4|)
+ q(t,|u_3-u_2|)
+ q(t,|u_3-u_4|)
\\
&-&q(t,|u_2-u_4|)-q(t,|u_1-u_3|)
\Big]\,,
\end{eqnarray}
\begin{eqnarray}
\label{eq:CFT_EE_3int}
   S_{A_1\cup A_2\cup A_3}&=&\frac{c\pi}{6\tau_0}
\Big[
3t
+ q(t,|u_1-u_2|)+
q(t,|u_1-u_4|)+
q(t,|u_1-u_6|)+q(t,|u_3-u_2|)
\nonumber
\\
&+&
q(t,|u_3-u_4|)
\nonumber
+
q(t,|u_3-u_6|)+
q(t,|u_5-u_2|)+
q(t,|u_5-u_4|)
\nonumber
\\
&+&
q(t,|u_5-u_6|)
-
q(t,|u_2-u_4|)-q(t,|u_2-u_6|)-q(t,|u_4-u_6|)
\nonumber
\\
&-&
q(t,|u_1-u_3|)-q(t,|u_3-u_5|)
-
q(t,|u_1-u_5|)
\Big]\,, 
\end{eqnarray}
with
\begin{equation}
 q(t,\ell) = \frac{\ell}{2} - \max\!\left(t, \frac{\ell}{2}  \right)=\begin{cases}
0& t < \frac{\ell}{2}, \\
\frac{\ell}{2}-t & t > \frac{\ell}{2}.
\end{cases}
\end{equation}
Using \eqref{eq:CFT_EE_2int} and \eqref{eq:CFT_EE_3int}, we obtain the CFT predictions of $I_2$ and $I_3$. While $I_2$ is non-trivial, namely \cite{Coser:2014gsa}
\begin{equation}
\label{eq:I2CFT}
  I_2(A_1:A_2) 
  = \frac{\pi c }{6 \tau_0} \left[
q(t,\vert u_3 - u_1\vert) + q(t, \vert u_4 - u_2\vert)
- q(t,\vert u_4 - u_1\vert) - q(t, \vert u_3 - u_2\vert)
\right]\,,
\end{equation}
we find that $I_3(A_1:A_2:A_3)$ is identically vanishing at any time, i.e. $I_3=0$. In Fig.\,\ref{fig:I2CFT}, we show the typical behavior of $I_2$ for some choices of the bipartition. We can summarize this behaviour by saying that $I_2$ is initially constant to zero until $t=\vert u_2-u_3\vert/2$, then it grows linearly until $t=(\vert u_2-u_3\vert+\min(\vert u_1-u_2\vert,\vert u_3-u_4\vert))/2 $ and then it becomes constant again until $t=(\vert u_2-u_3\vert/2+\max(\vert u_1-u_2\vert,\vert u_3-u_4\vert))/2  $. In the last part of its evolution, it decreases linearly before returning to zero at $t=(\vert u_2-u_3\vert+\vert u_1-u_2\vert+\vert u_3-u_4\vert)/2 $.

While, as already discussed in the main text, the early-time behavior in the top-right panel of Fig.\,\ref{fig:quandripartition_time_DBC} can be understood from the quasi-particle perspective through $I_3=0$, the behavior shown in the left panel of Fig.\,\ref{fig:I2CFT} is qualitatively similar to the early-time dynamics observed in the top-right panel of Fig.\,\ref{fig:tripartition_time_DBC_nonemptyB}, at least until boundary effects modify the evolution predicted by the quasi-particle picture in infinite systems.

\bibliographystyle{nb}
\bibliography{Refs1}

@article{Parker:2018yvk,
    author = "Parker, Daniel E. and Cao, Xiangyu and Avdoshkin, Alexander and Scaffidi, Thomas and Altman, Ehud",
    title = "{A Universal Operator Growth Hypothesis}",
    eprint = "1812.08657",
    archivePrefix = "arXiv",
    primaryClass = "cond-mat.stat-mech",
    doi = "10.1103/PhysRevX.9.041017",
    journal = "Phys. Rev. X",
    volume = "9",
    number = "4",
    pages = "041017",
    year = "2019"
}

@article{Chen:2020nlj,
    author = "Chen, Bowen and Czech, Bartlomiej and Wang, Zi-zhi",
    title = "{Query complexity and cutoff dependence of the CFT2 ground state}",
    eprint = "2004.11377",
    archivePrefix = "arXiv",
    primaryClass = "hep-th",
    doi = "10.1103/PhysRevD.103.026015",
    journal = "Phys. Rev. D",
    volume = "103",
    number = "2",
    pages = "026015",
    year = "2021"
}

@article{Bao:2024azn,
    author = "Bao, Ning and Furuya, Keiichiro and Naskar, Joydeep",
    title = "{Towards a complete classification of holographic entropy inequalities}",
    eprint = "2409.17317",
    archivePrefix = "arXiv",
    primaryClass = "hep-th",
    doi = "10.1007/JHEP03(2025)117",
    journal = "JHEP",
    volume = "03",
    pages = "117",
    year = "2025"
}

@article{Czech:2023xed,
    author = "Czech, Bartlomiej and Shuai, Sirui and Wang, Yixu and Zhang, Daiming",
    title = "{Holographic entropy inequalities and the topology of entanglement wedge nesting}",
    eprint = "2309.15145",
    archivePrefix = "arXiv",
    primaryClass = "hep-th",
    doi = "10.1103/PhysRevD.109.L101903",
    journal = "Phys. Rev. D",
    volume = "109",
    number = "10",
    pages = "L101903",
    year = "2024"
}

@article{HernandezCuenca:2019wgh,
    author = "Hern{\'a}ndez Cuenca, Sergio",
    title = "{Holographic entropy cone for five regions}",
    eprint = "1903.09148",
    archivePrefix = "arXiv",
    primaryClass = "hep-th",
    doi = "10.1103/PhysRevD.100.026004",
    journal = "Phys. Rev. D",
    volume = "100",
    number = "2",
    pages = "026004",
    year = "2019"
}

@article{Hubeny:2018trv,
    author = "Hubeny, Veronika E. and Rangamani, Mukund and Rota, Massimiliano",
    title = "{Holographic entropy relations}",
    eprint = "1808.07871",
    archivePrefix = "arXiv",
    primaryClass = "hep-th",
    doi = "10.1002/prop.201800067",
    journal = "Fortsch. Phys.",
    volume = "66",
    number = "11-12",
    pages = "1800067",
    year = "2018"
}

@article{He:2020xuo,
    author = "He, Temple and Hubeny, Veronika E. and Rangamani, Mukund",
    title = "{Superbalance of Holographic Entropy Inequalities}",
    eprint = "2002.04558",
    archivePrefix = "arXiv",
    primaryClass = "hep-th",
    doi = "10.1007/JHEP07(2020)245",
    journal = "JHEP",
    volume = "07",
    pages = "245",
    year = "2020"
}

@article{Hubeny:2018ijt,
    author = "Hubeny, Veronika E. and Rangamani, Mukund and Rota, Massimiliano",
    title = "{The holographic entropy arrangement}",
    eprint = "1812.08133",
    archivePrefix = "arXiv",
    primaryClass = "hep-th",
    doi = "10.1002/prop.201900011",
    journal = "Fortsch. Phys.",
    volume = "67",
    number = "4",
    pages = "1900011",
    year = "2019"
}

@article{Grado-White:2024gtx,
    author = "Grado-White, Brianna and Grimaldi, Guglielmo and Headrick, Matthew and Hubeny, Veronika E.",
    title = "{Testing holographic entropy inequalities in 2 + 1 dimensions}",
    eprint = "2407.07165",
    archivePrefix = "arXiv",
    primaryClass = "hep-th",
    reportNumber = "BRX-TH-6721",
    doi = "10.1007/JHEP01(2025)065",
    journal = "JHEP",
    volume = "01",
    pages = "065",
    year = "2025"
}

@article{Grimaldi:2025jad,
    author = "Grimaldi, Guglielmo and Headrick, Matthew and Hubeny, Veronika E.",
    title = "{A new characterization of the holographic entropy cone}",
    eprint = "2508.21823",
    archivePrefix = "arXiv",
    primaryClass = "hep-th",
    doi = "10.21468/SciPostPhys.20.4.122",
    journal = "SciPost Phys.",
    volume = "20",
    number = "4",
    pages = "122",
    year = "2026"
}

@article{Czech:2025jnw,
    author = "Czech, Bartlomiej and Shuai, Sirui and Wang, Yixu",
    title = "{Entropy Inequalities Constrain Holographic Erasure Correction}",
    eprint = "2502.12246",
    archivePrefix = "arXiv",
    primaryClass = "hep-th",
    doi = "10.1103/dl3c-h3hg",
    journal = "Phys. Rev. Lett.",
    volume = "135",
    number = "14",
    pages = "141603",
    year = "2025"
}

@article{Hayden:2021gno,
    author = "Hayden, Patrick and Parrikar, Onkar and Sorce, Jonathan",
    title = "{The Markov gap for geometric reflected entropy}",
    eprint = "2107.00009",
    archivePrefix = "arXiv",
    primaryClass = "hep-th",
    doi = "10.1007/JHEP10(2021)047",
    journal = "JHEP",
    volume = "10",
    pages = "047",
    year = "2021"
}

@article{Abt:2017pmf,
    author = "Abt, Raimond and Erdmenger, Johanna and Hinrichsen, Haye and Melby-Thompson, Charles M. and Meyer, Ren{\'e} and Northe, Christian and Reyes, Ignacio A.",
    title = "{Topological Complexity in AdS$_3$/CFT$_2$}",
    eprint = "1710.01327",
    archivePrefix = "arXiv",
    primaryClass = "hep-th",
    doi = "10.1002/prop.201800034",
    journal = "Fortsch. Phys.",
    volume = "66",
    number = "6",
    pages = "1800034",
    year = "2018"
}

@article{Boruch:2021hqs,
    author = "Boruch, Jan and Caputa, Pawel and Ge, Dongsheng and Takayanagi, Tadashi",
    title = "{Holographic path-integral optimization}",
    reportNumber = "YITP-21-25, IPMU21-0022",
    doi = "10.1007/JHEP07(2021)016",
    journal = "JHEP",
    volume = "07",
    pages = "016",
    year = "2021",
    note = "[Erratum: JHEP 09, 111 (2022)]"
}

@article{Dong:2016eik,
    author = "Dong, Xi and Harlow, Daniel and Wall, Aron C.",
    title = "{Reconstruction of Bulk Operators within the Entanglement Wedge in Gauge-Gravity Duality}",
    eprint = "1601.05416",
    archivePrefix = "arXiv",
    primaryClass = "hep-th",
    reportNumber = "NSF-KITP-16-005, NSF-KITP-16-005",
    doi = "10.1103/PhysRevLett.117.021601",
    journal = "Phys. Rev. Lett.",
    volume = "117",
    number = "2",
    pages = "021601",
    year = "2016"
}

@article{Concepcion:2026fhv,
    author = "Concepcion, Violet and Ritchie, Kyle",
    title = "{A Timelike Quantum Focusing Conjecture}",
    eprint = "2604.27054",
    archivePrefix = "arXiv",
    primaryClass = "hep-th",
    month = "4",
    year = "2026"
}

@article{Hashimoto:2026kjy,
    author = "Hashimoto, Koji and Tanahashi, Norihiro",
    title = "{Holography and Optimal Transport: Emergent Wasserstein Spacetime in Harmonic Oscillator, SYK and Krylov Complexity}",
    eprint = "2604.17649",
    archivePrefix = "arXiv",
    primaryClass = "hep-th",
    reportNumber = "KUNS-3098",
    month = "4",
    year = "2026"
}

@article{Alishahiha:2015rta,
    author = "Alishahiha, Mohsen",
    title = "{Holographic Complexity}",
    eprint = "1509.06614",
    archivePrefix = "arXiv",
    primaryClass = "hep-th",
    doi = "10.1103/PhysRevD.92.126009",
    journal = "Phys. Rev. D",
    volume = "92",
    number = "12",
    pages = "126009",
    year = "2015"
}

@article{Maric:2022rsc,
    author = "Mari{\'c}, Vanja and Fagotti, Maurizio",
    title = "{Universality in the tripartite information after global quenches}",
    eprint = "2209.14253",
    archivePrefix = "arXiv",
    primaryClass = "cond-mat.stat-mech",
    doi = "10.1103/PhysRevB.108.L161116",
    journal = "Phys. Rev. B",
    volume = "108",
    number = "16",
    pages = "L161116",
    year = "2023"
}

@article{Agon:2021lus,
    author = "Ag{\'o}n, C{\'e}sar A. and Bueno, Pablo and Casini, Horacio",
    title = "{Tripartite information at long distances}",
    eprint = "2109.09179",
    archivePrefix = "arXiv",
    primaryClass = "hep-th",
    doi = "10.21468/SciPostPhys.12.5.153",
    journal = "SciPost Phys.",
    volume = "12",
    number = "5",
    pages = "153",
    year = "2022"
}

@article{Casini:2008wt,
    author = "Casini, H. and Huerta, M.",
    title = "{Remarks on the entanglement entropy for disconnected regions}",
    eprint = "0812.1773",
    archivePrefix = "arXiv",
    primaryClass = "hep-th",
    doi = "10.1088/1126-6708/2009/03/048",
    journal = "JHEP",
    volume = "03",
    pages = "048",
    year = "2009"
}

@article{Maric:2020dpw,
    author = "Mari{\'c}, Vanja and Fagotti, Maurizio",
    title = "{Universality in the tripartite information after global quenches: (generalised) quantum XY models}",
    eprint = "2302.01322",
    archivePrefix = "arXiv",
    primaryClass = "cond-mat.stat-mech",
    doi = "10.1007/JHEP06(2023)140",
    journal = "JHEP",
    volume = "23",
    pages = "140",
    year = "2020"
}

@article{Caceffo:2023hns,
    author = "Caceffo, Fabio and Alba, Vincenzo",
    title = "{Negative tripartite mutual information after quantum quenches in integrable systems}",
    eprint = "2305.10245",
    archivePrefix = "arXiv",
    primaryClass = "cond-mat.stat-mech",
    doi = "10.1103/PhysRevB.108.134434",
    journal = "Phys. Rev. B",
    volume = "108",
    number = "13",
    pages = "134434",
    year = "2023"
}

@article{Asadi:2018ijf,
    author = "Asadi, M. and Ali-Akbari, M.",
    title = "{Holographic Mutual and Tripartite Information in a Symmetry Breaking Quench}",
    eprint = "1804.05604",
    archivePrefix = "arXiv",
    primaryClass = "hep-th",
    reportNumber = "IPM/P-2018-015",
    doi = "10.1016/j.physletb.2018.09.008",
    journal = "Phys. Lett. B",
    volume = "785",
    pages = "409--418",
    year = "2018"
}

@article{Allais:2011ys,
    author = "Allais, Andrea and Tonni, Erik",
    title = "{Holographic evolution of the mutual information}",
    eprint = "1110.1607",
    archivePrefix = "arXiv",
    primaryClass = "hep-th",
    doi = "10.1007/JHEP01(2012)102",
    journal = "JHEP",
    volume = "01",
    pages = "102",
    year = "2012"
}

@article{Balasubramanian:2011at,
    author = "Balasubramanian, V. and Bernamonti, A. and Copland, N. and Craps, B. and Galli, F.",
    title = "{Thermalization of mutual and tripartite information in strongly coupled two dimensional conformal field theories}",
    eprint = "1110.0488",
    archivePrefix = "arXiv",
    primaryClass = "hep-th",
    doi = "10.1103/PhysRevD.84.105017",
    journal = "Phys. Rev. D",
    volume = "84",
    pages = "105017",
    year = "2011"
}

@article{Agon:2018zso,
    author = "Ag{\'o}n, Cesar A. and Headrick, Matthew and Swingle, Brian",
    title = "{Subsystem Complexity and Holography}",
    eprint = "1804.01561",
    archivePrefix = "arXiv",
    primaryClass = "hep-th",
    doi = "10.1007/JHEP02(2019)145",
    journal = "JHEP",
    volume = "02",
    pages = "145",
    year = "2019"
}

@article{Hayden:2011ag,
    author = "Hayden, Patrick and Headrick, Matthew and Maloney, Alexander",
    title = "{Holographic Mutual Information is Monogamous}",
    eprint = "1107.2940",
    archivePrefix = "arXiv",
    primaryClass = "hep-th",
    reportNumber = "BRX-TH-638, BRX-TH-638",
    doi = "10.1103/PhysRevD.87.046003",
    journal = "Phys. Rev. D",
    volume = "87",
    number = "4",
    pages = "046003",
    year = "2013"
}

@article{Headrick:2007km,
    author = "Headrick, Matthew and Takayanagi, Tadashi",
    title = "{A Holographic proof of the strong subadditivity of entanglement entropy}",
    eprint = "0704.3719",
    archivePrefix = "arXiv",
    primaryClass = "hep-th",
    reportNumber = "SU-ITP-07-08, KUNS-2069, SU-ITP-07/08, KUNS-2069",
    doi = "10.1103/PhysRevD.76.106013",
    journal = "Phys. Rev. D",
    volume = "76",
    pages = "106013",
    year = "2007"
}

@article{Chapman:2018hou,
    author = "Chapman, Shira and Eisert, Jens and Hackl, Lucas and Heller, Michal P. and Jefferson, Ro and Marrochio, Hugo and Myers, Robert C.",
    title = "{Complexity and entanglement for thermofield double states}",
    eprint = "1810.05151",
    archivePrefix = "arXiv",
    primaryClass = "hep-th",
    doi = "10.21468/SciPostPhys.6.3.034",
    journal = "SciPost Phys.",
    volume = "6",
    number = "3",
    pages = "034",
    year = "2019"
}

@article{Alishahiha:2018lfv,
    author = "Alishahiha, Mohsen and Babaei Velni, Komeil and Mohammadi Mozaffar, M. Reza",
    title = "{Black hole subregion action and complexity}",
    eprint = "1809.06031",
    archivePrefix = "arXiv",
    primaryClass = "hep-th",
    reportNumber = "IPM-P-2018-069",
    doi = "10.1103/PhysRevD.99.126016",
    journal = "Phys. Rev. D",
    volume = "99",
    number = "12",
    pages = "126016",
    year = "2019"
}

@article{Caceres:2019pgf,
    author = "Caceres, Elena and Chapman, Shira and Couch, Josiah D. and Hernandez, Juan P. and Myers, Robert C. and Ruan, Shan-Ming",
    title = "{Complexity of Mixed States in QFT and Holography}",
    eprint = "1909.10557",
    archivePrefix = "arXiv",
    primaryClass = "hep-th",
    doi = "10.1007/JHEP03(2020)012",
    journal = "JHEP",
    volume = "03",
    pages = "012",
    year = "2020"
}

@article{DiGiulio:2020hlz,
    author = "Di Giulio, Giuseppe and Tonni, Erik",
    title = "{Complexity of mixed Gaussian states from Fisher information geometry}",
    eprint = "2006.00921",
    archivePrefix = "arXiv",
    primaryClass = "hep-th",
    doi = "10.1007/JHEP12(2020)101",
    journal = "JHEP",
    volume = "12",
    pages = "101",
    year = "2020"
}

@article{Fan:2025moc,
    author = "Fan, Yale and Hunter-Jones, Nicholas and Karch, Andreas and Mittal, Shivan",
    title = "{Sharp Transitions for Subsystem Complexity}",
    eprint = "2510.18832",
    archivePrefix = "arXiv",
    primaryClass = "hep-th",
    month = "10",
    year = "2025"
}

@article{Haah:2025hyf,
    author = "Haah, Jeongwan and Stanford, Douglas",
    title = "{Growth and collapse of subsystem complexity under random unitary circuits}",
    eprint = "2510.18805",
    archivePrefix = "arXiv",
    primaryClass = "quant-ph",
    month = "10",
    year = "2025"
}

@article{DiGiulio:2021oal,
    author = "Di Giulio, Giuseppe and Tonni, Erik",
    title = "{Subsystem complexity after a global quantum quench}",
    eprint = "2102.02764",
    archivePrefix = "arXiv",
    primaryClass = "hep-th",
    doi = "10.1007/JHEP05(2021)022",
    journal = "JHEP",
    volume = "05",
    pages = "022",
    year = "2021"
}

@book{Breuer:2007juk,
    author = "Breuer, Heinz-Peter and Petruccione, Francesco",
    title = "{The Theory of Open Quantum Systems}",
    doi = "10.1093/acprof:oso/9780199213900.001.0001",
    isbn = "978-0-19-170634-9, 978-0-19-921390-0",
    publisher = "Oxford University Press",
    month = "1",
    year = "2007"
}

@book{Bhatiabook,
    author = "Bhatia, R",
    title = "{Positive Definite Matrices}",
    isbn = "978-0691168258",
    publisher = "Princeton University Press",
    year = "2007"
}

@book{Agarwalbook,
    author = "Agarwal, G. S.",
    title = "{Quantum Optics}",
    doi = "doi.org/10.1017/CBO9781139035170",
    isbn = "9781107006409",
    publisher = "Cambridge University Press",
    year = "2012"
}

@article{Caputa:2024vrn,
    author = "Caputa, Pawel and Jeong, Hyun-Sik and Liu, Sinong and Pedraza, Juan F. and Qu, Le-Chen",
    title = "{Krylov complexity of density matrix operators}",
    eprint = "2402.09522",
    archivePrefix = "arXiv",
    primaryClass = "hep-th",
    reportNumber = "YITP-24-21, IFT-UAM/CSIC-24-25",
    doi = "10.1007/JHEP05(2024)337",
    journal = "JHEP",
    volume = "05",
    pages = "337",
    year = "2024"
}

@article{Baiguera:2025dkc,
    author = "Baiguera, Stefano and Balasubramanian, Vijay and Caputa, Pawel and Chapman, Shira and Haferkamp, Jonas and Heller, Michal P. and Halpern, Nicole Yunger",
    title = "{Quantum complexity in gravity, quantum field theory, and quantum information science}",
    eprint = "2503.10753",
    archivePrefix = "arXiv",
    primaryClass = "hep-th",
    reportNumber = "YITP-25-39",
    doi = "10.1016/j.physrep.2025.11.001",
    journal = "Phys. Rept.",
    volume = "1159",
    pages = "1--77",
    year = "2026"
}

@article{Rabinovici:2023yex,
    author = "Rabinovici, E. and S{\'a}nchez-Garrido, A. and Shir, R. and Sonner, J.",
    title = "{A bulk manifestation of Krylov complexity}",
    eprint = "2305.04355",
    archivePrefix = "arXiv",
    primaryClass = "hep-th",
    doi = "10.1007/JHEP08(2023)213",
    journal = "JHEP",
    volume = "08",
    pages = "213",
    year = "2023"
}

@article{Maldacena:1997re,
    author = "Maldacena, Juan Martin",
    title = "{The Large $N$ limit of superconformal field theories and supergravity}",
    eprint = "hep-th/9711200",
    archivePrefix = "arXiv",
    reportNumber = "HUTP-97-A097, HUTP-98-A097",
    doi = "10.4310/ATMP.1998.v2.n2.a1",
    journal = "Adv. Theor. Math. Phys.",
    volume = "2",
    pages = "231--252",
    year = "1998"
}

@article{Susskind:2014moa,
    author = "Susskind, Leonard",
    title = "{Entanglement is not enough}",
    eprint = "1411.0690",
    archivePrefix = "arXiv",
    primaryClass = "hep-th",
    doi = "10.1002/prop.201500095",
    journal = "Fortsch. Phys.",
    volume = "64",
    pages = "49--71",
    year = "2016"
}

@article{Rabinovici:2025otw,
    author = "Rabinovici, Eliezer and S{\'a}nchez-Garrido, Adri{\'a}n and Shir, Ruth and Sonner, Julian",
    title = "{Krylov Complexity}",
    eprint = "2507.06286",
    archivePrefix = "arXiv",
    primaryClass = "hep-th",
    reportNumber = "CERN-TH-2025-128",
    month = "7",
    year = "2025"
}

@article{Hartman:2013qma,
    author = "Hartman, Thomas and Maldacena, Juan",
    title = "{Time Evolution of Entanglement Entropy from Black Hole Interiors}",
    eprint = "1303.1080",
    archivePrefix = "arXiv",
    primaryClass = "hep-th",
    doi = "10.1007/JHEP05(2013)014",
    journal = "JHEP",
    volume = "05",
    pages = "014",
    year = "2013"
}

@article{Calabrese:2005in,
    author = "Calabrese, Pasquale and Cardy, John L.",
    title = "{Evolution of entanglement entropy in one-dimensional systems}",
    eprint = "cond-mat/0503393",
    archivePrefix = "arXiv",
    doi = "10.1088/1742-5468/2005/04/P04010",
    journal = "J. Stat. Mech.",
    volume = "0504",
    pages = "P04010",
    year = "2005"
}

@article{Coser:2014gsa,
    author = "Coser, Andrea and Tonni, Erik and Calabrese, Pasquale",
    title = "{Entanglement negativity after a global quantum quench}",
    eprint = "1410.0900",
    archivePrefix = "arXiv",
    primaryClass = "cond-mat.stat-mech",
    doi = "10.1088/1742-5468/2014/12/P12017",
    journal = "J. Stat. Mech.",
    volume = "1412",
    number = "12",
    pages = "P12017",
    year = "2014"
}

@article{Rabinovici:2020ryf,
    author = "Rabinovici, E. and S{\'a}nchez-Garrido, A. and Shir, R. and Sonner, J.",
    title = "{Operator complexity: a journey to the edge of Krylov space}",
    eprint = "2009.01862",
    archivePrefix = "arXiv",
    primaryClass = "hep-th",
    doi = "10.1007/JHEP06(2021)062",
    journal = "JHEP",
    volume = "06",
    pages = "062",
    year = "2021"
}

@article{Auzzi:2019vyh,
    author = "Auzzi, Roberto and Baiguera, Stefano and Legramandi, Andrea and Nardelli, Giuseppe and Roy, Pratim and Zenoni, Nicol{\`o}",
    title = "{On subregion action complexity in AdS$_{3}$ and in the BTZ black hole}",
    eprint = "1910.00526",
    archivePrefix = "arXiv",
    primaryClass = "hep-th",
    doi = "10.1007/JHEP01(2020)066",
    journal = "JHEP",
    volume = "01",
    pages = "066",
    year = "2020"
}

@article{Balasubramanian:2022tpr,
    author = "Balasubramanian, Vijay and Caputa, Pawel and Magan, Javier M. and Wu, Qingyue",
    title = "{Quantum chaos and the complexity of spread of states}",
    eprint = "2202.06957",
    archivePrefix = "arXiv",
    primaryClass = "hep-th",
    doi = "10.1103/PhysRevD.106.046007",
    journal = "Phys. Rev. D",
    volume = "106",
    number = "4",
    pages = "046007",
    year = "2022"
}

@article{Susskind:2014jwa,
    author = "Susskind, Leonard and Zhao, Ying",
    title = "{Switchbacks and the Bridge to Nowhere}",
    eprint = "1408.2823",
    archivePrefix = "arXiv",
    primaryClass = "hep-th",
    month = "8",
    year = "2014"
}

@article{Czech:2012bh,
    author = "Czech, Bartlomiej and Karczmarek, Joanna L. and Nogueira, Fernando and Van Raamsdonk, Mark",
    title = "{The Gravity Dual of a Density Matrix}",
    eprint = "1204.1330",
    archivePrefix = "arXiv",
    primaryClass = "hep-th",
    doi = "10.1088/0264-9381/29/15/155009",
    journal = "Class. Quant. Grav.",
    volume = "29",
    pages = "155009",
    year = "2012"
}

@article{Balasubramanian:2025hxg,
    author = "Balasubramanian, Vijay and Kang, Monica Jinwoo and Cummings, Charlie and Murdia, Chitraang and Ross, Simon F.",
    title = "{Purely Greenberger-Horne-Zeilinger{\textendash}like Entanglement is Forbidden in Holography}",
    eprint = "2509.03621",
    archivePrefix = "arXiv",
    primaryClass = "hep-th",
    doi = "10.1103/g5rw-nvnr",
    journal = "Phys. Rev. Lett.",
    volume = "136",
    number = "3",
    pages = "031602",
    year = "2026"
}

@article{Pedraza:2021fgp,
    author = "Pedraza, Juan F. and Russo, Andrea and Svesko, Andrew and Weller-Davies, Zachary",
    title = "{Sewing spacetime with Lorentzian threads: complexity and the emergence of time in quantum gravity}",
    eprint = "2106.12585",
    archivePrefix = "arXiv",
    primaryClass = "hep-th",
    reportNumber = "BRX-TH-6689",
    doi = "10.1007/JHEP02(2022)093",
    journal = "JHEP",
    volume = "02",
    pages = "093",
    year = "2022"
}

@article{Pedraza:2021mkh,
    author = "Pedraza, Juan F. and Russo, Andrea and Svesko, Andrew and Weller-Davies, Zachary",
    title = "{Lorentzian Threads as Gatelines and Holographic Complexity}",
    eprint = "2105.12735",
    archivePrefix = "arXiv",
    primaryClass = "hep-th",
    reportNumber = "BRX-TH-6683",
    doi = "10.1103/PhysRevLett.127.271602",
    journal = "Phys. Rev. Lett.",
    volume = "127",
    number = "27",
    pages = "271602",
    year = "2021"
}

@article{Mezei:2018jco,
    author = "Mezei, M{\'a}rk",
    title = "{Membrane theory of entanglement dynamics from holography}",
    eprint = "1803.10244",
    archivePrefix = "arXiv",
    primaryClass = "hep-th",
    doi = "10.1103/PhysRevD.98.106025",
    journal = "Phys. Rev. D",
    volume = "98",
    number = "10",
    pages = "106025",
    year = "2018"
}

@article{Jonay:2018yei,
    author = "Jonay, Cheryne and Huse, David A. and Nahum, Adam",
    title = "{Coarse-grained dynamics of operator and state entanglement}",
    eprint = "1803.00089",
    archivePrefix = "arXiv",
    primaryClass = "cond-mat.stat-mech",
    month = "2",
    year = "2018"
}

@article{Asplund:2015eha,
    author = "Asplund, Curtis T. and Bernamonti, Alice and Galli, Federico and Hartman, Thomas",
    title = "{Entanglement Scrambling in 2d Conformal Field Theory}",
    eprint = "1506.03772",
    archivePrefix = "arXiv",
    primaryClass = "hep-th",
    doi = "10.1007/JHEP09(2015)110",
    journal = "JHEP",
    volume = "09",
    pages = "110",
    year = "2015"
}

@article{Belin:2021bga,
    author = "Belin, Alexandre and Myers, Robert C. and Ruan, Shan-Ming and S{\'a}rosi, G{\'a}bor and Speranza, Antony J.",
    title = "{Does Complexity Equal Anything?}",
    eprint = "2111.02429",
    archivePrefix = "arXiv",
    primaryClass = "hep-th",
    reportNumber = "CERN-TH-2021-181, YITP-22-02",
    doi = "10.1103/PhysRevLett.128.081602",
    journal = "Phys. Rev. Lett.",
    volume = "128",
    number = "8",
    pages = "081602",
    year = "2022"
}

@article{Jian:2023mdh,
    author = "Jian, Shao-Kai and Zhang, Yuzhen",
    title = "{Subsystem complexity and measurements in holography}",
    eprint = "2312.04437",
    archivePrefix = "arXiv",
    primaryClass = "hep-th",
    doi = "10.1007/JHEP05(2024)241",
    journal = "JHEP",
    volume = "05",
    pages = "241",
    year = "2024"
}

@article{Bhattacharyya:2020iic,
    author = "Bhattacharyya, Arpan and Haque, S. Shajidul and Kim, Eugene H.",
    title = "{Complexity from the reduced density matrix: a new diagnostic for chaos}",
    eprint = "2011.04705",
    archivePrefix = "arXiv",
    primaryClass = "hep-th",
    doi = "10.1007/JHEP10(2021)028",
    journal = "JHEP",
    volume = "10",
    pages = "028",
    year = "2021"
}

@article{Balasubramanian:2026chr,
    author = "Balasubramanian, Vijay and Chan, William K. L. and Kang, Monica Jinwoo and Murdia, Chitraang and Ross, Simon F.",
    title = "{Constraints on four-party entanglement in holography}",
    eprint = "2606.00210",
    archivePrefix = "arXiv",
    primaryClass = "hep-th",
    month = "5",
    year = "2026"
}

@article{Patramanis:2021lkx,
    author = "Patramanis, Dimitrios",
    title = "{Probing the entanglement of operator growth}",
    eprint = "2111.03424",
    archivePrefix = "arXiv",
    primaryClass = "hep-th",
    doi = "10.1093/ptep/ptac081",
    journal = "PTEP",
    volume = "2022",
    number = "6",
    pages = "063A01",
    year = "2022"
}

@article{Caputa:2021sib,
    author = "Caputa, Pawel and Magan, Javier M. and Patramanis, Dimitrios",
    title = "{Geometry of Krylov complexity}",
    eprint = "2109.03824",
    archivePrefix = "arXiv",
    primaryClass = "hep-th",
    doi = "10.1103/PhysRevResearch.4.013041",
    journal = "Phys. Rev. Res.",
    volume = "4",
    number = "1",
    pages = "013041",
    year = "2022"
}

@article{Das:2024zuu,
    author = "Das, Rathindra Nath and Mori, Takato",
    title = "{Krylov Complexity of Purification}",
    eprint = "2408.00826",
    archivePrefix = "arXiv",
    primaryClass = "hep-th",
    reportNumber = "YITP-24-94",
    doi = "10.1103/qgcx-wxpd",
    journal = "Phys. Rev. Lett.",
    volume = "136",
    number = "3",
    pages = "030201",
    year = "2026"
}

@article{Caputa:2025ozd,
    author = "Caputa, Pawel and Di Giulio, Giuseppe and Loc, Tran Quang",
    title = "{Symmetry-resolved spread complexity}",
    eprint = "2509.12992",
    archivePrefix = "arXiv",
    primaryClass = "hep-th",
    reportNumber = "YITP-25-146",
    doi = "10.1007/JHEP02(2026)189",
    journal = "JHEP",
    volume = "02",
    pages = "189",
    year = "2026"
}

@article{Caputa:2025mii,
    author = "Caputa, Pawel and Di Giulio, Giuseppe and Loc, Tran Quang",
    title = "{Growth of block-diagonal operators and symmetry-resolved Krylov complexity}",
    eprint = "2507.02033",
    archivePrefix = "arXiv",
    primaryClass = "hep-th",
    reportNumber = "YITP-25-101",
    doi = "10.1103/9v9v-54zv",
    journal = "Phys. Rev. Res.",
    volume = "7",
    number = "4",
    pages = "043055",
    year = "2025"
}

@article{Lin:2022rbf,
    author = "Lin, Henry W.",
    title = "{The bulk Hilbert space of double scaled SYK}",
    eprint = "2208.07032",
    archivePrefix = "arXiv",
    primaryClass = "hep-th",
    doi = "10.1007/JHEP11(2022)060",
    journal = "JHEP",
    volume = "11",
    pages = "060",
    year = "2022"
}

@article{Camargo:2018eof,
    author = "Camargo, Hugo A. and Caputa, Pawel and Das, Diptarka and Heller, Michal P. and Jefferson, Ro",
    title = "{Complexity as a novel probe of quantum quenches: universal scalings and purifications}",
    eprint = "1807.07075",
    archivePrefix = "arXiv",
    primaryClass = "hep-th",
    doi = "10.1103/PhysRevLett.122.081601",
    journal = "Phys. Rev. Lett.",
    volume = "122",
    number = "8",
    pages = "081601",
    year = "2019"
}

@article{Ju:2026zbu,
    author = "Ju, Xin-Xiang and Zhao, Yang",
    title = "{The Holographic Multi-Entropy Cone}",
    eprint = "2606.15173",
    archivePrefix = "arXiv",
    primaryClass = "hep-th",
    month = "6",
    year = "2026"
}

@article{Naskar:2026zka,
    author = "Naskar, Joydeep",
    title = "{On a mixed-state extension of the holographic signal inequality}",
    eprint = "2605.26617",
    archivePrefix = "arXiv",
    primaryClass = "hep-th",
    month = "5",
    year = "2026"
}

@article{Banerjee:2025mzc,
    author = "Banerjee, Aditya",
    title = "{Non-Markovianity of subsystem dynamics in isolated quantum many-body systems}",
    eprint = "2501.18476",
    archivePrefix = "arXiv",
    primaryClass = "quant-ph",
    doi = "10.1103/q8h3-mkzr",
    journal = "Phys. Rev. B",
    volume = "112",
    number = "1",
    pages = "014302",
    year = "2025"
}

@article{Balasubramanian:2018hsu,
    author = "Balasubramanian, Vijay and DeCross, Matthew and Kar, Arjun and Parrikar, Onkar",
    title = "{Binding Complexity and Multiparty Entanglement}",
    eprint = "1811.04085",
    archivePrefix = "arXiv",
    primaryClass = "hep-th",
    doi = "10.1007/JHEP02(2019)069",
    journal = "JHEP",
    volume = "02",
    pages = "069",
    year = "2019"
}

@article{Fujiki:2026ucr,
    author = "Fujiki, Kosei and Harper, Jonathan and Takayanagi, Tadashi and Zenoni, Nicol{\`o}",
    title = "{The Entanglement Wedge Polygon}",
    eprint = "2606.21081",
    archivePrefix = "arXiv",
    primaryClass = "hep-th",
    reportNumber = "YITP-26-76",
    month = "6",
    year = "2026"
}

@article{Coppola:2025qxk,
    author = "Coppola, Michele and Ba{\~n}uls, Mari Carmen and Lenar{\v{c}}i{\v{c}}, Zala",
    title = "{Learning the non-Markovian features of subsystem dynamics}",
    eprint = "2507.14133",
    archivePrefix = "arXiv",
    primaryClass = "cond-mat.stat-mech",
    doi = "10.21468/SciPostPhys.19.6.149",
    journal = "SciPost Phys.",
    volume = "19",
    number = "6",
    pages = "149",
    year = "2025"
}

@article{Banerjee:2025uew,
    author = "Banerjee, Aditya",
    title = "{Quantum scarring enhances non-Markovianity of subsystem dynamics}",
    eprint = "2507.23757",
    archivePrefix = "arXiv",
    primaryClass = "quant-ph",
    doi = "10.1088/1367-2630/ae5ea1",
    journal = "New J. Phys.",
    volume = "28",
    number = "5",
    pages = "054502",
    year = "2026"
}

@article{Gerbershagen:2024qlz,
    author = "Gerbershagen, Marius and Hernandez, Juan and Khramtsov, Mikhail and Knysh, Maria",
    title = "{Holographic dual of Bures metric and subregion complexity}",
    eprint = "2412.08707",
    archivePrefix = "arXiv",
    primaryClass = "hep-th",
    doi = "10.1007/JHEP04(2025)059",
    journal = "JHEP",
    volume = "04",
    pages = "059",
    year = "2025"
}

@book{Nielsen:2012yss,
    author = "Nielsen, Michael A. and Chuang, Isaac L.",
    title = "{Quantum Computation and Quantum Information}",
    doi = "10.1017/cbo9780511976667",
    isbn = "978-0-521-63503-5",
    publisher = "Cambridge University Press",
    month = "6",
    year = "2012"
}

@book{amaribook,
    author = "Shun-ichi Amari",
    title = "{Information Geometry and Its Applications}",
    doi = "10.1007/978-4-431-55978-8",
    isbn = "978-4-431-55978-8",
    publisher = "Springer Tokyo",
    year = "2016"
}

@article{Ryu:2006bv,
    author = "Ryu, Shinsei and Takayanagi, Tadashi",
    title = "{Holographic derivation of entanglement entropy from AdS/CFT}",
    eprint = "hep-th/0603001",
    archivePrefix = "arXiv",
    reportNumber = "NSF-KITP-06-11, NSF-KITP-06-11",
    doi = "10.1103/PhysRevLett.96.181602",
    journal = "Phys. Rev. Lett.",
    volume = "96",
    pages = "181602",
    year = "2006"
}

@article{Dyson:1962es,
    author = "Dyson, F. J.",
    title = "{Statistical theory of the energy levels of complex systems. I}",
    doi = "10.1063/1.1703773",
    journal = "J. Math. Phys.",
    volume = "3",
    pages = "140--156",
    year = "1962"
}

@article{Serafini:2007fqx,
    author = "Serafini, Alessio and Dahlsten, Oscar C. O. and Plenio, Martin B.",
    title = "{Teleportation fidelities of squeezed states from thermodynamical state space measures}",
    eprint = "quant-ph/0610090",
    archivePrefix = "arXiv",
    doi = "10.1103/PhysRevLett.98.170501",
    journal = "Phys. Rev. Lett.",
    volume = "98",
    pages = "170501",
    year = "2007"
}

@article{Serafini:2007ohw,
    author = "Serafini, A. and Dahlsten, O. C. O. and Gross, D. and Plenio, M. B.",
    title = "{Canonical and micro-canonical typical entanglement of continuous variable systems}",
    eprint = "quant-ph/0701051",
    archivePrefix = "arXiv",
    doi = "10.1088/1751-8113/40/31/027",
    journal = "J. Phys. A",
    volume = "40",
    pages = "9551",
    year = "2007"
}

@article{Fukuda:2019ycv,
    author = "Fukuda, Motohisa and Koenig, Robert",
    title = "{Typical entanglement for Gaussian states}",
    eprint = "1903.04126",
    archivePrefix = "arXiv",
    primaryClass = "quant-ph",
    doi = "10.1063/1.5119950",
    journal = "J. Math. Phys.",
    volume = "60",
    pages = "112203",
    year = "2019"
}

@article{Bohigas:1983er,
    author = "Bohigas, O. and Giannoni, M. J. and Schmit, C.",
    title = "{Characterization of chaotic quantum spectra and universality of level fluctuation laws}",
    doi = "10.1103/PhysRevLett.52.1",
    journal = "Phys. Rev. Lett.",
    volume = "52",
    pages = "1--4",
    year = "1984"
}

@article{Tiutiakina:2023ilu,
    author = "Tiutiakina, Anastasiia and De Luca, Andrea and De Nardis, Jacopo",
    title = "{Frame potential of Brownian SYK model of Majorana and Dirac fermions}",
    eprint = "2306.11160",
    archivePrefix = "arXiv",
    primaryClass = "cond-mat.dis-nn",
    doi = "10.1007/JHEP01(2024)115",
    journal = "JHEP",
    volume = "01",
    pages = "115",
    year = "2024"
}

@article{Calabrese:2004eu,
    author = "Calabrese, Pasquale and Cardy, John L.",
    title = "{Entanglement entropy and quantum field theory}",
    eprint = "hep-th/0405152",
    archivePrefix = "arXiv",
    doi = "10.1088/1742-5468/2004/06/P06002",
    journal = "J. Stat. Mech.",
    volume = "0406",
    pages = "P06002",
    year = "2004"
}

@article{Holzhey:1994we,
    author = "Holzhey, Christoph and Larsen, Finn and Wilczek, Frank",
    title = "{Geometric and renormalized entropy in conformal field theory}",
    eprint = "hep-th/9403108",
    archivePrefix = "arXiv",
    reportNumber = "PUPT-1454, IASSNS-HEP-93-88",
    doi = "10.1016/0550-3213(94)90402-2",
    journal = "Nucl. Phys. B",
    volume = "424",
    pages = "443--467",
    year = "1994"
}

@article{Hubeny:2007xt,
    author = "Hubeny, Veronika E. and Rangamani, Mukund and Takayanagi, Tadashi",
    title = "{A Covariant holographic entanglement entropy proposal}",
    eprint = "0705.0016",
    archivePrefix = "arXiv",
    primaryClass = "hep-th",
    reportNumber = "DCPT-07-13, KUNS-2069",
    doi = "10.1088/1126-6708/2007/07/062",
    journal = "JHEP",
    volume = "07",
    pages = "062",
    year = "2007"
}

@article{Susskind:2014rva,
    author = "Susskind, Leonard",
    title = "{Computational Complexity and Black Hole Horizons}",
    eprint = "1403.5695",
    archivePrefix = "arXiv",
    primaryClass = "hep-th",
    doi = "10.1002/prop.201500092",
    journal = "Fortsch. Phys.",
    volume = "64",
    pages = "24--43",
    year = "2016",
    note = "[Addendum: Fortsch.Phys. 64, 44--48 (2016)]"
}

@article{Bao:2015bfa,
    author = "Bao, Ning and Nezami, Sepehr and Ooguri, Hirosi and Stoica, Bogdan and Sully, James and Walter, Michael",
    title = "{The Holographic Entropy Cone}",
    eprint = "1505.07839",
    archivePrefix = "arXiv",
    primaryClass = "hep-th",
    reportNumber = "CALT-TH-2015-020, IPMU15-0074, SLAC-PUB-16294, SU-ITP-15-08, CALT-TH 2015-020, IPMU15-0074, SLAC-PUB-16294, SU-ITP-15/08",
    doi = "10.1007/JHEP09(2015)130",
    journal = "JHEP",
    volume = "09",
    pages = "130",
    year = "2015"
}

@article{Cardy:2014rqa,
    author = "Cardy, John",
    title = "{Thermalization and Revivals after a Quantum Quench in Conformal Field Theory}",
    eprint = "1403.3040",
    archivePrefix = "arXiv",
    primaryClass = "cond-mat.stat-mech",
    reportNumber = "NSF-KITP-14-021",
    doi = "10.1103/PhysRevLett.112.220401",
    journal = "Phys. Rev. Lett.",
    volume = "112",
    pages = "220401",
    year = "2014"
}

@book{Serafini:2017rrn,
    author = "Serafini, Alessio",
    title = "{Quantum Continuous Variables}",
    doi = "10.1201/9781315118727",
    isbn = "978-1-315-11872-7, 978-1-003-25097-5",
    publisher = "CRC Press",
    month = "10",
    year = "2017"
}

@article{Weedbrook:2011wxo,
    author = "Weedbrook, Christian and Pirandola, Stefano and Garc{\'\i}a-Patr{\'o}n, Ra{\'u}l and Cerf, Nicolas J. and Ralph, Timothy C. and Shapiro, Jeffrey H. and Lloyd, Seth",
    title = "{Gaussian quantum information}",
    eprint = "1110.3234",
    archivePrefix = "arXiv",
    primaryClass = "quant-ph",
    doi = "10.1103/RevModPhys.84.621",
    journal = "Rev. Mod. Phys.",
    volume = "84",
    number = "2",
    pages = "621",
    year = "2012"
}

@article{DiGiulio:2021noo,
    author = "Di Giulio, Giuseppe and Tonni, Erik",
    title = "{Subsystem complexity after a local quantum quench}",
    eprint = "2106.08282",
    archivePrefix = "arXiv",
    primaryClass = "hep-th",
    doi = "10.1007/JHEP08(2021)135",
    journal = "JHEP",
    volume = "08",
    pages = "135",
    year = "2021"
}

@article{Perelomov:1971bd,
    author = "Perelomov, A. M.",
    title = "{Coherent states for arbitrary lie groups}",
    doi = "10.1007/BF01645091",
    journal = "Commun. Math. Phys.",
    volume = "26",
    pages = "222--236",
    year = "1972"
}

@article{Nielsen:2006cea,
    author = "Nielsen, Michael A. and Dowling, Mark R. and Gu, Mile and Doherty, Andrew C.",
    title = "{Quantum Computation as Geometry}",
    eprint = "quant-ph/0603161",
    archivePrefix = "arXiv",
    doi = "10.1126/science.1121541",
    journal = "Science",
    volume = "311",
    number = "5764",
    pages = "1133--1135",
    year = "2006"
}

@article{Jefferson:2017sdb,
    author = "Jefferson, Ro and Myers, Robert C.",
    title = "{Circuit complexity in quantum field theory}",
    eprint = "1707.08570",
    archivePrefix = "arXiv",
    primaryClass = "hep-th",
    doi = "10.1007/JHEP10(2017)107",
    journal = "JHEP",
    volume = "10",
    pages = "107",
    year = "2017"
}

@article{Caputa:2017yrh,
    author = "Caputa, Pawel and Kundu, Nilay and Miyaji, Masamichi and Takayanagi, Tadashi and Watanabe, Kento",
    title = "{Liouville Action as Path-Integral Complexity: From Continuous Tensor Networks to AdS/CFT}",
    eprint = "1706.07056",
    archivePrefix = "arXiv",
    primaryClass = "hep-th",
    reportNumber = "YITP-17-65, IPMU17-0091",
    doi = "10.1007/JHEP11(2017)097",
    journal = "JHEP",
    volume = "11",
    pages = "097",
    year = "2017"
}

@article{Stanford:2014jda,
    author = "Stanford, Douglas and Susskind, Leonard",
    title = "{Complexity and Shock Wave Geometries}",
    eprint = "1406.2678",
    archivePrefix = "arXiv",
    primaryClass = "hep-th",
    doi = "10.1103/PhysRevD.90.126007",
    journal = "Phys. Rev. D",
    volume = "90",
    number = "12",
    pages = "126007",
    year = "2014"
}

@book{LanczosBook,
	address = {Heidelberg, Germany},
	author = {V. S. Viswanath and Gerhard M{\"u}ller},
	publisher = {Springer Berlin},
	title = {The Recursion Method: Application to Many-Body Dynamics},
	year = {1994}}

@article{Nandy:2024evd,
    author = "Nandy, Pratik and Matsoukas-Roubeas, Apollonas S. and Mart{\'\i}nez-Azcona, Pablo and Dymarsky, Anatoly and del Campo, Adolfo",
    title = "{Quantum dynamics in Krylov space: Methods and applications}",
    eprint = "2405.09628",
    archivePrefix = "arXiv",
    primaryClass = "quant-ph",
    reportNumber = "RIKEN-iTHEMS-Report-24",
    doi = "10.1016/j.physrep.2025.05.001",
    journal = "Phys. Rept.",
    volume = "1125-1128",
    pages = "1--82",
    year = "2025"
}

@article{Auzzi:2019mah,
    author = "Auzzi, Roberto and Nardelli, Giuseppe and Schaposnik Massolo, Fidel I. and Tallarita, Gianni and Zenoni, Nicol{\`o}",
    title = "{On volume subregion complexity in Vaidya spacetime}",
    eprint = "1908.10832",
    archivePrefix = "arXiv",
    primaryClass = "hep-th",
    doi = "10.1007/JHEP11(2019)098",
    journal = "JHEP",
    volume = "11",
    pages = "098",
    year = "2019"
}

@article{Chen:2018mcc,
    author = "Chen, Bin and Li, Wen-Ming and Yang, Run-Qiu and Zhang, Cheng-Yong and Zhang, Shao-Jun",
    title = "{Holographic subregion complexity under a thermal quench}",
    eprint = "1803.06680",
    archivePrefix = "arXiv",
    primaryClass = "hep-th",
    doi = "10.1007/JHEP07(2018)034",
    journal = "JHEP",
    volume = "07",
    pages = "034",
    year = "2018"
}

@article{Alba:2017ekd,
    author = "Alba, Vincenzo and Calabrese, Pasquale",
    title = "{Entanglement and thermodynamics after a quantum quench in integrable systems}",
    eprint = "1608.00614",
    archivePrefix = "arXiv",
    primaryClass = "cond-mat.str-el",
    doi = "10.1073/pnas.1703516114",
    journal = "Proc. Nat. Acad. Sci.",
    volume = "114",
    number = "30",
    pages = "7947",
    year = "2017"
}

@article{Ruan:2020vze,
    author = "Ruan, Shan-Ming",
    title = "{Purification Complexity without Purifications}",
    eprint = "2006.01088",
    archivePrefix = "arXiv",
    primaryClass = "hep-th",
    doi = "10.1007/JHEP01(2021)092",
    journal = "JHEP",
    volume = "01",
    pages = "092",
    year = "2021"
}

@article{Carmi:2016wjl,
    author = "Carmi, Dean and Myers, Robert C. and Rath, Pratik",
    title = "{Comments on Holographic Complexity}",
    eprint = "1612.00433",
    archivePrefix = "arXiv",
    primaryClass = "hep-th",
    doi = "10.1007/JHEP03(2017)118",
    journal = "JHEP",
    volume = "03",
    pages = "118",
    year = "2017"
}

\end{document}